\documentclass[journal]{IEEEtran}
\usepackage{ifpdf}
\usepackage{url}
\ifpdf
\else
\fi

\usepackage{cite}
\usepackage{balance}
\usepackage{bm,comment,color}

\ifCLASSINFOpdf
  \usepackage[pdftex]{graphicx}
  \graphicspath{{../pdf/}{../jpeg/}}
  \DeclareGraphicsExtensions{.pdf,.jpeg,.png}
\else
  \usepackage[dvips]{graphicx}
  \graphicspath{{../eps/}}
  \DeclareGraphicsExtensions{.eps}
\fi
\usepackage{amsmath}
\usepackage{algpseudocode}
\algtext*{EndWhile}
\algtext*{EndIf}
\algtext*{EndFor}
\usepackage{algorithm}
\usepackage{array}
\usepackage{amsfonts} 
\usepackage{amssymb}
\usepackage{esint} 
\usepackage{units}
\usepackage{multirow}
\usepackage{comment}

\usepackage{color}

\usepackage{pgfplots}
\usepackage{tikz}
\usetikzlibrary{calc}
\makeatletter
\newcommand{\gettikzxy}[3]{%
  \tikz@scan@one@point\pgfutil@firstofone#1\relax
  \edef#2{\the\pgf@x}%
  \edef#3{\the\pgf@y}%
}
\usetikzlibrary{spy,backgrounds}
\pgfplotsset{compat=newest}
\usetikzlibrary{plotmarks}
\usetikzlibrary{arrows.meta}
\usepgfplotslibrary{patchplots}
\usepackage{grffile}
\newlength\fheight 
\newlength\fwidth 
\usepgfplotslibrary{fillbetween}

\usepackage{acro}

\DeclareAcronym{5g}{
short=5G,
long= fifth generation,
}
\DeclareAcronym{6g}{
short=6G,
long= sixth generation,
}
\DeclareAcronym{3d}{
short=3D,
long= three-dimensional,
}
\DeclareAcronym{3gpp}{
short=3GPP,
long= 3rd Generation Partnership Project,
}
\DeclareAcronym{ace}{
short=ACE,
long= array calibration error,
}
\DeclareAcronym{alb}{
short=ALB,
long= absolute lower bound,
}
\DeclareAcronym{aod}{
short=AOD,
long= angle-of-departure,
}
\DeclareAcronym{aoa}{
short=AOA,
long= angle-of-arrival,
}
\DeclareAcronym{adc}{
short=ADC,
long= analog-to-digital converter,
}
\DeclareAcronym{aeb}{
short=AEB,
long= angle error bound,
}
\DeclareAcronym{age}{
short=AGE,
long= array gain error,
}
\DeclareAcronym{ade}{
short=ADE,
long= antenna displacement error,
}
\DeclareAcronym{bs}{
short=BS,
long= base station,
}
\DeclareAcronym{cfo}{
short=CFO,
long= carrier frequency offset,
}
\DeclareAcronym{ceb}{
short=CEB,
long= clock error bound,
}
\DeclareAcronym{coa}{
short=COA,
long= curvature-of-arrival,
}
\DeclareAcronym{crb}{
short=CRB,
long= Cram\'er-Rao bound,
}
\DeclareAcronym{ccrb}{
short=CCRB,
long= constrained Cram\'er-Rao bound,
}
\DeclareAcronym{crlb}{
short=CRLB,
long= Cram\'er-Rao lower bound,
}
\DeclareAcronym{cp}{
short=CP,
long= cyclic prefix,
}
\DeclareAcronym{dac}{
short=DAC,
long= digital-to-analog converter,
}
\DeclareAcronym{dftsofdm}{
short=DFT-s-OFDM,
long= discrete-Fourier-transform spread OFDM,
}
\DeclareAcronym{dl}{
short=DL,
long= deep learning,
}
\DeclareAcronym{fim}{
short=FIM,
long= Fisher information matrix,
}
\DeclareAcronym{gps}{
short=GPS,
long= global positioning system,
}
\DeclareAcronym{gnss}{
short=GNSS,
long= global navigation satellite system,
}
\DeclareAcronym{hwi}{
short=HWI,
long= hardware impairment,
}
\DeclareAcronym{isac}{
short=ISAC,
long= integrated sensing and communication,
}
\DeclareAcronym{iqi}{
short=IQI,
long= in-phase and quadrature imbalance,
}
\DeclareAcronym{kpi}{
short=KPI,
long= key performance indicator,
}
\DeclareAcronym{kf}{
short=KF,
long= Kalman filter,
}
\DeclareAcronym{ekf}{
short=EKF,
long= extended Kalman filter,
}
\DeclareAcronym{ukf}{
short=UKF,
long= unscented Kalman filter,
}
\DeclareAcronym{ckf}{
short=CKF,
long= cubature Kalman filter,
}
\DeclareAcronym{pf}{
short=PF,
long= particle filter,
}
\DeclareAcronym{lb}{
short=LB,
long= lower bound,
}
\DeclareAcronym{lse}{
short=LSE,
long= least-square estimator,
}
\DeclareAcronym{lo}{
short=LO,
long= local oscillator,
}
\DeclareAcronym{los}{
short=LOS,
long= line-of-sight,
}
\DeclareAcronym{mc}{
short=MC,
long= mutual coupling,
}
\DeclareAcronym{mac}{
short=MAC,
long= medium access control,
}
\DeclareAcronym{meb}{
short=MEB,
long= mapping error bound,
}
\DeclareAcronym{ml}{
short=ML,
long= machine learning,
}
\DeclareAcronym{mcrb}{
short=MCRB,
long= misspecified Cram\'er-Rao bound,
}
\DeclareAcronym{mimo}{
short=MIMO,
long= multiple-input-multiple-output,
}
\DeclareAcronym{mm}{
short=MM,
long= mismatched model,
}
\DeclareAcronym{mpc}{
short=MPC,
long= multipath component,
}
\DeclareAcronym{mmwave}{
short=mmWave,
long= millimeter wave,
}
\DeclareAcronym{mmle}{
short=MMLE,
long= mismatched maximum likelihood estimation,
}
\DeclareAcronym{mems}{
short=MEMS,
long= micro-electro-mechanical system,
}
\DeclareAcronym{mle}{
short=MLE,
long= maximum likelihood estimation,
}
\DeclareAcronym{nlos}{
short=NLOS,
long= none-line-of-sight,
}
\DeclareAcronym{ofdm}{
short=OFDM,
long= orthogonal frequency-division multiplexing,
}
\DeclareAcronym{oeb}{
short=OEB,
long= orientation error bound,
}
\DeclareAcronym{otfs}{
short=OTFS,
long= orthogonal time-frequency space,
}
\DeclareAcronym{pdf}{
short=PDF,
long= probability density function,
}
\DeclareAcronym{papr}{
short=PAPR,
long= peak-to-average-power ratio,
}
\DeclareAcronym{pan}{
short=PAN,
long= power amplifier nonlinearity,
}
\DeclareAcronym{pa}{
short=PA,
long= power amplifier,
}
\DeclareAcronym{ps}{
short=PS,
long= phase shifter,
}
\DeclareAcronym{pn}{
short=PN,
long= phase noise,
}
\DeclareAcronym{poa}{
short=POA,
long= phase-of-arrival,
}
\DeclareAcronym{pwm}{
short=PWM,
long= planar wave model,
}
\DeclareAcronym{pdoa}{
short=PDOA,
long= phase-difference-of-arrival,
}
\DeclareAcronym{prs}{
short=PRS,
long= positioning reference signals,
}
\DeclareAcronym{peb}{
short=PEB,
long= position error bound,
}
\DeclareAcronym{qam}{
short=QAM,
long= quadrature amplitude modulation,
}
\DeclareAcronym{rnn}{
short=RNN,
long= recurrent neural network,
}
\DeclareAcronym{rl}{
short=RL,
long= reinforcement learning,
}
\DeclareAcronym{rfc}{
short=RFC,
long= radio-frequency chain,
}
\DeclareAcronym{rf}{
short=RF,
long= radio frequency,
}
\DeclareAcronym{ris}{
short=RIS,
long= reconfigurable intelligent surface,
}
\DeclareAcronym{sa}{
short=SA,
long= subarray,
}
\DeclareAcronym{ser}{
short=SER,
long= symbol error rate,
}
\DeclareAcronym{simo}{
short=SIMO,
long= single-input-multiple-output,
}
\DeclareAcronym{sota}{
short=SOTA,
long= state-of-the-art,
}
\DeclareAcronym{swm}{
short=SWM,
long= spherical wave model,
}
\DeclareAcronym{slam}{
short=SLAM,
long= simultaneous localization and mapping,
}
\DeclareAcronym{tm}{
short=TM,
long= true model,
}
\DeclareAcronym{toa}{
short=TOA,
long= time-of-arrival,
}
\DeclareAcronym{tof}{
short=TOF,
long= time-of-flight,
}
\DeclareAcronym{tdoa}{
short=TDOA,
long= time-difference-of-arrival,
}
\DeclareAcronym{thz}{
short=THz,
long= terahertz,
}
\DeclareAcronym{ue}{
short=UE,
long= user equipment,
}
\DeclareAcronym{ummimo}{
short=UM-MIMO,
long= ultra-massive multi-input-multi-output,
}
\DeclareAcronym{vlp}{
short=VLP,
long= visible light positioning,
}
\DeclareAcronym{veb}{
short=VEB,
long= velocity error bound,
}
\DeclareAcronym{vlc}{
short=VLC,
long= visible light communication,
}
\DeclareAcronym{ula}{
short=ULA,
long= uniform linear array,
}
\DeclareAcronym{upa}{
short=UPA,
long= uniform planar array,
}
\DeclareAcronym{wlan}{
short=WLAN,
long= wireless local area network,
}

\usepackage{tabu,longtable}

\ifCLASSOPTIONcompsoc
 \usepackage[caption=false,font=normalsize,labelfont=sf,textfont=sf]{subfig}
\else
 \usepackage[caption=false,font=footnotesize]{subfig}
\fi
\usepackage{stfloats}
\usepackage[hidelinks]{hyperref}
\usepackage{xcolor}
\long\def\comment#1{}

\DeclareMathOperator*{\argmin}{arg\,min}

\newfont{\bbb}{msbm10 scaled 700}

\newcommand{\hthickline}{\noalign{\hrule height 0.80pt}}

\newfont{\bb}{msbm10 scaled 1100}

\newcommand{\av}{{\bf a}}
\newcommand{\bv}{{\bf b}}
\newcommand{\cv}{{\bf c}}
\newcommand{\dv}{{\bf d}}

\newcommand{\fv}{{\bf f}}

\newcommand{\hv}{{\bf h}}

\newcommand{\nv}{{\bf n}}
\newcommand{\ov}{{\bf o}}
\newcommand{\pv}{{\bf p}}
\newcommand{\qv}{{\bf q}}
\newcommand{\rv}{{\bf r}}
\newcommand{\sv}{{\bf s}}
\newcommand{\tv}{{\bf t}}

\newcommand{\wv}{{\bf w}}
\newcommand{\vv}{{\bf v}}
\newcommand{\xv}{{\bf x}}
\newcommand{\yv}{{\bf y}}
\newcommand{\zv}{{\bf z}}

\newcommand{\Am}{{\bf A}}
\newcommand{\Bm}{{\bf B}}
\newcommand{\Cm}{{\bf C}}

\newcommand{\Em}{{\bf E}}
\newcommand{\Fm}{{\bf F}}

\newcommand{\Hm}{{\bf H}}

\newcommand{\Jm}{{\bf J}}

\newcommand{\Mm}{{\bf M}}

\newcommand{\Qm}{{\bf Q}}
\newcommand{\Rm}{{\bf R}}

\newcommand{\Tm}{{\bf T}}
\newcommand{\Um}{{\bf U}}
\newcommand{\Wm}{{\bf W}}
\newcommand{\Vm}{{\bf V}}
\newcommand{\Xm}{{\bf X}}

\newcommand{\Zm}{{\bf Z}}

\newcommand{\Hc}{{\cal H}}

\newcommand{\gammav}{\hbox{\boldmath$\gamma$}}

\newcommand{\etav}{\hbox{\boldmath$\eta$}}

\newcommand{\epsilonv}{\hbox{\boldmath$\epsilon$}}

\newcommand{\muv}{\hbox{\boldmath$\mu$}}

\newcommand{\phiv}{\hbox{\boldmath$\phi$}}

\newcommand{\omegav}{\hbox{\boldmath$\omega$}}

\newcommand{\varphiv}{\hbox{\boldmath$\varphi$}}
\newcommand{\vpv}{\boldsymbol{\varphi}}

\newcommand{\Sigmam}{\hbox{\boldmath$\Sigma$}}

\newcommand{\Xim}{\hbox{\boldmath$\Xi$}}

\newcommand{\trace}{{\hbox{tr}}}

\renewcommand{\arg}{{\hbox{arg}}}

\newcommand{\herm}{{\sf H}}

\makeatletter
\def\ps@IEEEtitlepagestyle{%
\def\@oddfoot{\mycopyrightnotice}%
\def\@evenfoot{}%
}
\def\mycopyrightnotice{%
{
\scriptsize {\copyright~2023 IEEE. Personal use of this material is permitted. Permission from IEEE must be obtained for all other uses. More details can be found in IEEE Post-publication policies.}
} 
\gdef\mycopyrightnotice{}
}

\makeatletter
\let\old@ps@headings\ps@headings
\let\old@ps@IEEEtitlepagestyle\ps@IEEEtitlepagestyle
\def\confheader#1{
\def\@oddhead{\strut\hfill#1\hfill\strut}%
}
\makeatother
\confheader{
\scriptsize \shortstack{This article has been accepted for publication in a future issue of this journal, but has not been fully edited. Content may change prior to final publication. 
\\ Citation information: 10.1109/TWC.2023.3339523, IEEE Transactions on Wireless Communications.}
}

\begin{document} 

\title{Modeling and Analysis of {OFDM-based 5G/6G Localization} under Hardware Impairments}



\author{

Hui~Chen,~\IEEEmembership{Member,~IEEE},
Musa~Furkan~Keskin,~\IEEEmembership{Member,~IEEE},
Sina~Rezaei~Aghdam,~\IEEEmembership{Member,~IEEE},
Hyowon~Kim,~\IEEEmembership{Member,~IEEE},
Simon~Lindberg,~\IEEEmembership{Member,~IEEE},
Andreas~Wolfgang,~\IEEEmembership{Member,~IEEE},
Traian~E.~Abrudan,~\IEEEmembership{Member,~IEEE},
Thomas~Eriksson,~\IEEEmembership{Senior~Member,~IEEE},
and~Henk~Wymeersch,~\IEEEmembership{Fellow,~IEEE}
\thanks{H.~Chen, M.~F.~Keskin, S.~R.~Aghdam, T.~Eriksson and H.~Wymeersch are with the Department of Electrical Engineering, Chalmers University of Technology, 412 58 Gothenburg, Sweden (email: {hui.chen; furkan; sinar; thomase; henkw}@chalmers.se).}
\thanks{H. Kim is with the Department of Electronics Engineering, Chungnam National University, 34134 Daejeon, South Korea (email: hyowon.kim@cnu.ac.kr).}
\thanks{S.~Lindberg and A.~Wolfgang are with Qamcom Research \& Technology, Gothenburg, Sweden (email: {simon.lindberg; andreas.wolfgang}@qamcom.se).}
\thanks{T.~E.~Abrudan is with Nokia Bell Labs, Finland (email: traian.abrudan@nokia-bell-labs.com).}
\thanks{This work was supported, in part, by the European Commission through the H2020 project Hexa-X (Grant Agreement no. 101015956) and by the Swedish Research Council (VR grant 2022-03007).}

}


\maketitle

\begin{abstract}
Localization is envisioned as a key enabler to satisfy the requirements of {communications} and context-aware services in the fifth/sixth generation (5G/6G) communication systems. User localization can be achieved based on delay and angle estimation using uplink/downlink pilot signals. However, hardware impairments (HWIs) (such as phase noise and mutual coupling) distort the signals at both the transmitter and receiver sides and thus affect the localization performance. While this impact can be ignored at lower frequencies with less severe HWIs, and less stringent localization requirements, modeling and analysis efforts are needed for high-frequency bands to assess degradation in localization accuracy due to HWIs. In this work, we model various types of impairments for a mmWave multiple-input-multiple-output communication system and conduct a misspecified Cram\'er-Rao bound analysis to evaluate HWI-induced performance losses in terms of angle/delay estimation and the resulting 3D position/orientation estimation error. We also investigate the effect of individual and overall HWIs on {communications} in terms of symbol error rate (SER). Our extensive simulation results demonstrate that each type of HWI leads to a different level of degradation in angle and delay estimation performance, and the prominent impairment factors on delay estimation will have a dominant negative effect on SER.
\end{abstract}

\IEEEpeerreviewmaketitle
\acresetall 

\section{Introduction}
\label{sec:intro}
Localization refers to the process of estimating the position {(and possibly orientation)} of a \ac{ue}, which is expected to have a tight interaction with {communications} in future wireless systems~\cite{chen2022tutorial}. {More specifically,} localization can benefit from a large array dimension and wide bandwidth of high-frequency signals (e.g., mmWave and sub-THz) {provided by communication infrastructure}~\cite{behravan2022positioning}. In return, the position and orientation information can improve spatial efficiency and optimize resource allocation for {communications}~\cite{di2014location}. As a result, high-accuracy context-aware applications such as the tactile internet, augmented reality, and smart cities will be supported in future wireless networks~\cite{kwon2021joint, xiao2020overview}.

In \acp{gnss} and traditional cellular networks, range-based algorithms, such as trilateration, are usually applied for estimating position. When moving to higher carrier frequencies, more antennas can be packed in a single array due to shorter wavelengths. As a consequence, in addition to delay estimation, \ac{aoa} and \ac{aod} information can be exploited for localization, and a variety of new localization techniques have recently emerged in the {fifth/sixth generation (5G/6G)} systems, e.g., \cite{wen20205g,ge20205g,nazari2022mmwave,han2015performance}, considering localization with minimal infrastructure requirements. 
When the \ac{ue} is equipped with an antenna array, orientation estimation is also possible~\cite{nazari2022mmwave}. In Doppler-assisted localization, although new unknowns (e.g., velocity) are introduced, localization performance can be improved because mobility forms a virtual array with a large aperture compared to the stationary scenarios~\cite{han2015performance}. 
Most localization works rely on idealized models of the received signals as a function of the channel parameters (angles, delays, Dopplers) induced by the propagation environment, based on the assumption of deterministic and sparse channels in high-frequency systems~\cite{abu2018error, nazari2022mmwave, chen2022tutorial, wen2018tensor, wen20205g, elzanaty2021reconfigurable}. In reality, however, pilot signals can be distorted due to the presence of \acp{hwi} such as \ac{pn}, \ac{cfo}, \ac{mc}, \ac{pan}, \ac{age}, \ac{ade}, \ac{iqi}, etc~\cite{schenk2008rf}. Consequently, when algorithm derivation is based on a mismatched model (i.e., without considering the \acp{hwi} in the channel model), the localization performance is unavoidably affected.

{The effect of \acp{hwi} on {communications} have been studied extensively in the literature~\cite{schenk2008rf, jacobsson2018massive, kolawole2018impact, shen2020beamforming}. In \cite{schenk2008rf}, different types of impairments have been accurately modeled and the effects on a \ac{mimo}-\ac{ofdm} system are discussed. In~\cite{jacobsson2018massive}, an aggregate statistical HWI model considering \ac{pan}, local oscillators with \ac{pn}, and finite-resolution \acp{adc} is derived and validated with numerical simulations. The residual additive transceiver hardware impairments, caused by direct current offset, \ac{mc}, \ac{iqi} and quantization noise, are discussed in~\cite{kolawole2018impact}, with the derived spectral efficiency to quantify the degradation caused by the HWIs.
In addition to modeling and analysis of the HWIs, research has also been conducted on impairment mitigation algorithms. By incorporating signal distortions caused by hardware impairments, beamforming optimization is performed to maximize the received SNR at the destination~\cite{shen2020beamforming}. A channel estimation algorithm is designed by taking into account the transceiver impairments in~\cite{wu2019efficient}, {showing superior performance than the conventional algorithms in terms of bit error rate and normalized mean-squared-error}. Contrary to model-based solutions, channel estimation under HWI can also be formulated as a deep learning problem~\cite{yassine2022mpnet}. 
Nevertheless, these works focus only on communication performance.

Research on localization and sensing (here, sensing includes detection, angle, and delay estimation, as well as tracking of passive targets) considering \acp{hwi} is recently drawing attention. 
The effect of \ac{pn} on monostatic sensing~\cite{keskin2023monostatic,gerstmair2019safe},
\ac{mc} on \ac{aoa} estimation~\cite{ye2009doa}, \ac{iqi} on mmWave localization~\cite{ghaseminajm2020localization}, and \ac{pan} on joint radar-communication systems \cite{bozorgi2021rf} have been studied. However, these works only consider one or two types of impairments and cannot provide a thorough analysis in real scenarios.
{In~\cite{tubail2022error, ceniklioglu2022error}, the impairments are modeled as additional Gaussian noise, with the variance determined by an ad hoc HWI factor, from which the error bounds for 3D localization are discussed. The effects of HWIs considering clock synchronization error are discussed in~\cite{ghaseminajm2022error}. However, the approaches in \cite{tubail2022error, ceniklioglu2022error,ghaseminajm2022error} fail to capture the contribution of each individual HWI.}
In~\cite{chen2022mcrb}, which forms the basis of the current paper, a simplified synchronized \ac{simo} uplink system is considered for 2D positioning, and the results show that different types of impairments affect angle and delay estimation in different ways. Nevertheless, the perfect synchronization assumption is impractical, and the impairments such as array calibration error and \ac{iqi} are not considered.
%
%
%
{Besides analyzing the effect of HWIs on localization or {communications} alone, more recent works consider the HWIs in joint localization and communication systems and use learning-based methods to mitigate the performance degradation~\cite{mateos2022end, sankhe2019no}. Nevertheless, only a limited number of impairment types are discussed (\ac{mc} and \ac{ade} in~\cite{mateos2022end}, \ac{iqi} and DC offset in~\cite{sankhe2019no}). In addition, no theoretical analysis is performed in these works, and the relative importance of each HWI on {communications} compared to localization is unknown. Hence, there is a need for a more systematic study that evaluates the effect of different types of HWI on both communication and localization performance.}
}

In this paper, we investigate the problem of estimating the 3D position and 3D orientation of a multiple-antenna UE using several multiple-antenna BSs (a typical uplink localization scenario) in a mmWave communication system under a wide variety of \acp{hwi}.
Specifically, we consider an \ac{ofdm}-based system by rigorously modeling the impact of various \acp{hwi} on the received observations, and assume that the corresponding channel estimation and localization algorithms have no knowledge about these \acp{hwi}, resulting in degradation of localization and communication performance.
The \ac{mcrb}~\cite{richmond2015parameter, fortunati2017performance, ozturk2022ris} is employed to quantify the estimation performance loss due to model mismatch. In addition, the effect of HWI on {communications} is evaluated numerically in terms of \ac{ser} based on the developed model for a hardware-impaired channel under the same HWI levels, which allows a fair comparison of the impact of HWI on communication and localization performance.
The contributions are summarized as follows:
\begin{itemize}
\item \textbf{Channel model with multiple HWIs:} Based on the ideal MIMO model ({\ac{mm}}) with perfect hardware, we develop a more general channel model for the considered mmWave system (\ac{tm}) that can accommodate a variety of HWI types (including \ac{pn}, \ac{cfo}, \ac{mc}, \ac{pan}, \ac{age}, \ac{ade}, and \ac{iqi}) in a 3D environment. To the best of the authors' knowledge, this is the first study to derive a comprehensive and realistic signal model for localization and communications that provides explicit modeling of major HWIs that are likely to affect 5G/6G communication systems at high-frequency operation.


\item \textbf{Analytical performance prediction of channel parameter estimation and localization under HWIs:} We leverage MCRB analysis to evaluate the effect of individual and combined HWIs on the estimation of channel parameters (\ac{aod}, \ac{aoa} and delay estimation) and on the corresponding localization performance (3D position and 3D orientation estimation). 
{Such analysis quantifies the impact of different types and levels of HWIs on localization {\acp{kpi}} (e.g., position and orientation estimation error bounds), which can serve as guidelines for the accuracy requirements of HWI calibration and mitigation strategies in 5G/6G wireless systems supporting emerging applications.}

\item \textbf{Performance evaluation and comparison with communications:} 
Extensive simulations are performed to verify the performance analysis of the effect of HWI on localization and communication performance\footnote{{Note that a communication system does not require localization functions. Specifically, communication tasks require the estimation of the end-to-end channel, while localization tasks aim to extract the geometrical parameters from the channel (e.g., angle and delays that are used for estimating locations). However, the channel (including \acp{hwi}) is identical for both communications and localizations, which motivates this study. }}
{The performance evaluation of both localization and {communications} in the face of HWIs serves two purposes: i) determine a reasonable level of HWIs based on the SERs under different levels of HWIs {(the evaluation of HWI-introduced localization error on communications is beyond the focus of this work)}; ii) demonstrate the different impacts of certain HWIs on the performance of communication and localization, and provide valuable insights on the system design considering various \acp{kpi} (e.g., array calibration error cannot be ignored although it has limited effect on communications).}
We notice that the dominant factors that affect delay estimation will also affect communications, whereas the impairments that only affect \ac{aoa}, \ac{aod} have a limited impact on {communications}.
\end{itemize}

The rest of this paper is organized as follows. Section~II reviews the system models with and without HWIs. Section~III describes the channel estimation and localization algorithms. Theoretical performance analysis is carried out in Section IV. Next, the simulation results are presented in Section~V, followed by the concluding remarks in Section~VI.

\textit{Notations and Symbols:} Italic letters denote scalars (e.g. $a$), bold lower-case letters denote vectors (e.g. $\av$), and bold upper-case letters denote matrices (e.g. $\Am$). $(\cdot)^\top$, $(\cdot)^\herm$, $(\cdot)^*$, $(\cdot)^{-1}$, $\trace(\cdot)$, and $\Vert{\cdot}\Vert$ represent the transpose, Hermitian transpose, conjugate, inverse, trace, and $\ell$-2 norm operations, respectively; $\Am \odot \Bm$, $\Am\otimes \Bm$, $\av \circ \bv$ are the Hadamard product, Kronecker product, and outer product, respectively; $[\cdot,\ \cdot,\ \cdots, \cdot]^\top$ denotes a column vector; $\trace(\cdot)$ returns the trace of a matrix, $[\cdot]_{i,j}$ is the element in the $i$th row, $j$th column of a matrix, and $[\cdot]_{a:b,c:d}$ is the submatrix constructed from the $a$th to the $b$th row, and the $c$th to $d$th column of a matrix. 

\begin{figure*}
    \centering
    \begin{tikzpicture}
    \node (image) [anchor=south west]{\includegraphics[width=0.9\linewidth]{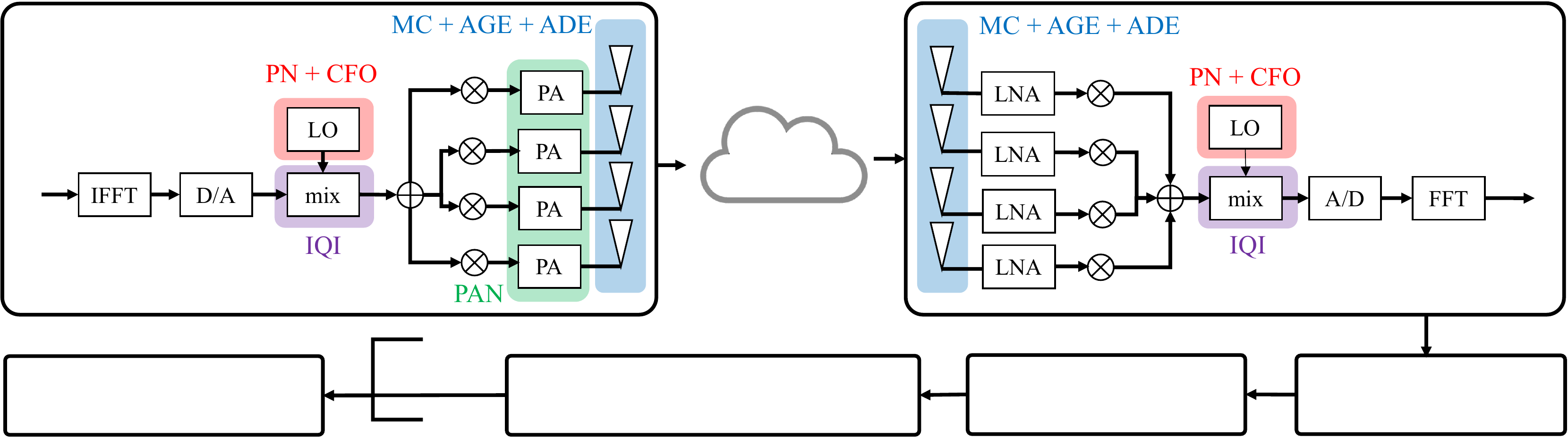}};
    \gettikzxy{(image.north east)}{\ix}{\iy};
    \node at (0.025*\ix,0.95*\iy)[rotate=0,anchor=north]{\scriptsize{UE}};
    \node at (0.021*\ix,0.59*\iy)[rotate=0,anchor=north]{\scriptsize{$\xv_g$}};
    \node at (0.96*\ix,0.95*\iy)[rotate=0,anchor=north]{\scriptsize{$l$th BS}};
    \node at (0.98*\ix,0.59*\iy)[rotate=0,anchor=north]{\scriptsize{$\yv_g$}};
    \node at (0.496*\ix,0.7*\iy)[rotate=0,anchor=north]{\scriptsize{\shortstack{Channel\\$\Hm_l$}}};
    \node at (0.11*\ix,0.21*\iy)[rotate=0,anchor=north]{\scriptsize{\shortstack{Estimated UE state:\\$\sv\! =\! [\pv_\text{U}^\top, B_\text{U}, \text{vec}(\Rm_\text{U})^\top]^\top$}}}; 
    \node at (0.29*\ix,0.28*\iy)[rotate=0,anchor=north]{\scriptsize{\shortstack{$\etav_1$\\$\cdots$}}};
    \node at (0.29*\ix,0.1*\iy)[rotate=0,anchor=north]{\scriptsize{$\etav_L$}};
    \node at (0.46*\ix,0.213*\iy)[rotate=0,anchor=north]{\scriptsize{\shortstack{Channel parameter extraction:\\$\etav_l = [\phi_\text{B}, \theta_\text{B}, \phi_\text{U}, \theta_\text{U}, \tau, \rho, \xi]^\top$}}};
    \node at (0.704*\ix,0.18*\iy)[rotate=0,anchor=north]{\scriptsize{Estimated channel: $\hat \Hm_l$}};
    \node at (0.907*\ix,0.218*\iy)[rotate=0,anchor=north]{\scriptsize{\shortstack{Received symbol: \\$\hat\yv_l = [\yv_1^\top, \ldots, \yv_g^\top]^\top$}}};
    \end{tikzpicture}
    \caption{Block diagram of the hardware impairments considered at transmitter and receiver (highlighted in shaded regions). When the localization algorithm does not have perfect knowledge of the generative model, it operates under model mismatch. PN (phase noise), CFO (carrier frequency offset), MC (mutual coupling), PAN (power amplifier non-linearity), AGE (array gain error), ADE (antenna displacement error), and IQI (in-phase and quadrature imbalance) are considered.}
    \label{fig:block_diagram}\vspace{-5mm}
\end{figure*}
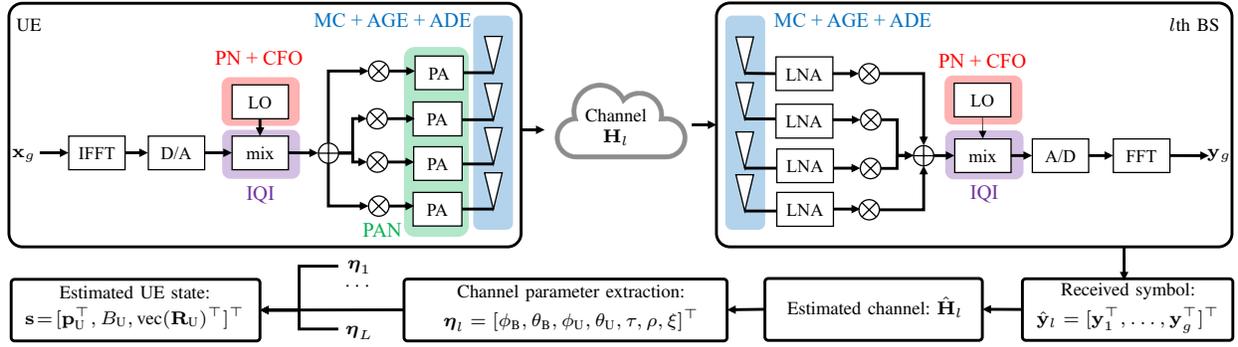



\section{System Model}
\label{sec-2-model}
In this section, we start with a \ac{mimo} channel model (\ac{hwi}-free model) and then describe the model considering the \ac{hwi}.

\subsection{Geometric Model}
\label{sec:geometric_model}
The block diagram of considered HWIs and localization procedures are shown in Fig.~\ref{fig:block_diagram}. An uplink \ac{mimo} system consisting of a \ac{ue} and $L$ \acp{bs} is considered. The BSs and UE are equipped with an \ac{upa} (antennas lie on the local YZ plane) driven by a single \ac{rfc}. The number of antenna elements at the $l$-th BS and the UE arrays is denoted as $N_{\text{B},l} = N_{\text{B},l,z}\times N_{\text{B},l,y}$ and $N_{\text{U}} = N_\text{U,z}\times N_\text{U,y}$ where $N_\text{z}$ and $N_\text{y}$ are the number of antennas on the Z and Y axes, respectively. The BSs are perfectly synchronized while a clock offset $B_\text{U}$ exists between the UE and the BSs. We denote the array center and orientation of the $l$-th BS as $\pv_{\text{B}, l} \in \mathbb{R}^3$ and $\ov_{\text{B},l} \in \mathbb{R}^3$ in the global coordinate system. Similarly, the position and orientation of the UE can be denoted as $\pv_\text{U}, \ov_\text{U}$. Since the orientation represented by an Euler angle vector is not unique, we use rotation matrices $\Rm_{\text{B}, l}\in \text{SO(3)}$ and $\Rm_\text{U}\in \text{SO(3)}$ in orientation estimation (more details can be found in~\cite{nazari2022mmwave, chen2022tutorial}).
{In localization, channel estimations are performed at each BS {(receiver)}, and all estimates are combined to estimate the UE {(transmitter)} \emph{state parameter vector} $\sv = [\pv_\text{U}^\top, B_\text{U}, \text{vec}(\Rm_\text{U})^\top]^\top \in \mathbb{R}^{13}$, containing the UE position $\pv_\text{U}$, clock offset $B_\text{U}$, and rotation matrix $\Rm_\text{U}$, as shown in Fig.~\ref{fig:block_diagram}.} {The single-UE setup can be extended to multi-UE scenarios by allocating orthogonal frequency and time resource blocks for positioning pilot signals~\cite{dwivedi2021positioning}.}

\subsection{Hardware Impairment-free Model}
\label{sec:hwi_free_model}
Considering the transmitted \ac{ofdm} symbol\footnote{For {communications}, different modulations (e.g., 16-QAM) can be adopted. While for positioning, constant modulus pilots are typically used~\cite{chen2022tutorial, wen20205g, ge20205g, nazari2022mmwave, abu2018error, nazari2022mmwave}.} at the $g$-th transmission and $k$-th subcarrier, $x_{g,k}$, with an average transmit power $\mathbb{E}\{ |x_{g,k}|^2\} = P/{N_\text{U}}$, its observation at a specific \ac{bs} (the index $l$ is omitted for convenience) 
can be formulated as
\begin{equation}
    y_{g,k} = \wv_{\text{}g}^\top \Hm_k \vv_{\text{}g} x_{g,k} + n_{g,k},
    \label{eq:ideal_far_field_model}
\end{equation}
where $\wv_{\text{}g} \in \mathbb{C}^{N_\text{B}}$ is the combiner at the BS for the \mbox{$g$-th} transmission and $\vv_{\text{}g} \in \mathbb{C}^{N_\text{U}}$ is the precoder at the UE, both with unit amplitude entries, {$n_{g,k}\in \mathcal{CN}(0, \wv_{\text{}g}^\herm \wv_{\text{}g}\sigma_n^2)$} is the noise vector with each entry following a complex normal distribution, with $\sigma_n^2=N_0 W$ ($N_0$ is the noise power spectral density (PSD) and $W=K\Delta_f$ is the total bandwidth with $K$  subcarriers and subcarrier spacing $\Delta_f$). We assume $\Hm_k$ remains the same during ${G}$ transmissions (within the coherence time).  
The channel matrix at subcarrier $k$ is given by 
\begin{equation}
\begin{split}
    \Hm_k & = \underbrace{\alpha d_k(\tau) \av_\text{B}(\varphiv_\text{B})\av^{\top}_\text{U}(\varphiv_\text{U})}_{\text{LOS~path}}
    \\ & + \underbrace{\sum_{p=1}^{P} \alpha_{p}d_{p,k}(\tau_p) \av_\text{B}(\varphiv_{\text{B},p})\av^{\top}_\text{U}(\varphiv_{\text{U},p})}_{\text{NLOS~paths}},
\end{split}
\end{equation} 
where for the \ac{los} path, $\alpha = \rho e^{-j\xi}$ {(with $\rho$ and $\xi$ as the amplitude and phase, respectively)} is the complex channel gain assumed to be identical for different subcarriers, $d_k(\tau) = e^{-j 2 \pi k \Delta_f \tau}$ ($\Delta_f$ is the subcarrier spacing) as a function of the path delay $\tau$, while $\av_\text{B}(\varphiv_\text{B})$ and $\av_\text{U}(\varphiv_\text{U})$ are the receiver and transmitter steering vectors as a function of the  \ac{aoa} $\varphiv_{\text{B}} = [\phi_{\text{B}}, \theta_{\text{B}}]^\top$ (azimuth angle $\phi_\text{B}$ and elevation angle $\theta_\text{B}$), and \ac{aod} $\phiv_{\text{U}} = [\phi_{\text{U}}, \theta_{\text{U}}]^\top$. 
 A steering vector $\av(\varphiv)$ of an $N$-element array is a function of the direction of the (incoming or outgoing) signal and the locations of the antenna elements, which can be expressed as~\cite{chen2022tutorial}
\begin{align}
    \av(\varphiv) & = 
    e^{j\frac{2\pi f_c}{c} \Zm^\top \tv(\boldsymbol{\varphi})},
    \label{eq:steering_vector_bs}
\end{align}
where we apply the $\exp$ operator element-wise, $\Zm\in \mathbb{R}^{3\times N}$ is the matrix containing the position of the $N$ antennas in the local coordinate system (all zeros in the first row of $ \Zm$) and $\tv(\varphiv)=[\cos(\theta) \cos(\phi), \cos(\theta) \sin(\phi), \sin(\theta) ]^\top$.  For the NLOS paths, each path can correspond to single or multi-bounce reflections, or diffuse scattering. Hence, the NLOS path will not be utilized for the positioning of the UE in this work. We further make the assumption that the LOS path is resolvable with respect to the NLOS paths (though the NLOS paths may be mutually unresolved).  This is a reasonable assumption\footnote{For example, with a bandwidth of 1 GHz and $8\times 8$ BS arrays, a delay resolution of 30 cm and an angle resolution of 22 degrees is achievable. Unless the UE is very close to a reflector, multipath can be resolved in the combined range-angle domain.} for 5G/6G systems, due to large bandwidth and a large number of antennas~\cite{abu2018error}. 
Without significant loss of generality, the channel matrix for the $k$th subcarrier can thus be simplified as
\begin{equation}
    \Hm_k = \alpha d_k(\tau) \av_\text{B}(\boldsymbol{\varphi}_\text{B})\av^{\top}_\text{U}(\varphiv_\text{U}).
    \label{eq:channel_matrix_far_field}
\end{equation}
Correspondingly,  the \emph{channel geometric parameter vector} of the \ac{los} path between a BS and the UE is defined as $\etav_{\text{ch}} = [\etav_1^\top, \ldots, \etav_L^\top]^\top$ with $\etav_l =  [\varphiv_{\text{B},l}^\top, \varphiv_{\text{U},l}^\top, \tau_l, \rho_l, \xi_l]^\top \in \mathbb{R}^{7}$ for the $l$th BS. For later analysis, we define a vector by removing all the nuisance parameters (i.e., complex channel gain for each path{, as we do not exploit the signal strength for localization}) as $\cv_{\text{ch}} = [\cv_1^\top, \ldots, \cv_L^\top]^\top$ with $\cv_l =  [\varphiv_{\text{B},l}^\top, \varphiv_{\text{U},l}^\top, \tau_l]^\top \in \mathbb{R}^{5}$.
The relationships between the channel parameters vector $\cv$ and the state parameters $\sv$ can be expressed as
\begin{align}
    \vpv_{\text{B}} & = 
    \begin{bmatrix}
        \phi_{\text{B}}\\
        \theta_{\text{B}}
    \end{bmatrix}=
    \begin{bmatrix}
        \arctan2( t_{\text{B},2}, t_{\text{B},1})\\
        \arcsin( t_{\text{B},3})
    \end{bmatrix},
    \\
    \vpv_{\text{U}} & = 
    \begin{bmatrix}
        \phi_{\text{U}}\\
        \theta_{\text{U}}
    \end{bmatrix}=
    \begin{bmatrix}
        \arctan2( t_{\text{U},2}, t_{\text{U},1})\\
        \arcsin( t_{\text{U},3})
    \end{bmatrix},
    \\
    \tau & = \frac{\Vert \pv_\text{U} - \pv_\text{B} \Vert}{c} + B_\text{U},
\end{align}
where $c$ is the speed of light, $ \tv_\text{B} = [ t_\text{B,1},  t_\text{B,2},  t_\text{B,3}]^\top$ and $ \tv_\text{U} = [ t_\text{U,1},  t_\text{U,2},  t_\text{U,3}]^\top$ are the direction vectors in the local coordinate system that can be expressed using global direction vectors and rotation matrices as

\begin{equation}
    \tv_{\text{B}} =  \Rm_\text{B}^{-1}\frac{\pv_\text{U} - \pv_\text{B}}{\Vert \pv_\text{U} - \pv_\text{B} \Vert},\ \ 
    \tv_{\text{U}} =  \Rm_\text{U}^{-1}\frac{\pv_\text{B} - \pv_\text{U}}{\Vert \pv_\text{B} - \pv_\text{U} \Vert}.
\end{equation}




Finally, by concatenating all the received symbols into a column, we obtain the received symbol block $\yv \in \mathbb{R}^{{G}K}$ as $\yv = [\yv_1^\top, \ldots, \yv_g^\top, \ldots, \yv^\top_{G}]^\top$, 
where $\yv_g=[y_{g, 1}, \ldots, y_{g, K}]^\top$ can be expressed as
\begin{align}
    \yv_g = \alpha(\wv_{\text{}g}^\top  \av_\text{}(\varphiv_\text{B})\av^{\top}_\text{}(\varphiv_\text{U})\vv_g) \dv(\tau) \odot \xv_{g} + \nv_{g},
    \label{eq:ideal_model_y_g}
\end{align}
in which $\dv(\tau)=[d_1(\tau), \ldots, d_K(\tau)]^\top$, $\xv_g=[x_{g, 1}, \ldots, x_{g, K}]^\top$, and $\nv_g=[n_{g, 1}, \ldots, n_{g, K}]^\top$. 


\subsection{Hardware Impairments}
\label{sec:hwi_model}
In this work, several types of HWIs are considered as shown in Fig.~\ref{fig:block_diagram}.
We study the effects of \ac{pn}, \ac{cfo}, \ac{mc}, \ac{pan}, \ac{age}, \ac{ade}, and \ac{iqi}. Note that the impairments such as \ac{pn}, \ac{cfo}, \ac{mc}, \ac{age}, \ac{ade} and \ac{iqi} exist both at the transmitter and the receiver, while the PAN appears only at the transmitter. 
{The HWIs are usually compensated during offline  calibration or online with dedicated signals and routines, depending on whether the impairment is static or time-variant. In this work, we model all the \acp{hwi} except the PAN as random perturbations around the nominal values that correspond to the \emph{residual errors} for time-variant impairments (i.e., PN and CFO) after calibration~\cite{wang2021generalized, almradi2015spectral, lin2006joint}, and corresponding to an \emph{ensemble of devices} for static impairments (i.e., MC~\cite{ye20082}, AGE~\cite{van2018improved}, ADE~\cite{yassine2022mpnet}, and IQI~\cite{ghaseminajm2020localization}). The PAN is fixed in order to evaluate the effect of different types of pilot signals. The imperfections of \ac{adc}, \ac{dac}, low-noise amplifier, and mixer are not considered.} 

\subsubsection{Phase Noise and Carrier Frequency Offset} Imperfect \acp{lo} in the up-conversion and down-conversion processes add PN to the carrier wave phase. In addition, when the down-converting \ac{lo} in the receiver does not perfectly synchronize with the received signal’s carrier~\cite{mohammadian2021rf}, CFO occurs. In general, both PN and CFO are estimated and compensated by the receiver~\cite{hajiabdolrahim2020extended}, so we only consider the residual PN and residual CFO at the receiver. With PN and CFO, the observation, $y_{g,k}$, is modified as in~\cite{lin2006joint}
\begin{align}
    y_{g,k} & \to  \fv^\top_k \Em_g \Xim_g \Fm^{\mathsf{H}} \yv_g,
    \label{eq:PN_CFO}
    \\
    \Em_g & = e^{j\frac{2\pi \epsilon gK_{\text{tot}}}{{K}}}\text{diag}([1, e^{j\frac{2 \pi \epsilon}{K}}, \ldots, e^{j\frac{2 \pi (K-1) \epsilon}{K}}]),
    \label{eq:cfo}
    \\
    \Xim_g & =\text{diag}([e^{j\nu_{g,1}}, \ldots, e^{j\nu_{g,K}}]),
\end{align}
where $\yv_g$ is the received signals of the ideal model without PN or CFO (i.e., from \eqref{eq:ideal_far_field_model}), 
$\Fm = [\fv_1, \fv_2, \ldots, \fv_K]$ is the FFT matrix.
The CFO matrix $\Em_g$ considers both inter-OFDM symbol phase changes as well as inter-carrier interference~\cite{lin2006joint, roman2006blind}. More specifically, $K_{\text{tot}}=K+K_\text{cp}$ with $K_\text{cp}$ as the length of the cyclic prefix, and $\epsilon$ is the residual CFO with $\epsilon \sim \mathcal{N}(0, \sigma_\text{CFO}^2)$. $\Xim_g$ is the residual\footnote{Note that $\nu_{g,k}$ and $\epsilon$ represent residual PN and CFO that remains after the carrier synchronization process processing (e.g., \cite{salim2014channel,chung2021phase}). Hence, $\nu_{g,k}$ is assumed to be independent across time.} \ac{pn} matrix with $\nu_{g,k} \sim \mathcal{N}(0, \sigma_\text{PN}^2)$. In \eqref{eq:PN_CFO}, the vector $\yv_g$ is converted to the time domain by $\Fm^{\mathsf{H}} \yv_g$, where the successive \ac{pn} samples, as well as the CFO, are applied. Finally, $\fv^\top_k$ extracts the $k$-th subcarrier after applying an FFT  to $\Em_g \Xim_g \Fm^{\mathsf{H}} \yv_g$. Note that the residual CFO $\epsilon$ is fixed for each realization (e.g., one localization measurement with $G$ transmission), while the \ac{pn} $\nu_{g,k}$ is different for all the subcarriers and OFDM symbols.


\subsubsection{Mutual Coupling} \label{sec:coupling}
\Ac{mc} refers to the electromagnetic interaction between the antenna elements in an array~\cite{ye2009doa}. 
{For a \ac{upa}, we adopt the \ac{mc} model as in~\cite{ye20082} by assuming the antenna is only affected by the coupling of the surrounding elements. As a result, the MC matrix can be expressed as 
\begin{equation}
    \Cm_\text{}
    = 
    \begin{bmatrix}
        \Cm_1 & \Cm_2 & 0 & \cdots & 0\\
        \Cm_2 & \Cm_1 & 0 & \cdots & 0\\
        \vdots & \ddots & \ddots & \ddots & \vdots\\
        0 & \cdots & \Cm_2 & \Cm_1 & \Cm_2\\
        0 & \cdots & 0 & \Cm_2 & \Cm_1
    \end{bmatrix}.
\end{equation}
Here, $\Cm_\text{}\in \mathbb{C}^{N_\text{z}N_\text{y} \times N_\text{z}N_\text{y}}$ is the MC matrix with sub-matrices $\Cm_1 = \text{Toeplitz}([1, c_x, 0\ldots, 0]) \in \mathbb{C}^{N_\text{y}\times N_\text{y}}$ and $\Cm_2 = \text{Toeplitz}([c_x, c_{xy}, 0, \ldots, 0]) \in \mathbb{C}^{N_\text{y}\times N_\text{y}}$, {where $c_x$ and $c_{xy}$ are the coupling coefficients between a specific antenna and its surrounding antennas with $0.5$ and $\sqrt{2}/2$ wavelengths distance, respectively}~\cite{ye20082}. For convenience, we use one variable $\sigma_\text{MC}$ to denote the severity of the MC such that $c_x\sim \mathcal{CN}(0, \sigma_\text{MC}^2)$ and $c_{xy}\sim \mathcal{CN}(0, \sigma_\text{MC}^2/4)$.} The MC leads to the following substitution of the channel matrix
\begin{equation}
    \Hm_k\to \Cm_\text{B}\Hm_k \Cm_\text{U}^{\top}.
    \label{eq:MC}
\end{equation}

\subsubsection{Power Amplifier Nonlinearity}
For the PA nonlinearity, we consider a $Q$-th order memoryless polynomial nonlinear model~\cite{schenk2008rf} with a clipping point $x_\text{clip} \in \mathbb{R}$ as
\begin{equation}
{h}_{\text{PA}}(\check{{x}}_t) =
\begin{cases}
      \sum_{q = 0}^{Q-1} \beta_{q+1} \check{{x}} |\check{{x}}|^{q} & |\check x| \le x_{\text{clip}},\\
      \sum_{q = 0}^{Q-1} \beta_{q+1} \frac{\check{{x}}}{|\check{{x}}|} |{{x}_\text{clip}}|^{q+1} & |\check x| > x_{\text{clip}},
\end{cases}
\label{eq:PA}
\end{equation}
where $\check{x}_t = x_t/R$ denotes the voltage of the transmitted time-domain signal ($R$ is the load impedance in Ohms) in the time domain and $\beta_1, \dots, \beta_{Q}$
are complex-valued parameters. 
We assume that~\eqref{eq:PA} models both the effect of digital pre-distortion and power amplifier, and we use non-oversampled signals as input to PA for tractable localization performance analysis\footnote{In order to fully characterize the effect of PAN, an oversampled model is needed, which also captures the intersymbol interference introduced by the nonlinearity, in addition to the symbol distortion (see (25) in~\cite{moghaddam2021statistical}).}. Note that the PA affects the time domain signals and each antenna at the Tx has a separate PA, and the PA model in~\eqref{eq:PA} does not consider the out-of-band emissions (only the in-band distortion). For simplicity, the models are the same for different PAs and $\hv_{\text{PA}}(\check{\xv}_t)$ returns the time domain signal vector (by operating point-wise on each of the elements) with PA nonlinearity introduced.

\subsubsection{{Array Calibration Error (AGE and ADE)}}
{The array calibration errors are caused by the variations in array gain and antenna displacement.} To model the \ac{age}, we define the complex excitation coefficient of the $n$-th antenna at direction $\varphiv$ as~\cite{van2018improved}
\begin{equation}
    b_n(\varphiv) = (1+\delta_{a})e^{j \delta_p},
    \label{eq:AG}
\end{equation}
where $\delta_a\in \mathcal{N}(0, \sigma_\text{AA}^2)$, and $\delta_p\in \mathcal{N}(0, \sigma_\text{AP}^2)$ are the relative amplitude error and phase error, respectively. Regarding the \ac{ade}, we assume the $n$-th antenna position has a displacement on the 2D plane of the local coordinate system as 
\begin{equation}
  \tilde \zv_n = \zv_n + [0, \delta_{n,y}, \delta_{n,z}]^\top,
  \label{eq:AD}
\end{equation}
with {$\zv_n \in \mathbb{R}^{3}$} is the ideal position of the $n$th antenna in the local coordinate system,  $\delta_{n,y}, \delta_{n,z} \in \mathcal{N}(0, \sigma^2_\text{ADE})$ are the displacement error. The steering vector is then modified as
\begin{equation}
    \av(\varphiv) \rightarrow \bv(\varphiv)\odot e^{j\frac{2\pi}{\lambda}\tilde \Zm^\top \tv},
\end{equation}
where $\tilde \Zm = [\tilde \zv_1, \ldots, \tilde \zv_N]$ contains the geometry information of all the antennas. The array calibration error is fixed for a certain array and varies across different devices.

\subsubsection{{In-phase and quadrature imbalance}}
\Ac{iqi} operates on the time domain signal and the transmitted symbol vector is modified as~\cite{ghaseminajm2020localization}
\begin{equation}
    \xv_g \rightarrow \Fm (\alpha_\text{U} \Fm^\herm \xv_g + \beta_\text{U} \Fm^\herm\xv_g^*) = \alpha_\text{U}\xv_g + \beta_\text{U}\xv_g^*,
\end{equation}
where the FFT matrix $\Fm$ and IFFT matrix $\Fm^\herm$ switch between time and frequency domain, $\alpha_\text{U} = \frac{1}{2}+\frac{1}{2}m_\text{U} e^{j\psi_\text{U}}$, $\beta_\text{U} = \frac{1}{2}-\frac{1}{2}m_\text{U} e^{j\psi_\text{U}}$ with $m_\text{U}$ and $\psi_\text{U}$ as the amplitude and phase imbalance parameters at the UE side. We assume that the IQI is compensated in the system, leading to a residual impairment and the imbalance parameters can be modeled as $m_\text{U}\sim \mathcal{N}(1, \sigma_\text{IA}^2)$ and $\phi_\text{U}\sim \mathcal{N}(0, \sigma_\text{IP}^2)$. Similarly, the IQI at the receiving BS can be expressed as 
\begin{equation}
    \yv_g \rightarrow \alpha_\text{B} \yv_g +  \beta_\text{B} \yv_g^*.
    \label{eq:iqi}
\end{equation}
More accurate frequency-dependent IQI models can be found in~\cite{narasimhan2009digital, minn2010pilot}, which is beyond the scope of this work.

\subsection{Hardware-impaired Model}
Considering all types of \acp{hwi} described in Sec.~\ref{sec:hwi_model} and substituting~\eqref{eq:PN_CFO}--\eqref{eq:iqi} into~\eqref{eq:ideal_model_y_g}, the observation can be rewritten in the frequency domain.

\subsubsection{Transmit Signal under HWI}
The precoded transmitted signal across subcarriers and antennas is modified from 
$\Xm_g = \mathbf{x}_g \vv^\top_{g} \in \mathbb{C}^{K\times N_\text{U}}$
to 
\begin{align}
\check{{\Xm}}_{g} & =  \Fm \mathbf{h}_{\text{PA}}(
    \underbrace{\mathbf{E}_{\text{U}}\mathbf{\Xi}_{\text{U}}
    ({\alpha}_{\text{U}} \mathbf{F}^\herm \mathbf{x}_g+
    {\beta}_{\text{U}} \mathbf{F}^\herm \mathbf{x}_g^*){\vv}^\top_{g}}_{\text{precoded time domain signal before PA}}).
    \label{eq:hwi_3}
\end{align}
\subsubsection{Channel under HWI}
The channel is modified from $\Hm_k=\alpha d_k(\tau) \av_\text{}(\varphiv_\text{B})\av^{\top}_\text{}(\varphiv_\text{U}) \in \mathbb{C}^{N_\text{B} \times N_\text{U}}$ in~\eqref{eq:channel_matrix_far_field} to 
\begin{align}
    \begin{split}
        \check{\mathbf{H}}& = \alpha d_k(\tau)
    \mathbf{C}_{\text{B}}(\underbrace{\bm{b}_\text{B}(\bm{\varphi}_{\text{B}}) \odot e^{j\frac{2\pi}{\lambda}{\tilde{\bf{Z}}_\text{B}^\top {\tv}_\text{B}(\bm{\varphi}_{\text{B}})}}}_{\text{steering vector\ } \tilde \av_\text{B}(\boldsymbol{\varphi}_\text{B})})
    \\
    & \times (\underbrace{\bm{b}_\text{U}(\bm{\varphi}_{\text{U}}) \odot e^{j\frac{2\pi}{\lambda}{\tilde{\bf{Z}}_\text{U}^\top {\tv}_\text{U}(\bm{\varphi}_{\text{U}})}}}_{\text{steering vector\ } \tilde \av_\text{U}(\boldsymbol{\varphi}_\text{U})})
    \mathbf{C}_{\text{U}}^\top.
    \end{split}
    \label{eq:hwi_2}
\end{align}
\subsubsection{Received Signal under HWI}
The received signal is modified from $\yv_g \in \mathbb{C}^{K\times 1}$ to 
\eqref{eq:hwi_1}\footnote{{The signal model considering all the HWIs provides a more accurate analysis compared to simply adding the effect of individuals as they do not hold a linear relationship. In addition, equations (21) to (23) provide flexibility in evaluating the contribution of individual HWIs compared to the overall HWIs to identify dominant components that degrade the system performance.}} as

\begin{equation}
\begin{split}
    \check{\mathbf{y}}_g &= \mathbf{F}(\alpha_{\text{B}}(\mathbf{E}_{\text{B},g}\mathbf{\Xi}_{\text{B},g}\mathbf{F}^\herm
    ({\check{\Xm}_{g}\check{\mathbf{H}}^{\top} \wv_g \odot \dv(\tau))}) 
    \\ + &  
    \beta_{\text{B}} (\mathbf{E}_{\text{B},g}\mathbf{\Xi}_{\text{B},g}\mathbf{F}^\herm
    (\check{\Xm}_{g}\check{\mathbf{H}}^{\top} \wv_g \odot \dv(\tau)))^*) + 
    {\mathbf{n}_{g}}.
\end{split}
    \label{eq:hwi_1}
\end{equation}

 \subsection{Summary of the Models}
To summarize, we have defined a \ac{mm} in~\eqref{eq:ideal_far_field_model} without considering the \ac{hwi}, which will be used for algorithm development. With \acp{hwi} introduced, the impaired model defined in~\eqref{eq:hwi_1} will be used as the \ac{tm}. In the following section, we will evaluate the impact of using the \ac{mm} to process data generated by \ac{tm} on localization performance. For the sake of convenience in performance analysis, we use $\muv_g(\etav)$ and $\bar{\muv}_g(\etav)$ to denote the noise-free observation of \eqref{eq:ideal_far_field_model} and \eqref{eq:hwi_1}, respectively.

\section{Localization Algorithm}
\label{sec:localization_algorithm}
Based on the models described above, a two-stage localization\footnote{In contrast, the direct localization estimates the state vector $\sv$ from the observed signal vector $\yv$ directly. Considering the high complexity involved, we adopt two-stage localization in this work.} problem can be formulated such that the channel parameter vectors $\hat \etav_{\text{ch}} = [\etav_1^\top, \ldots, \etav_L^\top]^\top$ are firstly estimated based on the received signals $\hat \yv_1, \ldots, \hat \yv_L$ from all the BSs, and then the state vector $\hat \sv$ is determined from $\hat \etav_{\text{ch}}$.

\subsection{Mismatched Maximum Likelihood Estimator} 
The \ac{mle} can be employed when the observation $\yv$ is generated from the same model used by the algorithm. If $\yv \sim f_\text{TM}( \yv|\bar\etav)$, 
the \ac{mle} of the UE position and channel gain is 
\begin{align}
        \hat \etav_\text{MLE}
        & = \arg \max_{\bar{\boldsymbol{\eta}}} \ln f_\text{TM}( \yv|\bar\etav),
\end{align}
where $\ln f_\text{TM}( \yv|\bar\etav)$ is the log-likelihood of the \ac{tm}.
However, if {the estimator does not know the existing mismatch, the \ac{mle} boils down to \ac{mmle} (i.e., $\yv \sim f_\text{TM}( \yv|\bar\etav)$ while the estimator uses $f_\text{MM}( \yv|\etav) \neq f_\text{TM}( \yv|\bar\etav)$)}, given by
\begin{align}
        \hat\etav_\text{MMLE}
        & = \arg \max_{\boldsymbol{\eta}} \ln f_\text{MM}( \yv|\etav).
        \label{eq:mmle}
\end{align}
More specifically, equation~\eqref{eq:mmle} formulates the \ac{mmle} for channel parameters extraction, which can also be implemented in position and orientation estimation with known or approximated likelihood function. A practical approach is to use the gradient descent method with an initial point, which will be detailed in the following subsections.  
\subsection{Channel Parameters Estimation}
\label{sec:channel_estimation_stage_1}
The channel parameters estimation will be performed with a coarse estimation using ESPRIT, which provides a good initial point for a refined estimation using~\eqref{eq:mmle}.
\subsubsection{{Coarse Estimation using ESPRIT}}
We aim to obtain an initial estimate of the channel parameters with a low complexity, which can be solved using tensor-based beamspace ESPRIT\footnote{While this work considers only the LOS channel, the ESPRIT also works for the scenarios with NLOS paths.} algorithm~\cite{wen2018tensor}. To implement the beamspace ESPRIT algorithm, we reformulate a beamspace channel matrix $\Hm^{(b)}$ {(for tensor decomposition)} based on~\eqref{eq:ideal_far_field_model} as
\begin{equation}
\Hm^{(b)}_k = \alpha d_k(\tau) \Wm^{\herm}\av_{\text{B}}(\varphiv_\text{B})\av^\top_{\text{U}}(\varphiv_\text{U})\Vm
\label{eq:beamspace_channel}
\end{equation}
where $\Wm = \Tm_1\otimes \Tm_2 \in \mathbb{C}^{N_1N_2\times M_1M_2}$ and $\Vm = (\Tm_3\otimes \Tm_4)^* \in \mathbb{C}^{N_3N_4\times M_3M_4}$ are the combining matrix and precoder matrix and the total number of transmissions ${G} = M_1M_2M_3M_4$. {Here, $\Tm_i$ is the codebook containing $M_i$ steering vectors of the $i$-th dimension (i.e., azimuth/elevation of AOA/AOD).} 
Since the first row of the antenna position matrix $\tilde \Zm$ is all zeros (see Sec.~\ref{sec:geometric_model} and~\eqref{eq:steering_vector_bs}), we can express the steering vector as
\begin{equation}
    \av_\text{B}(\varphiv_\text{B}) = \av^{(M_1)}(\omega_1)\otimes \av^{(M_2)}(\omega_2),
\end{equation}
with
\begin{align}
    \omega_1 & = \pi\sin(\phi_\text{B})\cos(\theta_\text{B}), \ \ \omega_2 = \pi\sin(\theta_\text{B}),
    \label{eq:omega_1_2}
    \\
    \av^{(M_1)}_\text{B}(\omega_1) & 
    = e^{j\frac{2\pi f_c \sin(\phi_\text{B})\cos(\theta_\text{B})}{c}{\tilde\zv}_\text{B,2} } 
    = e^{j\frac{2}{\lambda_c}\omega_1{\tilde\zv}_\text{B,2}},
    \\
    \av^{(M_2)}_\text{B}(\omega_2) & 
    = e^{j\frac{2\pi f_c \sin(\theta_\text{B})}{c}{\tilde\zv}_\text{B,3}}
    = e^{j\frac{2}{\lambda_c}\omega_2{\tilde\zv}_\text{B,3}}.
\end{align}
Here, ${\tilde\zv}^\top_\text{B,2}\in \mathbb{C}^{1\times N_\text{B}}$ and ${\tilde\zv}^\top_\text{B,3}\in \mathbb{C}^{1\times N_\text{B}}$ are the second and third row of the matrix $\tilde\Zm$, respectively.
{The combining matrix can then be defined in terms of a grid of the spatial frequencies $\bar \omegav_1 = [\bar \omega_{1,1}, \ldots, \bar \omega_{1,M_1}]$ and $\bar \omegav_2 = [\bar \omega_{2,1}, \ldots, \bar \omega_{2,M_2}]$ as
\begin{align}
    \Tm_1 & = [\av^{(N_1)}(\bar\omega_{1,1}), \ldots, \av^{(N_1)}(\bar\omega_{1,M_1})]^\top \in \mathbb{C}^{N_1\times M_1},\\
    \Tm_2 & = [\av^{(N_2)}(\bar\omega_{2,1}), \ldots, \av^{(N_2)}(\bar\omega_{2,M_2})]^\top \in \mathbb{C}^{N_2\times M_2},
\end{align}
where $\bar\omega_{1, m}$ and $\bar\omega_{2, m}$ are decided by beamforming directions (detailed in Sec.~\ref{sec-4-simulation}).}
A similar reasoning applies to the steering vectors $\av^{(M_3)}_\text{U}(\omega_3)$ and $\av^{(M_4)}_\text{U}(\omega_4)$ at UE to define $\Tm_3$ and $\Tm_4$, with 
\begin{equation}
    \omega_3 = \pi\sin(\phi_\text{U})\cos(\theta_\text{U}), \ \ \omega_4 = \pi\sin(\theta_\text{U}).
    \label{eq:omega_3_4}
\end{equation}

We further define $\bv^{(M_n)}(\omega_n) = \Tm_n^\herm\av^{N_n}(\omega_n) \in \mathbb{C}^{M_n}$ for $n\in\{1, 2, 3, 4\}$ and $\bv^{(M_5)}(\omega_5) = \dv(\tau)$ ($\omega_5 = 2\pi\Delta_f\tau$), and the 
beamspace channel matrix in~\eqref{eq:beamspace_channel} can be represented by a tensor $\boldsymbol{\Hc}\in \mathbb{C}^{M_1\times M_2\times \cdots \times M_5}$ as~\cite{wen20205g}
\begin{equation}
    \boldsymbol{\Hc}^{(b)} = \alpha 
    \bv^{(M_1)}(\omega_1) \circ 
    \ldots \circ
    \bv^{(M_5)}(\omega_5).
    \label{eq:tensor_channel}
\end{equation}

{In practice, the estimated beamspace channel matrix can be estimated with known pilot signals as $\text{vec}(\hat \Hm_k^{(b)}) = [\hat y_{1,k}/x_{1,k}, \ldots, \hat y_{{G},k}/x_{{G},k}]^\top$.} By rearranging the estimated channel into a tensor $\hat{\boldsymbol{\Hc}}^{(b)}$ shown in~\eqref{eq:tensor_channel}, the beamspace tensor-based ESPRIT method can then be used to estimate $\omega_1$ to $\omega_5$ and obtain the AOA, AOD, and delay accordingly~\cite{wen20205g, wen2018tensor}.


\subsubsection{Fine Estimation using MMLE}
From ESPRIT, we can obtain an initial estimate of the channel parameters $\hat \etav_0$. The refinement of the initial estimate can be formulated as an optimization problem, based on~\eqref{eq:mmle}, as
\begin{equation}
    \hat \etav = \arg \min_{\boldsymbol{\eta}} \Vert \yv - \muv(\etav) \Vert^2.
    \label{eq:mmle_channel_estimation}
\end{equation}
Since $\alpha$ appears linearly in the noise-free observation $\muv$,
we further define $\gammav(\etav) = \muv(\cv)/\alpha$ with $\cv = [\varphiv_\text{B}^\top, \varphiv_\text{U}^\top, \tau]^\top$. By setting $\partial \Vert \yv - \muv(\etav) \Vert^2/\partial \alpha = 0$, we can have
\begin{equation}
    \hat \cv = \arg \min_{\cv} \Vert \yv - \frac{\gammav^\herm(\cv)\yv}{\Vert \gammav^\herm(\cv) \Vert^2} \gammav(\cv) \Vert^2.
\end{equation}
In this way, the nuisance parameters can be removed, which reduces the dimension of the unknown parameters.


\subsection{Localization Algorithm}
\label{sec:localization_stage_2}
\subsubsection{Coarse Estimation}
Given the estimated geometric parameter vector $\cv_l$ ($1\le l \le L$) for all the channels, the least squares solution for coarse estimation of position and orientation as~\cite{zheng2022coverage}
\begin{equation}
\label{eq_hat_Ru_LS}
\hat{\mathbf{R}}_\text{U,LS} = \left\{
    \begin{array}{rl}
    \mathbf{U}\mathbf{V}^\mathsf{T}, & \text{if det}(\mathbf{U}\mathbf{V}^\mathsf{T}) = 1,  \\
    \mathbf{U}\mathbf{J}\mathbf{V}^\mathsf{T}, & \text{if det}(\mathbf{U}\mathbf{V}^\mathsf{T}) = -1,  
    \end{array} \right.
\end{equation}
\begin{equation}
    [\hat\pv_\text{U,LS}, \hat B_{\text{U,LS}}]^\top = (\Qm_3^\top \Qm_3)^{-1}\Qm_3^\top \qv,
    \label{eq:LS_position}
\end{equation}
where $\mathbf{J} = \text{diag}([1, 1, -1])$, $\Um$ and $\Vm$ are the unitary basis matrices of the singular value decomposition of the matrix $\Qm_1\Qm_2^\top$, together with $\Qm_3, \qv$ are given by

\begin{align}
    \Qm_1 & = -[\Rm_{\text{B},1} \tv(\hat \varphiv_{\text{B},1}), \ldots, \Rm_{\text{B},L} \tv(\hat \varphiv_{\text{B},L})],
    \\
    \Qm_2 & = [ \tv(\hat \varphiv_{\text{U},1}), \ldots,  \tv(\hat \varphiv_{\text{U},L})],
    \\
    \Qm_3 & = 
    \begin{bmatrix}
        \mathbf{I}_{3} & \Rm_{\text{B},1} \tv(\hat \varphiv_{\text{B},1})\\
        \vdots & \vdots\\
        \mathbf{I}_{3} & \Rm_{\text{B},L} \tv(\hat \varphiv_{\text{B},L})
    \end{bmatrix},
    \\
    \qv & = 
    \begin{bmatrix}
        \pv_{\text{B}}^{(1)}+ \Rm_{\text{B},1}\hat\tau_1  \tv(\hat \varphiv_{\text{B},1})\\
        \vdots\\
        \pv_{\text{B},L} + \Rm_{\text{B},L}\hat\tau_L  \tv(\hat \varphiv_{\text{B},L})]^\top
    \end{bmatrix}.
\end{align}

The estimator for position and clock offset in~\eqref{eq:LS_position} does not require the orientation of the UE $\Rm_\text{U}$, which is still sufficient as a coarse estimate, as will be shown in the simulation section.


\subsubsection{MMLE}
Once the initial position and orientation results are obtained, joint position and orientation estimation using MMLE can be formulated as
\begin{equation}
    {\hat\sv} 
    = \argmin_{\sv} \sum_{l=1}^L(\cv_l(\sv) - \hat\cv_l)^\top \Sigmam^{-1}_{c_l}( \cv_l(\sv) - \hat\cv_l),
    \label{eq:pseudotrue_state_final}
\end{equation}
which can be solved using the manifold optimization toolbox Manopt~\cite{boumal2014manopt}.
Note that the covariance matrix may not be accurately obtained in practice. We formulate localization as an MMLE problem with two purposes: (a) to evaluate the performance improvement with known covariance matrices compared to the coarse estimation; (b) to validate the derived bound, which will be detailed in Sec.~\ref{sec:lower_bounds_analysis}.

\section{Lower Bound Analysis}
\label{sec:lower_bounds_analysis}

In the next, we derive the CRB for MM, as well as the \ac{mcrb} for the mismatched estimator in~\eqref{eq:mmle}.

\subsection{CRB Analysis for the Mismatched Model}
Based on the defined channel parameter vector ${\boldsymbol\etav}$ and state vector $\sv$, the signal model in~\eqref{eq:ideal_far_field_model} and $\yv \sim f_{\text{\ac{mm}}}( \yv|\etav)$, the channel estimation CRB of the \ac{mm} for the $l$th channel can be obtained as $\bm{\mathcal{I}}(\etav_{l})^{-1} \in \mathbb{R}^{7\times 7}$ with~\cite{kay1993fundamentals}
\begin{align}
    \bm{\mathcal{I}}({\boldsymbol\eta_l}) & 
    = \frac{2}{\sigma_n^2}\sum^{G}_{g=1} \sum^K_{k=1}\mathrm{Re}\left\{
    \left(\frac{\partial\mu_{g,k}}{\partial{\boldsymbol\eta}_{l}}\right)^{\mathsf{H}} 
    \left(\frac{\partial\mu_{g,k}}{\partial{\boldsymbol\eta}_{l}}\right)\right\}.
    \label{eq:FIM_measurement}
\end{align}
Here, $\mathrm{Re}\{\cdot\}$ extracts the real part of a complex variable. 
Consequently, the {\ac{fim}} of all the channel parameters $\etav_{\text{ch}}$ can be formulated as
\begin{equation}
    \bm{\mathcal{I}}({\boldsymbol\eta_{\text{ch}}}) = \text{blkdiag}(\bm{\mathcal{I}}({\boldsymbol\eta_1}), \ldots, \bm{\mathcal{I}}({\boldsymbol\eta_L})).
\end{equation}
where $\text{blkdiag}(\cdot)$ returns the block diagonal matrix created by aligning the input matrices.
The FIM of the state vector $\bm{\mathcal{I}}(\sv) \in \mathbb{R}^{13\times 13}$ can then be formulated as
\begin{equation}
    \bm{\mathcal{I}}(\sv) = \Mm(\Mm^\top\ \Jm_{\mathrm{S}}^\top {\bm{\mathcal{I}}}({\cv_{\text{ch}}}) \Jm_{\mathrm{S}}\ \Mm)^{-1}\Mm^\top,
    \label{eq:CRB_from_FIM}
\end{equation}
where ${\bm{\mathcal{I}}}({\cv}_{\text{ch}})\in \mathbb{R}^{5L\times5L}$ is the equivalent FIM of non-nuisance parameters $\cv_\text{ch}$ obtained from $\bm{\mathcal{I}}({\boldsymbol\eta}_{\text{ch}})$, $\Jm_{\mathrm{S}} \triangleq \frac{\partial {\cv_{\text{ch}}}}{\partial \sv}$ is the Jacobian matrix using a denominator-layout notation, $\Mm = \text{blkdiag}(\bm{\text{I}}_{4\times 4}, \bar \Mm)$ with $\bar \Mm$
as~\cite{nazari2022mmwave}
\begin{equation}
    \bar\Mm = \frac{1}{\sqrt{2}}
    \begin{bmatrix}
        -\rv_3 & \mathbf{0}_{3\times1} & \rv_2 \\
        \mathbf{0}_{3\times1} & -\rv_3 & -\rv_1\\
        \rv_1 & \rv_2 & \mathbf{0}_{3\times1}
    \end{bmatrix},
\end{equation}
where $\rv_1$, $\rv_2$, and $\rv_3$ are the first, second, and third columns of the UE rotation matrix $\Rm_\text{U}$.

Based on $\bm{\mathcal{I}}({\boldsymbol\eta})$ in~\eqref{eq:FIM_measurement}, we can define the \ac{aod} error bound (ADEB), \ac{aoa} error bound (AAEB), and delay error bound (DEB) of the link between the UE and the $l$th BS) as
\begin{align}
\mathrm{AAEB} & = \sqrt{\trace([\bm{\mathcal{I}}({\boldsymbol\eta}_l)^{-1}]_{1:2, 1:2})}
\label{eq:AAEB},\\
\mathrm{ADEB} & = \sqrt{\trace([\bm{\mathcal{I}}({\boldsymbol\eta}_l)^{-1}]_{3:4, 3:4})}
\label{eq:ADEB},\\
\mathrm{DEB} & = \sqrt{([\bm{\mathcal{I}}({\boldsymbol\eta}_l)^{-1}]_{5, 5})}
\label{eq:DEB}.
\end{align}


Similarly, based on $\bm{\mathcal{I}}({\sv})$, we can define the position error bound (PEB), clock offset error bound (CEB) and orientation error bound (OEB) as
\begin{align}
\mathrm{PEB} & = \sqrt{\trace([\bm{\mathcal{I}}({\sv})^{-1}]_{1:3, 1:3})}
\label{eq:PEB},\\
\mathrm{CEB} & = \sqrt{([\bm{\mathcal{I}}({\sv})^{-1}]_{4, 4})}
\label{eq:CEB},\\
\mathrm{OEB} & = \sqrt{\trace([\bm{\mathcal{I}}({\sv})^{-1}]_{5:13, 5:13})}.
\label{eq:OEB}
\end{align}
The bounds from~\eqref{eq:AAEB}--\eqref{eq:OEB} will be used to evaluate the channel estimation and localization performance. In the next subsections, we will first formulate the MCRB for channel estimation, and then the mismatched lower bound for position and orientation estimation will be derived.  

\subsection{Misspecified CRB of Channel Parameters}
For a given channel model, the model is said to be mismatched or misspecified when $\yv \sim f_
    {\text{TM}}( \yv|\etav)$, while the estimation is based on 
     the assumption that $\yv \sim f_
    {\text{MM}}( \yv|\etav)$), where $f_
    {\text{TM}}( \yv|\etav)\neq f_
    {\text{MM}}( \yv|\etav)$. 

The \ac{lb} of using a mismatched estimator can be obtained as~\cite{fortunati2017performance}
\begin{align}
    \text{LB}(\bar {\boldsymbol\eta}, {\boldsymbol\eta}_0) 
    & = \underbrace{\Am_{{\boldsymbol\eta}_0}^{-1}\Bm_{{\boldsymbol\eta}_0}\Am_{{\boldsymbol\eta}_0}^{-1}}_{=\text{MCRB}({\boldsymbol\eta}_0)} + \underbrace{(\bar{\boldsymbol\eta} - {\boldsymbol\eta}_0)(\bar{\boldsymbol\eta} - {\boldsymbol\eta}_0)^\top}_{=\text{Bias}({\boldsymbol\eta}_0)}, 
    \label{eq:LB_channel_parameters}
\end{align}
where $\bar{\etav}$ is the true channel parameter vector, ${\boldsymbol\eta}_0$ is the pseudo-true parameter vector (which will be introduced soon), and $\Am_{{\boldsymbol\eta}_0}, \Bm_{{\boldsymbol\eta}_0}$ are two possible generalizations of the FIMs.
The LB is a bound in the sense that 
\begin{align}
    \mathbb{E} \{ (\hat{\boldsymbol{\eta}}_{\text{MMLE}} -\bar{\etav})(\hat{\boldsymbol{\eta}}_{\text{MMLE}} -\bar{\etav})^\top \} \succeq \text{LB}(\bar {\boldsymbol\eta}, {\boldsymbol\eta}_0),
\end{align}
where the expectation is with respect to $f_
    {\text{TM}}( \yv|\boldsymbol \eta )$.  What remains is the formal definition and computation of the pseudo-true parameter $\boldsymbol{\eta}_0$ and  $\Am_{{\boldsymbol\eta}_0}, \Bm_{{\boldsymbol\eta}_0}$.

\subsubsection{Pseudo-true Parameter}
{Assume the \ac{pdf} of the \ac{tm}, where the observation data come from, is $f_\text{TM}(\yv|\bar {\boldsymbol\eta})$, where $\yv$ is the received signals and $\bar {\boldsymbol\eta} \in \mathbb{R}^7$ (7 unknowns for this case) is the vector containing all the channel parameters. Similarly, the \ac{pdf} of the \ac{mm} for the received signal $\yv$ can be noted as $f_\text{MM}(\yv, {\boldsymbol\eta})$.}
The pseudo-true parameter vector is defined as the point that minimizes the Kullback-Leibler divergence between $f_\text{TM}(\yv|\bar {\boldsymbol\eta})$ and $f_\text{MM}(\yv| {\boldsymbol\eta})$ as
\begin{align}
    {\boldsymbol\eta}_0 = \arg \min_{\boldsymbol\eta} D_\text{KL}(f_\text{TM}(\yv|\bar {\boldsymbol\eta})\Vert f_\text{MM}(\yv| {\boldsymbol\eta})).
\end{align}
We define $\epsilonv(\etav) \triangleq \bar\muv(\bar {\boldsymbol\eta}) - \muv({\boldsymbol\eta})$, and the pseudo-true parameter can be obtained as~\cite{ozturk2022ris}
\begin{equation}
    {\boldsymbol\eta}_0 
    = \arg \min_{{\boldsymbol\eta}} \Vert \epsilonv(\etav) \Vert^2 
    = \arg \min_{{\boldsymbol\eta}} \Vert \bar\muv(\bar {\boldsymbol\eta}) - \muv({\boldsymbol\eta})\Vert^2  
    \label{eq:pseudotrue_final}.
\end{equation}
Hence, ${\boldsymbol\eta}_0$ can be found by solving \eqref{eq:mmle_channel_estimation} with the observation $\yv = \bar\muv(\bar {\boldsymbol\eta})$, which can be accomplished using the same algorithm in Sec.~\ref{sec:localization_algorithm},
initialized with the true value $\bar {\boldsymbol\eta}$. 

\subsubsection{MCRB Component Matrices}
The matrices $\Am_{{\boldsymbol\eta}_0}$ and $\Bm_{{\boldsymbol\eta}_0}$ can be obtained based on the pseudo-true parameter vector ${\boldsymbol\eta}_0$ as~\cite{ozturk2022ris}
\begin{align}
    &[\Am_{{\boldsymbol\eta}_0}]_{i,j}
        = \left. 
            \int \frac{\partial^2 \text{ln}f_\text{MM}(\yv|{\boldsymbol\eta})}{\partial{\eta}_i \partial{\eta}_j} f_\text{TM}(\yv|\bar {\boldsymbol\eta})\text{d}\yv
        \right|_{{\boldsymbol\eta} = {\boldsymbol\eta}_0} \notag \\
        &=  \left. \frac{2}{\sigma_n^2}\text{Re}\left[\frac{\partial^2\muv({\boldsymbol\eta})}{\partial \eta_i \partial \eta_j}\epsilonv({\boldsymbol\eta}) -  \frac{\partial\muv({\boldsymbol\eta})}{\partial \eta_j}
        \left(\frac{\partial\muv({\boldsymbol\eta})}{\partial \eta_i} \right)^{\mathsf{H}}\right]
        \right|_{{\boldsymbol\eta} = {\boldsymbol\eta}_0}\label{eq:matrix_A}
\end{align}
and
\begin{align}
        &[\Bm_{{\boldsymbol\eta}_0} ]_{i,j} 
        = \left. \int 
        \frac{\partial \text{ln}f_\text{MM}(\yv|{\boldsymbol\eta})}{\partial{\eta}_i} 
        \frac{\partial \text{ln}f_\text{MM}(\yv|{\boldsymbol\eta})}{\partial{\eta}_j}
        f_\text{TM}(\yv|\bar {\boldsymbol\eta})\text{d}\yv
        \right|_{{\boldsymbol\eta} = {\boldsymbol\eta}_0} \notag \\
        & =  
        \frac{4}{\sigma_n^4}
        \text{Re} \left[
        \frac{\partial\muv({\boldsymbol\eta})}{\partial \eta_i}\epsilonv({\boldsymbol\eta})
        \right]
        \text{Re} \left[
        \frac{\partial\muv({\boldsymbol\eta})}{\partial \eta_j}\epsilonv({\boldsymbol\eta})
        \right] \notag  \\
        & \left. + 
        \frac{2}{\sigma_n^2}\text{Re}\left[  \frac{\partial\muv({\boldsymbol\eta})}{\partial \eta_j}
        \left(\frac{\partial\muv({\boldsymbol\eta})}{\partial \eta_i} \right)^{\mathsf{H}}
        \right]
        \right|_{{\boldsymbol\eta} = {\boldsymbol\eta}_0}.
    \label{eq:matrix_B}
\end{align}

\subsection{Absolute Lower Bound (ALB) for Localization}
Another way to interpret the LB specified in~\eqref{eq:LB_channel_parameters} is that the estimated channel parameter vector from an efficient estimator follows a nonzero-mean multi-variable Gaussian distribution as
\begin{equation}
    \hat\etav_{l} \sim \mathcal{N}({\etav_{0,l}}, \Am_{{\boldsymbol\eta}_{0,l}}^{-1}\Bm_{{\boldsymbol\eta}_{0,l}}\Am_{{\boldsymbol\eta}_{0,l}}^{-1}),
\end{equation}
while the assumed distribution of the MMLE is
\begin{equation}
    \hat \etav_{l} \sim \mathcal{N}(\etav_l(\bar\sv), \mathbf I(\etav_l)^{-1}),
\end{equation}
where $\bar \sv$ is the true state of the UE. As a result, the position and orientation estimation (from the channel parameter vectors of all the paths) of the two-stage localization problem is another mismatched problem and the bound follows as
\begin{equation}
    \text{LB}(\bar \sv, \sv_0) = \text{MCRB}(\sv_0) + \underbrace{(\bar \sv - \sv_0)(\bar \sv - \sv_0)^\top}_{\text{Absolute lower bound (ALB)}}.
    \label{eq:state_ALB}
\end{equation}
Similar to~\eqref{eq:LB_channel_parameters}, $\bar{\sv}$ is the true state parameter vector, ${\sv}_0$ is the pseudo-true state parameter vector. 

It is possible to derive the localization LB constrained MCRB~\cite{fortunati2016constrained}; {however, considering the high complexity when involving 3D orientation estimation, we will focus on the bias term, defined as the absolute lower bound (ALB) of the localization performance as $\text{ALB} = (\bar\sv-\sv_0)(\bar\sv-\sv_0)^\top$, which can sufficiently evaluate the effect of HWIs on localization as will be shown in Sec.~\ref{sec:localization_results_vs_bounds}}
Following a similar derivation in~\eqref{eq:pseudotrue_final}. The pseudo-true parameters for state vector $\sv$ can be obtained as
\begin{equation}
    {\sv}_0 
    = \argmin_{\bar\sv} \sum_l(\etav_{0,l} - \etav_l(\bar \sv))^\top \bm{\mathcal{I}}(\etav_l)( \etav_{0, l} - \etav_l(\bar \sv)),
\end{equation}
where $\etav_{0,l} = \arg \min_{{\boldsymbol\eta}} \Vert \bar\muv(\bar{\boldsymbol\eta}_l) - \muv({\boldsymbol\eta_l})\Vert^2 $ is the biased mapping obtained by calculating the pseudo-true parameters of the $l$th channel from~\eqref{eq:pseudotrue_final}, and 
$\bm{\mathcal{I}}(\etav_l)$ is the inverse of the covariance matrix that can be obtained from~\eqref{eq:FIM_measurement}. 
\subsection{Summary of Different Bounds}
In this section, we introduced different types of lower bounds. For channel geometric parameters, the CRB and LB are derived for AOA, AOD, and delay. For state parameters, the CRB and ALB are derived for the position, orientation, and clock offset. 
{Considering the error bound calculation requires the inverse of the \ac{fim}, it is challenging to derive a closed-form expression accounting for all the \acp{hwi} for performance analysis, and hence only numerical results for localization are presented.} 
All types of the lower bounds are summarized in Table~\ref{table:summary_of_lower_bounds}, which will be used in Sec.~\ref{sec-4-simulation} Numerical Results. 

\vspace{-.2cm}
\begin{table}[ht]
\scriptsize
\centering
\caption{Summary of Different Lower Bounds}
\vspace{-0.cm}
\renewcommand{\arraystretch}{0.9}
\begin{tabular} {c | c | c | c | c}
    \hthickline
    \textbf{Types} & \multicolumn{3}{c|}{\textbf{Parameters}} & \textbf{Remarks}\\
    \hline
    \hline
    \textbf{} & AOA & AOD & Delay & Channel Parameters \\
    \hline
    \textbf{CRB} & AAEB & ADEB & DEB & \eqref{eq:AAEB}-\eqref{eq:DEB}\\
    \hline
    \textbf{LB} & AALB & ADLB & DLB & \eqref{eq:LB_channel_parameters}\\
    \hline
    \hline
    \textbf{} & Position & Orientation & Clock Offset & State Parameters\\
    \hline
    \textbf{CRB} & PEB & OEB & CEB & \eqref{eq:PEB}-\eqref{eq:OEB}\\
    \hline
    \textbf{ALB} & PALB & OALB & CALB & \eqref{eq:state_ALB}\\
    \hline
\end{tabular}
\renewcommand{\arraystretch}{1}
\label{table:summary_of_lower_bounds}\vspace{-5mm}
\end{table}



\section{Numerical Results}
\label{sec-4-simulation}
\subsection{Default Parameters}
We consider a 3D MIMO uplink scenario with one UE and two BSs. The carrier frequency, bandwidth, and subcarrier spacing of the system are set as $f_c = \unit[60]{GHz}$, $W = \unit[200]{MHz}$, and $\Delta f=\unit[240]{KHz}$, respectively. We utilize 12\% of the total number of subcarriers $K_{\text{com}}=833$ for localization, resulting in $K=100$ subcarriers as pilot signals. The amplitude of the channel gain is calculated as $\rho = \frac{\lambda}{4\pi c \tau}$. The rest of the system parameters\footnote{The PA parameters are estimated from the measurements of the RF WebLab, which can be remotely accessed at \url{www.dpdcompetition.com}. Part of the parameters come from the \href{https://hexa-x.eu/wp-content/uploads/2022/02/Hexa-X_D3.1_v1.4.pdf}{Hexa-X Deliverable 3.1}.} can be found in Table~\ref{table:Simulation_parameters}. Note that the selection of these parameters is to show the performance of the estimator in comparison to the derived bound, and the analysis of each HWI type is also discussed in the simulation results. 

Regarding the evaluation of communication performance, only the first BS is considered, and 16-\ac{qam} modulation is adopted. Different from localization, where BS-UE beam sweeping is needed, we evaluate the effect on {communications} with fixed precoder and combiner vectors across different transmissions. By considering all HWIs, we assume the channel can be perfectly estimated (with a sufficient number of pilots) as $\hat\Hm = \check \Hm = \hat\av_\text{B} \hat \av_\text{U}$ with $\hat\av_\text{B} = \sqrt{\alpha}\Cm_\text{B}\tilde\av_\text{B}(\varphiv_\text{B})$ and $\hat\av_\text{U} = \sqrt{\alpha}\tilde\av_\text{U}(\varphiv_\text{U})\Cm_\text{U}^\top$ from~\eqref{eq:hwi_2}. In order to maximize the SNR with the amplitude constraints of the precoder and combiner, we choose $\wv$ and $\vv$ respectively as the conjugate of $\hat \av_\text{B}$ and $\hat \av_\text{U}$ with each of the elements normalized to a unit amplitude. For each realization ($500$ in total), 20 OFDM symbols are sent with data drawn randomly from 16-\ac{qam}, and \ac{ser} is used to evaluate the effect of HWIs on {communications}.

For localization, the pilot signal $x_{g,k}$ is chosen with random phase and a constant amplitude $|x_{g,k}|^2 = {P/{N_\text{U}}}$. 
{To assist the beamspace ESPRIT algorithm, we set the number of sweeping beams as $M_1=4$, $M_2=4$, $M_3=3$, $M_4=3$ with a total number of transmission $G = 144$. For a specific spatial frequency vector $\bar \omegav_n$ ($n\in \{1,2,3,4\}$), we assume the sweeping range as $(M_n-1)\Delta_\omega$ centered at the location prior $\mathring\omega_n = \omega_n + \delta_\omega$, where $\omega_n$ is defined in~\eqref{eq:omega_1_2},~\eqref{eq:omega_3_4}, and $\delta_\omega$ is the error). More specifically, we choose $\bar \omega_{n,m} = \omega_n + \delta_\omega + \frac{2m-M_n-1}{2}\Delta_\omega$, with $\Delta_\omega=0.15$ and $\delta_\omega = 0.05$ in the simulation.}
The sweeping priority is set to `BS-first' by default, which means that the UE can change its precoder vector when the BS finishes the $M_1M_2 = 16$ different sweeping beams. Different error bounds (i.e., CRBs, LBs, ALBs defined in~\eqref{eq:AAEB}--\eqref{eq:OEB}, ~\eqref{eq:LB_channel_parameters}, and~\eqref{eq:state_ALB}) are utilized as localization performance metrics.


\begin{table}[ht]
\scriptsize
\centering
\caption{Default Simulation Parameters}
\vspace{-0.3cm}
\renewcommand{\arraystretch}{1.}
\begin{tabular} {c | c | c }
    \hthickline
    \textbf{Parameters} & \textbf{True Model} & \textbf{Mismatched Model}\\
    \hline
    BS Positions & \multicolumn{2}{c}{$\pv^{1}_\text{B} = [0, 0, 3]^\top$, $\pv^{2}_\text{B} = [0, 10, 3]^\top$} \\
    \hline
    BS Orientations & \multicolumn{2}{c}{$\ov^{2}_\text{B} = [-30^\circ, 15^\circ, 0^\circ]^\top$, $\ov^{1}_\text{B} = [0^\circ, 15^\circ, 0^\circ]^\top$} \\
    \hline
    BS Antennas & \multicolumn{2}{c}{$N^1_\text{B} = N^2_\text{B} = 8\times 8$} \\
    \hline
    UE Position & \multicolumn{2}{c}{$\pv_\text{U} = [8, 4, 0]^\top$, $\ov_\text{U} = [180^\circ, 0^\circ, 0^\circ]^\top$} \\
    \hline
    UE Orientation & \multicolumn{2}{c}{$\ov_\text{U} = [180^\circ, 0^\circ, 0^\circ]^\top$}\\
    \hline
    UE Antennas & \multicolumn{2}{c}{$N_\text{U} = 4\times 4$}   
    \\
    \hline
    Load Impedance & \multicolumn{2}{c}{$R = \unit[50]{\Omega}$} \\
    \hline
    Noise PSD & \multicolumn{2}{c}{ $N_0 = \unit[-173.855]{dBm/Hz}$}  \\
    \hline
    Noise Figure & \multicolumn{2}{c}{$\unit[10]{dB}$} 
    \\
    \hthickline
    Phase Noise & $ \sigma_\text{PN} = 2.0^\circ$ & $\sigma_\text{PN} = 0^\circ$ \\
    \hline
    Carrier Freq. Offset & $\sigma_\text{CFO} = 2e{-4}$
    & $\sigma_\text{CFO} = 0$ \\
    \hline
    Mutual Coupling & $\sigma_\text{MC} = 0.001$ & $\sigma_\text{MC} = 0$ 
    \\
    \hline
     & $\beta_1=$ $0.9798$+$0.0286j$ &
    \\
    Power Amplifier & $\beta_2=$ $0.0122$-$0.0043j$ & n/a
    \\
     & $\beta_3=$ $-0.0007$+$0.0001j$ &
    \\
    \hline
    PA Clipping Voltage & {$x_\text{clip}=\unit[1]{V}$} & n/a
    \\
    \hline
    {Array Gain Error} & $\sigma_\text{GA} = \sigma_\text{GP} = 0.002$ & $\sigma_\text{RA} = \sigma_\text{RP} = 0$ 
    \\
    \hline
    Antenna Disp. Error & $\sigma_\text{AD} = \unit[5]{um}$ ($1e{-3}\lambda$)& $\sigma_\text{AD} = 0$ 
    \\
    \hline
    IQ Imbalance & $\sigma_\text{IA} = \sigma_\text{IP} = {0.02}$ & $\sigma_\text{IA} = \sigma_\text{IP} = 0$ 
    \\
    \hthickline
\end{tabular}
\renewcommand{\arraystretch}{1}
\label{table:Simulation_parameters}
\end{table}


\subsection{The Effect of HWIs on {Communications}}

\subsubsection{The Effect of HWIs on SER}
{To have a tangible performance evaluation of HWIs on {communications}, adding Gaussian noise to the transmitted and received signals is usually used (see~\cite[(7.9)]{schenk2008rf}). We approximate the HWI-introduced effect on {communications} as random noise, which shows a good alignment between the Monte Carlo-based numerical evaluation (that considers the true characteristics of HWIs without any approximation) and the theoretical SER evaluation (with random noise approximation of HWIs)}\footnote{{Minimum Euclidean distance decoding is adopted}, and the SER of M-QAM can be calculated as $\text{SER}_M = 1-(1-\frac{2\sqrt{M}-1}{\sqrt{M}}\text{Q}(\sqrt{\frac{3 \text{SNR}}{M-1}}))^2$ ~\cite[(6,23)]{goldsmith2005wireless}, where $\text{Q}(\cdot)$ is the Q-function and $\text{SNR}$ is effective SNR considering both approximated HWI noise and background noise.}. 
{The rationale behind adopting this approach in our context is grounded in several key factors. First, we work with OFDM systems with hundreds of subcarriers, allowing us to leverage the central limit theorem to effectively model HWIs in the frequency domain as Gaussian noise. Second, our focus primarily involves residual HWIs, which, by their very nature, are inherently small. It’s worth noting that in a broader context, where HWIs may assume different characteristics, a more careful approach to modeling might be needed to comprehensively evaluate their impact on communication systems.}
Considering that the effects of some HWIs depend on the amplitude of the symbol (e.g., PAN), we also obtain the minimum and maximum noise levels across different symbols to evaluate the lower bound and upper bound of the SER. The SERs of 16-QAM with different transmit power for different HWI coefficients are visualized in Fig.~\ref{fig:sim_commun_SER}, where the black solid curve is the benchmark SER without HWIs. By default, $c_\text{HWI} = 1$, and the HWI level is the same as the parameters in Table~\ref{table:Simulation_parameters}. A value of $c_\text{HWI} = 2$ indicates that the standard deviations (e.g., $\sigma_\text{PN}, \sigma_\text{CFO}$) of all the impairments (except for PAN) are multiplied by $2$. 

We can see from the figure that the analytical SERs with approximated noise levels are close to the numerical SERs {under certain HWI levels (see markers and solid curves for $c_\text{HWI}=0.1/1/2/3$)}, and both are within the lower and upper bounds (shaded areas). However, the approximated SERs are less accurate when the impairment level is high (see {cross markers for $c_\text{HWI}=5$} at high transmit powers). {Considering the communication systems may not work properly at such a high HWI level (where SER $>0.1$), the approximation of the HWI-introduced effect on communications as Gaussian noise in this work is reasonable in a practical setting}. We can also see from Fig.~\ref{fig:sim_commun_SER} that the selected impairment level (i.e., $c_\text{HWI}=1$) has limited effects on {communications}. Nevertheless, we will show the localization performance will be affected by the same level of HWIs in Sec.~\ref{sec:effect_of_hwi_on_localization}.

\begin{figure}[h]
\centering
%
%
\newcommand\ms{0.8}
\newcommand\lw{0.3}
\definecolor{mycolor1}{rgb}{0.00000,1.00000,1.00000}%
\definecolor{mycolor2}{rgb}{1.00000,0.00000,1.00000}%


\begin{tikzpicture}[font = \footnotesize]
\begin{axis}[%
width=6.9cm,
height=4.8cm,
at={(0in,0in)},
scale only axis,
xmin=-15,
xmax=5,
xlabel style={yshift=-0.0 ex},
xlabel={$P$ [dBm]},
ymode=log,
ymin=1e-10,
ymax=1,
ylabel style={yshift=0 ex},
ylabel={SER (16-QAM)}, axis background/.style={fill=white},
xmajorgrids,
ymajorgrids,
legend columns = 1, 
legend style={font=\tiny, at={(0, 0)}, anchor=south west, legend cell align=left, align=left, draw=white!15!black}
]

\addplot [color=black, mark=+, mark options={solid, black}]
  table[row sep=crcr]{%
-15	0.616770738277983\\
-12	0.455746367950762\\
-9	0.260185481749494\\
-6	0.0909036669918748\\
-3	0.0125531863141259\\
0	0.000295174723457636\\
3	2.17252571266258e-07\\
6	1.64757096854373e-13\\
9	0\\
12	0\\
15	0\\
};
\addlegendentry{Anal. without HWI}

\addplot [color=green, only marks, mark=diamond, mark options={solid, green}]
  table[row sep=crcr]{%
-15	0.624735\\
-12	0.4689275\\
-9	0.27468875\\
-6	0.1022125\\
-3	0.0165625\\
0	0.0006725\\
3	2.5e-06\\
6	0\\
9	0\\
12	0\\
15	0\\
};
\addlegendentry{Numer. HWI ($c_\text{HWI}$=$0.1$)}

\addplot [color=green]
  table[row sep=crcr]{%
-15	0.617515125204392\\
-12	0.45774942805767\\
-9	0.264436531729406\\
-6	0.0965684646974645\\
-3	0.0156288621305681\\
0	0.000652276327125278\\
3	3.45884375230199e-06\\
6	1.37518818554838e-09\\
9	8.26005930321116e-14\\
12	0\\
15	0\\
};
\addlegendentry{Approx. HWI ($c_\text{HWI}$=$0.1$)}

\addplot [color=blue, only marks, mark=o, mark options={solid, blue}]
  table[row sep=crcr]{%
-15	0.6268675\\
-12	0.47120125\\
-9	0.279285\\
-6	0.1093525\\
-3	0.02133375\\
0	0.00146125\\
3	3.375e-05\\
6	0\\
9	0\\
12	0\\
15	0\\
};
\addlegendentry{Numer. HWI ($c_\text{HWI}$=$1$)}

\addplot [color=blue]
  table[row sep=crcr]{%
-15	0.618503192186044\\
-12	0.460425828391916\\
-9	0.269840900918926\\
-6	0.103977575294444\\
-3	0.020270828848776\\
0	0.00154355048830346\\
3	3.57138382279798e-05\\
6	4.21510472747855e-07\\
9	6.57399334969e-09\\
12	2.40429898212824e-10\\
15	3.95261601227048e-11\\
};
\addlegendentry{Approx. HWI ($c_\text{HWI}$=$1$)}

\addplot [color=red, only marks, mark=square, mark options={solid, red}]
  table[row sep=crcr]{%
-15	0.63103375\\
-12	0.4830175\\
-9	0.30098125\\
-6	0.13709125\\
-3	0.04263125\\
0	0.0098775\\
3	0.0031325\\
6	0.0015425\\
9	0.0007075\\
12	0.000485\\
15	0.00039125\\
};
\addlegendentry{Numer. HWI ($c_\text{HWI}$=$2$)}

\addplot [color=red]
  table[row sep=crcr]{%
-15	0.622436087791632\\
-12	0.471327677342691\\
-9	0.29209857835157\\
-6	0.134901586353048\\
-3	0.0435271419188652\\
0	0.0107050454990176\\
3	0.00293866297255907\\
6	0.00110752755570309\\
9	0.000424560026935583\\
12	0.000297370284128351\\
15	0.00016835739345078\\
};
\addlegendentry{Approx. HWI ($c_\text{HWI}$=$2$)}

\addplot [color=mycolor1, only marks, mark=triangle, mark options={solid, mycolor1}]
  table[row sep=crcr]{%
-15	0.64335875\\
-12	0.50578375\\
-9	0.34287125\\
-6	0.20660875\\
-3	0.10435875\\
0	0.0602725\\
3	0.04226\\
6	0.0358\\
9	0.032155\\
12	0.02767875\\
15	0.033295\\
};
\addlegendentry{Numer. HWI ($c_\text{HWI}$=$3$)}

\addplot [color=mycolor1]
  table[row sep=crcr]{%
-15	0.631654210157399\\
-12	0.494123351833703\\
-9	0.340207995368778\\
-6	0.214576486066541\\
-3	0.113682214527113\\
0	0.0667702907600619\\
3	0.0455842026151033\\
6	0.0374851111468808\\
9	0.033427238086229\\
12	0.0268216720211664\\
15	0.0323698932832588\\
};
\addlegendentry{Approx. HWI ($c_\text{HWI}$=$3$)}

\addplot [color=mycolor2, only marks, mark=+, mark options={solid, mycolor2}]
  table[row sep=crcr]{%
-15	0.690095\\
-12	0.59067625\\
-9	0.4925125\\
-6	0.3943975\\
-3	0.342115\\
0	0.31522375\\
3	0.2956\\
6	0.2929275\\
9	0.27860875\\
12	0.27500625\\
15	0.27249875\\
};
\addlegendentry{Numer. HWI ($c_\text{HWI}$=$5$)}

\addplot [color=mycolor2]
  table[row sep=crcr]{%
-15	0.671763597835234\\
-12	0.584485659349383\\
-9	0.507045477332981\\
-6	0.436822564111683\\
-3	0.430142489012012\\
0	0.395425806104887\\
3	0.386175340178631\\
6	0.391047409033463\\
9	0.377189035259085\\
12	0.383858157691694\\
15	0.379025003771476\\
};
\addlegendentry{Approx. HWI ($c_\text{HWI}$=$5$)}

\addplot [name path = A, color=green, line width=\lw pt, forget plot, opacity=0.1]
  table[row sep=crcr]{%
-15	0.615746873738333\\
-12	0.45569089841177\\
-9	0.259655228927956\\
-6	0.0907509645833825\\
-3	0.0125667163435749\\
0	0.00031899194647278\\
3	3.53017645893239e-07\\
6	1.00297548044637e-12\\
9	0\\
12	0\\
15	0\\
};
\addplot [name path = B, color=green, line width=\lw pt, forget plot, opacity=0.1]
  table[row sep=crcr]{%
-15	0.618186740726178\\
-12	0.459681724577448\\
-9	0.267735779431147\\
-6	0.100385138570738\\
-3	0.0168380231218399\\
0	0.000802243139002901\\
3	6.61532308343649e-06\\
6	7.10215108945533e-09\\
9	2.94342328288622e-12\\
12	1.77635683940025e-15\\
15	0\\
};
\addplot[fill = green, area legend, fill opacity=0.1] fill between[of=A and B];

\addplot [name path = A, color=blue, line width=\lw pt, forget plot, opacity=0.1]
  table[row sep=crcr]{%
-15	0.616083209302574\\
-12	0.457760002168733\\
-9	0.262691921506294\\
-6	0.0961950075417868\\
-3	0.0157931091771299\\
0	0.000681341419514703\\
3	4.01004688455053e-06\\
6	2.25508678397546e-09\\
9	2.4580337765201e-13\\
12	0\\
15	0\\
};
\addplot [name path = B, color=blue, line width=\lw pt, forget plot, opacity=0.1]
  table[row sep=crcr]{%
-15	0.619434256757882\\
-12	0.461555843920473\\
-9	0.272462623436988\\
-6	0.107363915304938\\
-3	0.0218477122873224\\
0	0.00200971250771875\\
3	6.4996071281187e-05\\
6	1.36909322334677e-06\\
9	4.28664783669319e-08\\
12	3.64832364319057e-09\\
15	9.73896074896174e-10\\
};
\addplot[fill = blue, area legend, fill opacity=0.1] fill between[of=A and B];

\addplot [name path = A, color=red, line width=\lw pt, forget plot, opacity=0.1]
  table[row sep=crcr]{%
-15	0.618512992347192\\
-12	0.463683551339138\\
-9	0.275988836397762\\
-6	0.112713983683098\\
-3	0.0273006467482081\\
0	0.00330791712650313\\
3	0.000278437338095516\\
6	2.6904110881909e-05\\
9	3.74680402859262e-06\\
12	9.69216895296832e-07\\
15	4.47742252940309e-07\\
};
\addplot [name path = B, color=red, line width=\lw pt, forget plot, opacity=0.1]
  table[row sep=crcr]{%
-15	0.624111832102431\\
-12	0.47417584162697\\
-9	0.298632975529787\\
-6	0.142849187726499\\
-3	0.0498474956829649\\
0	0.0142011958326349\\
3	0.00494329727713616\\
6	0.0022012175669861\\
9	0.000990790303817324\\
12	0.000747942458369288\\
15	0.000486750365569333\\
};
\addplot[fill = red, area legend, fill opacity=0.1] fill between[of=A and B];

\addplot [name path = A, color=mycolor1, line width=\lw pt, forget plot, opacity=0.1]
  table[row sep=crcr]{%
-15	0.62349801410921\\
-12	0.472571670953708\\
-9	0.298032668864143\\
-6	0.147447676689721\\
-3	0.0527291672936011\\
0	0.0168402026726285\\
3	0.00564267803353247\\
6	0.00281228533230493\\
9	0.00176081662787153\\
12	0.00105437279633236\\
15	0.0011446544436774\\
};
\addplot [name path = B, color=mycolor1, line width=\lw pt, forget plot, opacity=0.1]
  table[row sep=crcr]{%
-15	0.634267730518193\\
-12	0.502070748630758\\
-9	0.354182618431027\\
-6	0.236242655501247\\
-3	0.135910454959452\\
0	0.0883195768856193\\
3	0.0659238388267194\\
6	0.0563855740048248\\
9	0.0510223759473172\\
12	0.0434825850253474\\
15	0.0511494643284338\\
};
\addplot[fill = mycolor1, area legend, fill opacity=0.1] fill between[of=A and B];

\addplot [name path = A, color=mycolor2, line width=\lw pt, forget plot, opacity=0.1]
  table[row sep=crcr]{%
-15	0.638452364754202\\
-12	0.511254043280615\\
-9	0.373513183666042\\
-6	0.253064137842824\\
-3	0.182896626763448\\
0	0.134483962175911\\
3	0.111400840378318\\
6	0.103639446672689\\
9	0.0936803760743438\\
12	0.0942474125484434\\
15	0.0884894720737328\\
};
\addplot [name path = B, color=mycolor2, line width=\lw pt, forget plot, opacity=0.1]
  table[row sep=crcr]{%
-15	0.68218042701311\\
-12	0.604092993268681\\
-9	0.537560961108009\\
-6	0.476485190629031\\
-3	0.478437464836714\\
0	0.448427328892912\\
3	0.444458833481582\\
6	0.446139902800246\\
9	0.432158335688738\\
12	0.440629090566694\\
15	0.437653609594342\\
};
\addplot[fill = mycolor2, area legend, fill opacity=0.1] fill between[of=A and B];

\end{axis}

\end{tikzpicture}%
\vspace*{-1.2cm}
\caption{{The effect of different HWI levels on SER. The boundaries of the shadow areas indicate the upper and lower bounds for SER.}}
\label{fig:sim_commun_SER}
\end{figure}
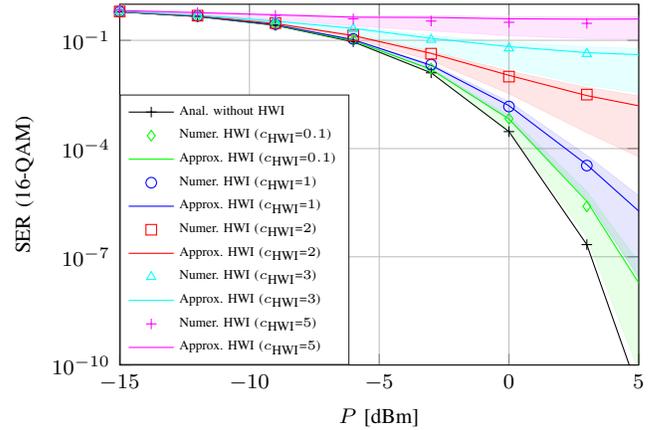

\begin{figure}[h]
\centering
%
\newcommand\ms{0.8}
\newcommand\lw{1.0}
\definecolor{mycolor1}{rgb}{0.00000,1.00000,1.00000}%
\begin{tikzpicture}[font=\footnotesize, spy using outlines={rectangle, magnification=2.5, size=0.3cm, connect spies}]
\begin{axis}[%
width=6cm,
height=3.2cm,
at={(0in,0in)},
scale only axis,
xmin=-10,
xmax=10,
xlabel style={yshift=-0.0 ex},
xlabel={$P$ [dBm]},
ymode=log,
ymin=1e-10,
ymax=1,
ylabel style={yshift=0 ex},
ylabel={SER (16QAM)},
axis background/.style={fill=white},
xmajorgrids,
ymajorgrids,
legend columns=1, 
legend style={font=\tiny, at={(0, 0)}, anchor=south west, legend cell align=left, align=left, draw=white!15!black}
]
\addplot [color=green, line width=\lw pt, mark=diamond, mark options={solid, green}]
  table[row sep=crcr]{%
-10	0.329848345240201\\
-8	0.201630385927599\\
-6	0.0964710539742797\\
-4	0.0322827374375819\\
-2	0.0064914010267042\\
0	0.000645670411711596\\
2	2.54570153140898e-05\\
4	3.29395865250603e-07\\
6	1.33573463401149e-09\\
8	2.10564898850407e-12\\
10	2.22044604925031e-15\\
};
\addlegendentry{PAN}

\addplot [color=blue, line width=\lw pt, mark=o, mark options={solid, blue}]
  table[row sep=crcr]{%
-10	0.328925068898253\\
-8	0.200274385752016\\
-6	0.0949489776268799\\
-4	0.0310617790415498\\
-2	0.0059565055150852\\
0	0.000528949970261428\\
2	1.60411474139366e-05\\
4	1.383980273717e-07\\
6	2.03319583391703e-10\\
8	6.79456491070596e-14\\
10	0\\
};
\addlegendentry{PN}

\addplot [color=mycolor1, line width=\lw pt, mark size=\ms pt, mark=triangle, mark options={solid, rotate=270, mycolor1}]
  table[row sep=crcr]{%
-10	0.327664597153582\\
-8	0.199462602729708\\
-6	0.0939975979299681\\
-4	0.030555726493588\\
-2	0.00552489611468721\\
0	0.000506786746632071\\
2	1.14678782171529e-05\\
4	7.81402755833938e-08\\
6	1.06682440659256e-10\\
8	1.35447209004269e-14\\
10	0\\
};
\addlegendentry{IQI}

\addplot [color=red, line width=\lw pt, mark=square, mark options={solid, red}]
  table[row sep=crcr]{%
-10	0.326712773585771\\
-8	0.197216722432494\\
-6	0.0916889577972865\\
-4	0.0284353221412676\\
-2	0.00495762011154288\\
0	0.000331389392794645\\
2	5.15333990647182e-06\\
4	8.45396375126484e-09\\
6	5.81756864903582e-13\\
8	0\\
10	0\\
};
\addlegendentry{CFO}

\addplot [color=black, dashed, line width=\lw pt, mark size=2.5 pt, mark=+, mark options={solid, black}]
  table[row sep=crcr]{%
-10	0.326449998395406\\
-8	0.196699164110864\\
-6	0.0909297520387533\\
-4	0.0280441135332605\\
-2	0.00465623771922086\\
0	0.000296197231960393\\
2	4.19275434604405e-06\\
4	5.54618706516408e-09\\
6	1.72972747236599e-13\\
8	0\\
10	0\\
};
\addlegendentry{MC+AGE+ADE}

\addplot [color=black, line width=\lw pt]
  table[row sep=crcr]{%
-10	0.326440583133852\\
-8	0.196677457022276\\
-6	0.0909036669918748\\
-4	0.0280281480560545\\
-2	0.00464891091095554\\
0	0.000295174723457636\\
2	4.15664518271797e-06\\
4	5.42932743208269e-09\\
6	1.64757096854373e-13\\
8	0\\
10	0\\
};
\addlegendentry{Without HWI}


\begin{scope}
    \spy[black, size=1.5cm] on (3.7, 1.55) in node [fill=none] at (5.0, 2.55);
\end{scope}

\end{axis}

\end{tikzpicture}%
\vspace*{-1.cm}
\caption{The effect of individual HWIs on SER using approximated equivalent HWI noise. Under current simulation parameters, the PN, PAN, CFO and IQI increase the SER, whereas the MC, AGE and ADE have negligible effects on {communications}.}
\label{fig:sim_commun_individual}
\end{figure}
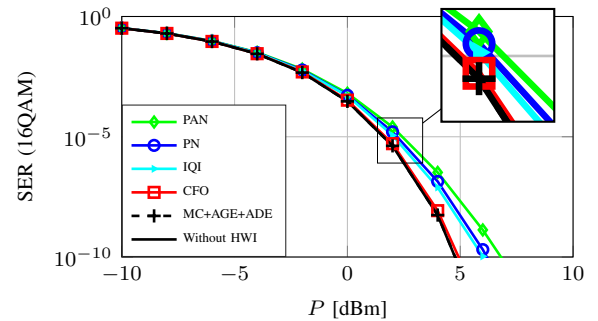


\subsubsection{The Effect of Individual HWIs on SER}
We are also interested in the effect of individual HWIs on {communications}. By considering PN, CFO, PAN, and IQI one by one, the SERs under HWI are shown in Fig.~\ref{fig:sim_commun_individual}. Benchmarked by the solid black curve without HWIs, these factors degrade SERs. We also performed simulations by including MC, AGE, ADE at the same time, as shown in the dashed curve with cross markers, and found their effects on {communications} are negligible under the current simulation setup.

\subsubsection{Insights into the Impact of HWI on {Communications}}
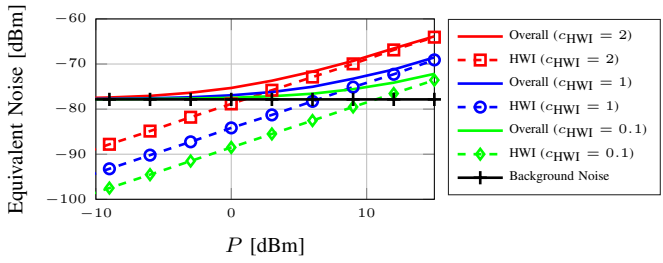
\begin{figure}[h]
\centering
%
\newcommand\ms{0.8}
\newcommand\lw{1.0}

\pgfplotsset{every tick label/.append style={font=\tiny}}

\begin{tikzpicture}[font=\footnotesize]
\begin{axis}[%
width=4.5cm,
height=2.4cm,
at={(0cm,0cm)},
scale only axis,
xmin=-10,
xmax=15,
xlabel style={yshift=0 ex},
xlabel={$P$ [dBm]},
ymin=-100,
ymax=-60,
ylabel style={yshift=0 ex},
ylabel={Equivalent Noise [dBm]},
axis background/.style={fill=white},
xmajorgrids,
ymajorgrids,
legend columns=1, 
legend style={font=\tiny, at={(1.04, 1)}, anchor=north west, legend cell align=left, align=left, draw=white!15!black}
]
\addplot [color=red, line width=1.0pt]
  table[row sep=crcr]{%
-15	-77.7348761018583\\
-12	-77.6336940327339\\
-9	-77.4211527097006\\
-6	-77.0602075239105\\
-3	-76.3710405607225\\
0	-75.3072404068125\\
3	-73.7019537393803\\
6	-71.6833973646499\\
9	-69.3110864520982\\
12	-66.5418813008339\\
15	-63.8669435788683\\
};
\addlegendentry{Overall ($c_\text{HWI} = 2$)}

\addplot [color=red, dashed, mark=square, mark options={solid, red}, line width=1.0pt]
  table[row sep=crcr]{%
-15	-93.8149615030985\\
-12	-90.8901086509082\\
-9	-87.8005603649831\\
-6	-84.8724091199303\\
-3	-81.7973725342897\\
0	-78.8469617865327\\
3	-75.8118963428734\\
6	-72.8887758258097\\
9	-69.9671978322538\\
12	-66.8777615345624\\
15	-64.0460937029184\\
};
\addlegendentry{HWI ($c_\text{HWI} = 2$)}

\addplot [color=blue, line width=1.0pt]
  table[row sep=crcr]{%
-15	-77.813934251058\\
-12	-77.7812350836043\\
-9	-77.718075909762\\
-6	-77.5978140312669\\
-3	-77.3708888547569\\
0	-76.9340712183701\\
3	-76.216089093957\\
6	-75.0483156629028\\
9	-73.2576601482282\\
12	-71.2131695673334\\
15	-68.5405166035526\\
};
\addlegendentry{Overall ($c_\text{HWI} = 1$)}

\addplot [color=blue, dashed, mark=o, mark options={solid, blue}, line width=1.0pt]
  table[row sep=crcr]{%
-15	-99.1025735444762\\
-12	-96.1960417338383\\
-9	-93.2316212742407\\
-6	-90.2023504602666\\
-3	-87.2555129496177\\
0	-84.1663494498703\\
3	-81.2847575007912\\
6	-78.2781374832354\\
9	-75.1161268937895\\
12	-72.2752588163991\\
15	-69.0844162250831\\
};
\addlegendentry{HWI ($c_\text{HWI} = 1$)}

\addplot [color=green, line width=1.0pt]
  table[row sep=crcr]{%
-15	-77.8232578139309\\
-12	-77.8151906116117\\
-9	-77.7949484038651\\
-6	-77.743464709825\\
-3	-77.6653962078956\\
0	-77.489590315322\\
3	-77.1499984226497\\
6	-76.5779517549924\\
9	-75.5829734772294\\
12	-74.1360206619677\\
15	-72.1835894481848\\
};
\addlegendentry{Overall ($c_\text{HWI} = 0.1$)}

\addplot [color=green, dashed, mark=diamond, mark options={solid, green}, line width=1.0pt]
  table[row sep=crcr]{%
-15	-103.469564211586\\
-12	-100.511332231514\\
-9	-97.5011358694292\\
-6	-94.5326716713431\\
-3	-91.5105311905103\\
0	-88.4944769965373\\
3	-85.50564948193\\
6	-82.5355485287052\\
9	-79.5166545467093\\
12	-76.5423408434657\\
15	-73.5625757478847\\
};
\addlegendentry{HWI ($c_\text{HWI} = 0.1$)}

\addplot [color=black, line width=1.0pt, mark size=2.5pt, mark=+, mark options={solid, black}]
  table[row sep=crcr]{%
-15	-77.8446732160197\\
-12	-77.8446732160197\\
-9	-77.8446732160197\\
-6	-77.8446732160197\\
-3	-77.8446732160197\\
0	-77.8446732160197\\
3	-77.8446732160197\\
6	-77.8446732160197\\
9	-77.8446732160197\\
12	-77.8446732160197\\
15	-77.8446732160197\\
};
\addlegendentry{Background Noise}

\end{axis}

\end{tikzpicture}%
\vspace{-1.2cm}
\caption{Visualization of overall system noise, equivalent HWI noise, and background noise with different transmit power $P$. The background noise has a large effect on {communications} in low transmit power, whereas the HWIs contribute more in high transmit power.}
\label{fig:sim_commun_hwi_noise}
\end{figure}

To gain further insight into the effects of HWI on {communications}, we separate the overall system noise into equivalent HWI noise and background noise. We can see from Fig.~\ref{fig:sim_commun_hwi_noise} that the equivalent HWI noise is determined by the HWI level and has an approximately linear relationship with the transmit power (when working within the linear region of the PA). In addition to the fixed background noise, the overall equivalent noise level keeps increasing and is dominated by the HWIs at high transmit power. 

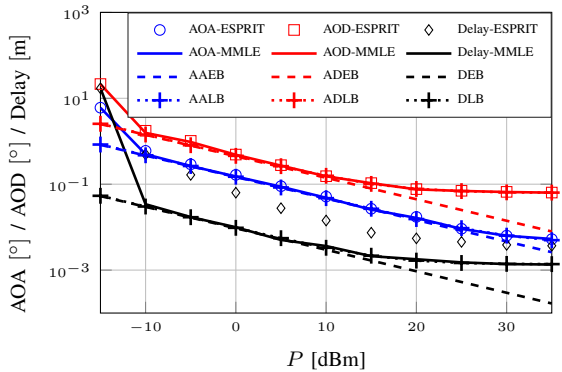
\begin{figure}
\centering
%

\pgfplotsset{every tick label/.append style={font=\tiny}}

\begin{tikzpicture}[font=\footnotesize]
\begin{axis}[%
width=6cm,
height=4.cm,
at={(0cm, 0cm)},
scale only axis,
xmin=-15,
xmax=35,
xlabel style={yshift=0 ex},
xlabel={$P$ [dBm]},
ymode=log,
ymin=1e-04,
ymax=1000,
ylabel style={yshift=0 ex},
ylabel={\!\!\!\!\!\!\!\! AOA $[^\circ]$\ /\ AOD $[^\circ]$\ /\ Delay $[\text{m}]$},
axis background/.style={fill=white},
xmajorgrids,
ymajorgrids,
legend columns=3, 
legend style={font=\tiny, at={(1, 1)}, anchor=north east, legend cell align=left, align=left, draw=white!15!black}
]
\addplot [color=blue, only marks, mark size=2pt, mark=o, mark options={solid, blue}]
  table[row sep=crcr]{%
-15	6.04602024934994\\
-10	0.59825150211909\\
-5	0.290188519506362\\
0	0.165047106969703\\
5	0.0898280256992932\\
10	0.0514623687721921\\
15	0.0268095673261553\\
20	0.0173250836267981\\
25	0.00919065712867203\\
30	0.00634825813200306\\
35	0.00526674276588774\\
};
\addlegendentry{AOA-ESPRIT}

\addplot [color=red, only marks, mark size=2pt, mark=square, mark options={solid, red}]
  table[row sep=crcr]{%
-15	21.4626121755985\\
-10	1.75261019109287\\
-5	1.01867635141198\\
0	0.489722448254146\\
5	0.277473511272137\\
10	0.153530930872806\\
15	0.10589597749886\\
20	0.0768684103370579\\
25	0.0702747574992961\\
30	0.0663018209747137\\
35	0.0638954530261008\\
};
\addlegendentry{AOD-ESPRIT}

\addplot [color=black, only marks, mark size=2pt, mark=diamond, mark options={solid, black}]
  table[row sep=crcr]{%
-15	17.0958172634971\\
-10	0.493343514362658\\
-5	0.164686643426564\\
0	0.0634284506528163\\
5	0.0276295524867484\\
10	0.0142658006733206\\
15	0.00741650906323063\\
20	0.0054815234113658\\
25	0.00452940621550783\\
30	0.00386221872331136\\
35	0.00368927115625654\\
};
\addlegendentry{Delay-ESPRIT}

\addplot [color=blue, line width=1.0pt]
  table[row sep=crcr]{%
-15	6.02580756075001\\
-10	0.499226439438803\\
-5	0.264883480716925\\
0	0.150936021791062\\
5	0.0873071172348558\\
10	0.0482791479013298\\
15	0.0255589997583025\\
20	0.0163867591339002\\
25	0.00910725970400661\\
30	0.00630313794269385\\
35	0.00530889494567289\\
};
\addlegendentry{AOA-MMLE}

\addplot [color=red, line width=1.0pt]
  table[row sep=crcr]{%
-15	21.3413443230748\\
-10	1.56505861839018\\
-5	0.961033103897953\\
0	0.484198454106307\\
5	0.270578207794734\\
10	0.152890401653928\\
15	0.105517587613411\\
20	0.0758135605538942\\
25	0.0696272663378516\\
30	0.0657921735102487\\
35	0.0635358048638214\\
};
\addlegendentry{AOD-MMLE}

\addplot [color=black, line width=1.0pt]
  table[row sep=crcr]{%
-15	17.0789271847216\\
-10	0.0341382789176001\\
-5	0.0171786573187065\\
0	0.00994111642414604\\
5	0.00503932936636499\\
10	0.0036301567223058\\
15	0.00211050844062806\\
20	0.00178821950811182\\
25	0.00151178622925227\\
30	0.00138116726953766\\
35	0.00134805485442369\\
};
\addlegendentry{Delay-MMLE}

\addplot [color=blue, dashed, line width=1.0pt]
  table[row sep=crcr]{%
-15	0.824858033161192\\
-10	0.46385175946177\\
-5	0.260843013107607\\
0	0.146682805657575\\
5	0.0824858033161192\\
10	0.046385175946177\\
15	0.0260843013107607\\
20	0.0146682805657575\\
25	0.00824858033161192\\
30	0.0046385175946177\\
35	0.00260843013107607\\
};
\addlegendentry{AAEB}

\addplot [color=red, dashed, line width=1.0pt]
  table[row sep=crcr]{%
-15	2.50705690328188\\
-10	1.40982170131915\\
-5	0.792801003801922\\
0	0.445824767090212\\
5	0.250705690328188\\
10	0.140982170131915\\
15	0.0792801003801922\\
20	0.0445824767090212\\
25	0.0250705690328188\\
30	0.0140982170131915\\
35	0.00792801003801922\\
};
\addlegendentry{ADEB}

\addplot [color=black, dashed, line width=1.0pt]
  table[row sep=crcr]{%
-15	0.0528150276044225\\
-10	0.0297000726130359\\
-5	0.0167015781914642\\
0	0.00939198761295821\\
5	0.00528150276044225\\
10	0.00297000726130359\\
15	0.00167015781914642\\
20	0.000939198761295819\\
25	0.000528150276044223\\
30	0.000297000726130359\\
35	0.000167015781914642\\
};
\addlegendentry{DEB}

\addplot [color=blue, line width=1.0pt, dotted, mark=+, mark options={solid, blue}, mark size=3pt]
  table[row sep=crcr]{%
-15	0.83838649104029\\
-10	0.471490364364313\\
-5	0.265162261613765\\
0	0.149140212974947\\
5	0.0839419495732236\\
10	0.0473314736852802\\
15	0.0267946564806373\\
20	0.0154406351857627\\
25	0.00947756731006815\\
30	0.00633932029517547\\
35	0.00497286319078655\\
};
\addlegendentry{AALB}

\addplot [color=red, line width=1.0pt, dotted, mark=+, mark options={solid, red}, mark size=3pt]
  table[row sep=crcr]{%
-15	2.54778313994436\\
-10	1.43397917009256\\
-5	0.808341478898183\\
0	0.457108739641626\\
5	0.262827948228591\\
10	0.156799289562748\\
15	0.101715656964481\\
20	0.0775814252164161\\
25	0.0690727987994221\\
30	0.0645899040695847\\
35	0.0637118510446538\\
};
\addlegendentry{ADLB}

\addplot [color=black, line width=1.0pt, dotted, mark=+, mark options={solid, black}, mark size=3pt]
  table[row sep=crcr]{%
-15	0.0537027426351361\\
-10	0.0302219011581906\\
-5	0.0170292592663076\\
0	0.00964333801487445\\
5	0.00553516457913712\\
10	0.00330941875567469\\
15	0.00216432537519992\\
20	0.00166361966312199\\
25	0.00146571879329349\\
30	0.00139561173868715\\
35	0.0013777689418204\\
};
\addlegendentry{DLB}

\end{axis}

\end{tikzpicture}%
\vspace*{-1.2cm}
\caption{Comparison between channel parameters estimation results (ESPRIT and MMLE) and different lower bounds (CRB of the MM and the LB of the mismatched estimator) in terms of AOA, AOD and delay. Due to the HWIs, the performance starts to saturate when the transmit power exceeds~\unit[30]{dBm}.}
\label{fig:sim_vs_crb}
\end{figure}

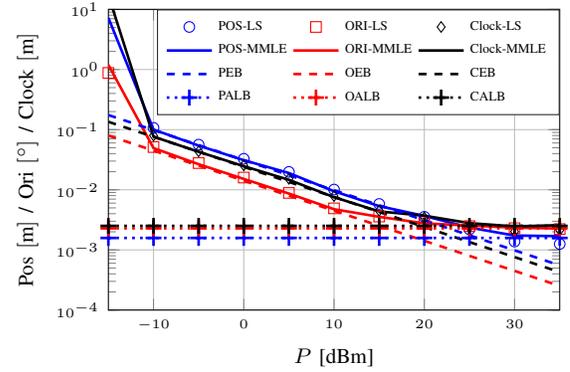
\begin{figure}[h]
\centering
%

\pgfplotsset{every tick label/.append style={font=\tiny}}

\begin{tikzpicture}[font=\footnotesize]
\begin{axis}[%
width=6cm,
height=4.cm,
at={(0cm, 0cm)},
scale only axis,
xmin=-15,
xmax=35,
xlabel style={yshift=0 ex},
xlabel={$P$ [dBm]},
ymode=log,
ymin=1e-04,
ymax=10,
ylabel style={yshift=0 ex},
ylabel={Pos $[\text{m}]$\ /\ Ori $[^\circ]$\ /\ Clock $[\text{m}]$},
axis background/.style={fill=white},
xmajorgrids,
ymajorgrids,
legend columns=3, 
legend style={font=\tiny, at={(1, 1)}, anchor=north east, legend cell align=left, align=left, draw=white!15!black}
]
\addplot [color=blue, only marks, mark size=2pt, mark=o, mark options={solid, blue}]
  table[row sep=crcr]{%
-15	10.1043854650696\\
-10	0.106981996796056\\
-5	0.0563382047357935\\
0	0.03277972942572\\
5	0.0198492347835457\\
10	0.0100966987857285\\
15	0.00581529774887493\\
20	0.00355477250684631\\
25	0.00219834332457156\\
30	0.00137678685224591\\
35	0.00125465966742659\\
};
\addlegendentry{POS-LS}

\addplot [color=red, only marks, mark size=2pt, mark=square, mark options={solid, red}]
  table[row sep=crcr]{%
-15	0.878289601019868\\
-10	0.0513338097288178\\
-5	0.0277879871741601\\
0	0.0159699303903712\\
5	0.008862577274306\\
10	0.00490998634412979\\
15	0.0035889028661028\\
20	0.00276136123020071\\
25	0.00248824140842209\\
30	0.00229754124260689\\
35	0.00222111002578567\\
};
\addlegendentry{ORI-LS}

\addplot [color=black, only marks, mark size=2pt, mark=diamond, mark options={solid, black}]
  table[row sep=crcr]{%
-15	18.1716064378826\\
-10	0.0797171459327886\\
-5	0.0435565137034668\\
0	0.0255386259929106\\
5	0.0152295077121525\\
10	0.0078546303742025\\
15	0.00450168541448982\\
20	0.00345057944004049\\
25	0.00238484242398682\\
30	0.00202144027600615\\
35	0.00212644961591765\\
};
\addlegendentry{Clock-LS}

\addplot [color=blue, line width=1.0pt]
  table[row sep=crcr]{%
-15	7.25307654682912\\
-10	0.100252243034642\\
-5	0.0542705153410365\\
0	0.0310554384801823\\
5	0.0189948234118757\\
10	0.00944987689254342\\
15	0.00535329482769373\\
20	0.00367848988423865\\
25	0.00235520612745945\\
30	0.00173436119152688\\
35	0.0016933298431584\\
};
\addlegendentry{POS-MMLE}

\addplot [color=red, line width=1.0pt]
  table[row sep=crcr]{%
-15	1.1970674370302\\
-10	0.0493546577543515\\
-5	0.0265945067539938\\
0	0.0151501666770208\\
5	0.00865010568993563\\
10	0.00475135489722841\\
15	0.00355263786570641\\
20	0.00276448892590848\\
25	0.0025452593549171\\
30	0.00234652429963439\\
35	0.00227110704789197\\
};
\addlegendentry{ORI-MMLE}

\addplot [color=black, line width=1.0pt]
  table[row sep=crcr]{%
-15	17.330622530903\\
-10	0.0764727187417064\\
-5	0.0427856715451033\\
0	0.0247059752712882\\
5	0.0146438195240023\\
10	0.00753551584671955\\
15	0.00436810825861497\\
20	0.00370771105729542\\
25	0.002792179896671\\
30	0.00245982842049348\\
35	0.00258015134858211\\
};
\addlegendentry{Clock-MMLE}

\addplot [color=blue, dashed, line width=1.0pt]
  table[row sep=crcr]{%
-15	0.17503493881371\\
-10	0.0984293794471132\\
-5	0.0553509076759533\\
0	0.0311261027729843\\
5	0.017503493881371\\
10	0.00984293794471132\\
15	0.00553509076759534\\
20	0.00311261027729842\\
25	0.0017503493881371\\
30	0.000984293794471132\\
35	0.000553509076759533\\
};
\addlegendentry{PEB}

\addplot [color=red, dashed, line width=1.0pt]
  table[row sep=crcr]{%
-15	0.0797224652509474\\
-10	0.0448312367566593\\
-5	0.0252104570876621\\
0	0.0141768818473303\\
5	0.00797224652509474\\
10	0.00448312367566593\\
15	0.00252104570876621\\
20	0.00141768818473303\\
25	0.000797224652509474\\
30	0.000448312367566593\\
35	0.000252104570876621\\
};
\addlegendentry{OEB}

\addplot [color=black, dashed, line width=1.0pt]
  table[row sep=crcr]{%
-15	0.135179399615961\\
-10	0.0760169627184751\\
-5	0.0427474795520526\\
0	0.0240386742998486\\
5	0.0135179399615961\\
10	0.00760169627184751\\
15	0.00427474795520527\\
20	0.00240386742998486\\
25	0.0013517939961596\\
30	0.000760169627184751\\
35	0.000427474795520526\\
};
\addlegendentry{CEB}

\addplot [color=blue, line width=1.0pt, dotted, mark=+, mark options={solid, blue}, mark size=3pt]
  table[row sep=crcr]{%
-15	0.00158146584033585\\
-10	0.00158146584033585\\
-5	0.00158146584033585\\
0	0.00158146584033585\\
5	0.00158146584033585\\
10	0.00158146584033585\\
15	0.00158146584033585\\
20	0.00158146584033585\\
25	0.00158146584033585\\
30	0.00158146584033585\\
35	0.00158146584033585\\
};
\addlegendentry{PALB}

\addplot [color=red, line width=1.0pt, dotted, mark=+, mark options={solid, red}, mark size=3pt]
  table[row sep=crcr]{%
-15	0.0022845895153903\\
-10	0.0022845895153903\\
-5	0.0022845895153903\\
0	0.0022845895153903\\
5	0.0022845895153903\\
10	0.0022845895153903\\
15	0.0022845895153903\\
20	0.0022845895153903\\
25	0.0022845895153903\\
30	0.0022845895153903\\
35	0.0022845895153903\\
};
\addlegendentry{OALB}

\addplot [color=black, line width=1.0pt, dotted, mark=+, mark options={solid, black}, mark size=3pt]
  table[row sep=crcr]{%
-15	0.00251071622290944\\
-10	0.00251071622290944\\
-5	0.00251071622290944\\
0	0.00251071622290944\\
5	0.00251071622290944\\
10	0.00251071622290944\\
15	0.00251071622290944\\
20	0.00251071622290944\\
25	0.00251071622290944\\
30	0.00251071622290944\\
35	0.00251071622290944\\
};
\addlegendentry{CALB}

\end{axis}

\end{tikzpicture}%
\vspace*{-1.2cm}
\caption{Comparison between localization results (position, orientation, and clock offset estimation) and different lower bounds (CRB of the MM and the LB of the mismatched estimator). We noticed that refined results using MMLE attain the ALBs.}
\label{fig:sim_vs_crb_localization}
\end{figure}


\subsection{The Effect of HWIs on Localization}
\label{sec:effect_of_hwi_on_localization}
Before analyzing the HWIs in detail, we first establish the validity of the derived bounds by comparing them against the performance of practical algorithms. 

\subsubsection{Channel Estimation Results}
For convenient analysis, we adopt one specific realization of the HWIs for the system. The results of channel parameters estimation using ESPRIT (circle, square, and diamond markers) and MMLE (solid curves) are shown in Fig.~\ref{fig:sim_vs_crb}. The estimators are benchmarked by the CRBs of the ideal/mismatched model (CRB-MM, dashed curves) and the LB using a mismatched model (dotted curves with cross markers). Note that the average transmit power $P$ is calculated without considering the nonlinearity of the power amplifier (calculated before the PA). When the transmit power $P$ is low, the LB is determined by the MCRB (since the bias part is constant, see~\eqref{eq:LB_channel_parameters}) and has a similar performance as CRBs. {This indicates that in low transmit power, the mismatched model will not significantly affect the performance, as the expected accuracy is low and limited by the noise.} With the increase of transmit power, the contribution of MCRB decreases due to an increased SNR, and eventually, the mismatched localization is lower bounded by the \ac{alb} (bias part in~\eqref{eq:LB_channel_parameters}). This indicates that the localization performance can no longer be improved by increasing transmit power, which cannot be ignored in scenarios requiring high-accuracy localization performance\footnote{{Note that the analysis here is under the same level of residual noise (e.g., PN, CFO, IQI). In practice, the impairment levels  depend on specific HWI calibration algorithms and transmit power.}}. Regarding the estimators, the ESPRIT (using a mismatched model) provides low-complexity results with limited performance in delay estimation. However, the refined results using MMLE can reach the LB (solid curves align well with the dotted curve).

\subsubsection{Localization Results}
\label{sec:localization_results_vs_bounds}
Based on the estimated channel parameters, we are able to estimate the UE position and orientation. Similar to the channel estimation results, two estimators (LS and MMLE) and two bounds (CRB and LB) are evaluated. The results for localization are shown in Fig.~\ref{fig:sim_vs_crb_localization}.
From the figure, we can see that at low transmit powers, the LB and CRBs coincide, implying that the HWIs are not the main source of error. At higher transmit powers ($\unit[10]{dBm}$ for OEB, and $\unit[20]{dBm}$ for PEB), LB deviates from the CRBs, and the positioning performance is thus more severely affected by HWIs. The MMLE (solid curves) in high SNR is close to the ALB (dotted curves with cross markers), indicating the validity of the MCRB analysis.

Now that the validity of the bounds has been established, we rely solely on the bounds to evaluate the effect of HWIs on localization. First, the impairments are studied individually, then the impact of the waveform type is evaluated, and finally, the impairment levels are varied. 

\subsubsection{The Effect of Individual Impairments}

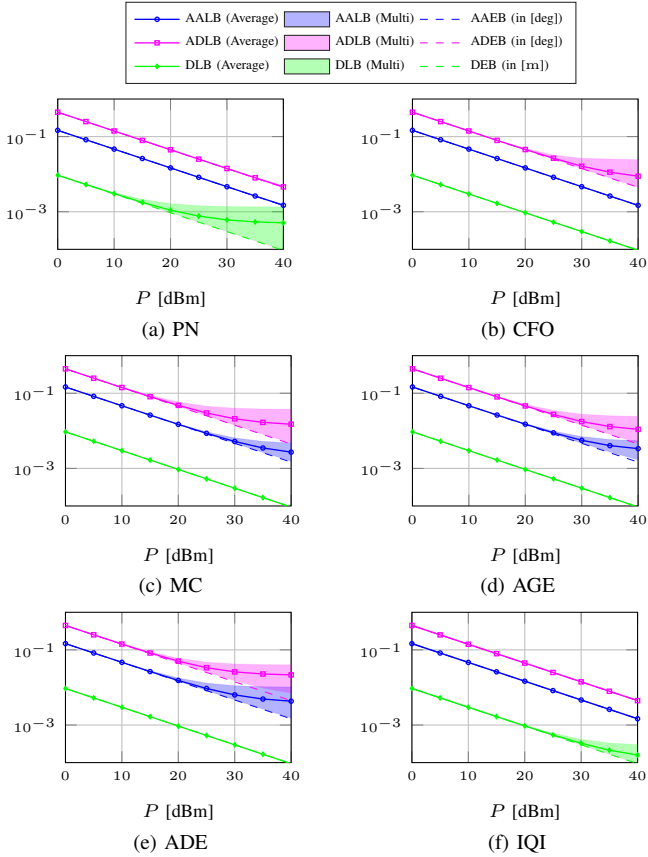
\begin{figure}[h]
\begin{minipage}[b]{0.48\linewidth}
    \centering
%
%
\newcommand\ms{0.8}
\newcommand\lw{0.3}
\definecolor{mycolor1}{rgb}{0.00000,1.00000,1.00000}%
\definecolor{mycolor2}{rgb}{1.00000,0.00000,1.00000}%

\pgfplotsset{every tick label/.append style={font=\tiny}}

\begin{tikzpicture}[font=\footnotesize]
\begin{axis}[%
width=3cm,
height=2cm,
at={(0cm,0cm)},
scale only axis,
xmin=0,
xmax=40,
xlabel style={font=\scriptsize, yshift=-0.5 ex},
xlabel={$P$ [dBm]},
ymode=log,
ymin=1e-04,
ymax=1,
ylabel style={font=\tiny, yshift=-0.5 ex},
axis background/.style={fill=white},
xmajorgrids,
ymajorgrids,
legend columns=3, 
legend style={font=\tiny, at={(0.30, 1.1)}, anchor=south west, legend cell align=left, align=left, draw=white!15!black}
]

\addplot [color=blue, line width=0.5 pt, mark size=\ms pt, mark=o, mark options={solid, blue}]
  table[row sep=crcr]{%
-10	0.464132898681694\\
-5	0.261001152673548\\
0	0.146772119956836\\
5	0.0825364531115454\\
10	0.0464146827540196\\
15	0.0261015592900344\\
20	0.0146799667512639\\
25	0.00825881125913688\\
30	0.00464802177501604\\
35	0.00262181224181218\\
40	0.00148874785589717\\
};
\addlegendentry{AALB (Average)}

\addplot [name path = A, color=blue, line width=\lw pt, forget plot, opacity=0.1]
  table[row sep=crcr]{%
-10	0.464168565847138\\
-5	0.261021971121214\\
0	0.146783285237351\\
5	0.0825423172584355\\
10	0.046421394886368\\
15	0.0261078030881306\\
20	0.0146887106007918\\
25	0.00828257768930167\\
30	0.0046608296957737\\
35	0.00265111079139474\\
40	0.00155997116824001\\
};

\addplot [name path = B, color=blue, line width=\lw pt, forget plot, opacity=0.1]
  table[row sep=crcr]{%
-10	0.464099830599144\\
-5	0.260982702138872\\
0	0.146761788290143\\
5	0.082531311531074\\
10	0.0464112344756824\\
15	0.0260989630318906\\
20	0.0146768656233387\\
25	0.00825376653038682\\
30	0.00464201548924833\\
35	0.00261025793796046\\
40	0.001468434824387\\
};

\addplot[fill = blue, area legend, fill opacity=0.3] fill between[of=A and B];
\addlegendentry{AALB (Multi)}

\addplot [color=blue, dashed, line width= \lw pt]
  table[row sep=crcr]{%
-10	0.46385175946177\\
-5	0.260843013107607\\
0	0.146682805657575\\
5	0.0824858033161191\\
10	0.0463851759461769\\
15	0.0260843013107607\\
20	0.0146682805657575\\
25	0.00824858033161192\\
30	0.0046385175946177\\
35	0.00260843013107607\\
40	0.00146682805657575\\
};
\addlegendentry{AAEB (in [deg])}

\addplot [color=mycolor2, line width=0.5 pt, mark size=\ms pt, mark=square, mark options={solid, mycolor2}]
  table[row sep=crcr]{%
-10	1.4106755754317\\
-5	0.793283323396787\\
0	0.446099067524751\\
5	0.250863022547039\\
10	0.141075954650845\\
15	0.0793343200443781\\
20	0.0446246614534316\\
25	0.0251217166199649\\
30	0.0141523073413055\\
35	0.00800014339122866\\
40	0.00459679050630916\\
};
\addlegendentry{ADLB (Average)}

\addplot [name path = C, color=mycolor2, line width=\lw pt, forget plot, opacity=0.1]
  table[row sep=crcr]{%
-10	1.41077405661686\\
-5	0.793337732099115\\
0	0.446148555289558\\
5	0.250896012478459\\
10	0.141103285727023\\
15	0.079363855735759\\
20	0.0446905756625815\\
25	0.0253803855102106\\
30	0.0142832974241871\\
35	0.00827341421718322\\
40	0.00490404741468044\\
};
\addplot [name path = D, color=mycolor2, line width=\lw pt, forget plot, opacity=0.1]
  table[row sep=crcr]{%
-10	1.41058328213759\\
-5	0.793229109195571\\
0	0.44606612965667\\
5	0.250841725948333\\
10	0.141060125217765\\
15	0.0793233523153941\\
20	0.044607224239751\\
25	0.0250857344509209\\
30	0.0141066225482003\\
35	0.00793378480596841\\
40	0.00446530316196951\\
};

\addplot[fill = mycolor2, area legend, fill opacity=0.3] fill between[of=C and D];
\addlegendentry{ADLB (Multi)}

\addplot [color=mycolor2, dashed, line width= \lw pt]
  table[row sep=crcr]{%
-10	1.40982170131915\\
-5	0.792801003801922\\
0	0.445824767090212\\
5	0.250705690328188\\
10	0.140982170131915\\
15	0.0792801003801922\\
20	0.0445824767090212\\
25	0.0250705690328188\\
30	0.0140982170131915\\
35	0.00792801003801921\\
40	0.00445824767090212\\
};
\addlegendentry{ADEB (in [deg])}

\addplot [color=green, line width=0.5 pt, mark size=\ms pt, mark=diamond, mark options={solid, green}]
  table[row sep=crcr]{%
-10	0.0297241997047896\\
-5	0.0167227186904901\\
0	0.00941730178503654\\
5	0.00531929778455921\\
10	0.00303237151931722\\
15	0.00177412199972842\\
20	0.0011041397250153\\
25	0.000766676323583828\\
30	0.00060771613293375\\
35	0.000537491715899352\\
40	0.000508471448771188\\
};
\addlegendentry{DLB (Average)}

\addplot [name path = E, color=green, line width=\lw pt, forget plot, opacity=0.1]
  table[row sep=crcr]{%
-10	0.0297488298146728\\
-5	0.0167662315881587\\
0	0.0094942873962101\\
5	0.00545222869841775\\
10	0.00326353785266847\\
15	0.002147187096343\\
20	0.00164258253757964\\
25	0.00144846357288771\\
30	0.0013781910069902\\
35	0.00135618991110306\\
40	0.00134980802785997\\
};
\addplot [name path = F, color=mycolor2, line width=\lw pt, forget plot, opacity=0.1]
  table[row sep=crcr]{%
-10	0.0297172382074809\\
-5	0.0167112631692381\\
0	0.00939759172793874\\
5	0.00528476098317435\\
10	0.00297196482790422\\
15	0.00167128196816703\\
20	0.000940288480555695\\
25	0.000529017798876185\\
30	0.000298152375123081\\
35	0.000168756256612251\\
40	9.67678885280831e-05\\
};
\addplot[fill = green, area legend, fill opacity=0.3] fill between[of=E and F];
\addlegendentry{DLB (Multi)}

\addplot [color=green, dashed, line width= \lw pt]
  table[row sep=crcr]{%
-10	0.0297000726130359\\
-5	0.0167015781914642\\
0	0.00939198761295821\\
5	0.00528150276044225\\
10	0.00297000726130359\\
15	0.00167015781914642\\
20	0.000939198761295819\\
25	0.000528150276044224\\
30	0.000297000726130359\\
35	0.000167015781914642\\
40	9.39198761295817e-05\\
};
\addlegendentry{DEB (in [$\unit[]{m}$])}

\end{axis}

\end{tikzpicture}%
    \vspace{-1.cm}
    \centerline{\ \ \ \footnotesize{(a) PN}} \medskip
\end{minipage}
\hspace{0.2cm}
\begin{minipage}[b]{0.48\linewidth}
  \centering
%
%
\newcommand\ms{0.8}
\newcommand\lw{0.3}
\definecolor{mycolor1}{rgb}{0.00000,1.00000,1.00000}%
\definecolor{mycolor2}{rgb}{1.00000,0.00000,1.00000}%

\pgfplotsset{every tick label/.append style={font=\tiny}}

\begin{tikzpicture}[font=\footnotesize]
\begin{axis}[%
width=3cm,
height=2cm,
at={(0cm,0cm)},
scale only axis,
xmin=0,
xmax=40,
xlabel style={font=\scriptsize, yshift=-0.5 ex},
xlabel={$P$ [dBm]},
ymode=log,
ymin=1e-04,
ymax=1,
ylabel style={font=\tiny, yshift=0 ex},
ylabel={}, axis background/.style={fill=white},
xmajorgrids,
ymajorgrids,
legend columns=3, 
legend style={font=\scriptsize, at={(-0.3,1.53)}, anchor=south west, legend cell align=left, align=left, draw=white!15!black}
]

\addplot [color=blue, line width=0.5 pt, mark size=\ms pt, mark=o, mark options={solid, blue}]
  table[row sep=crcr]{%
-10	0.464220347339174\\
-5	0.26105083822848\\
0	0.146800657558956\\
5	0.0825538247518773\\
10	0.0464269375537712\\
15	0.0261058115774317\\
20	0.0146820188995721\\
25	0.00825932944836969\\
30	0.00464745845192511\\
35	0.00261935899949563\\
40	0.00148276360042292\\
};

\addplot [name path = A, color=blue, line width=\lw pt, forget plot, opacity=0.1]
  table[row sep=crcr]{%
-10	0.465173325613365\\
-5	0.261588076830155\\
0	0.147105151053375\\
5	0.0827292896995155\\
10	0.0465370906503689\\
15	0.0261612984658376\\
20	0.0147172585108381\\
25	0.00828410882047913\\
30	0.00467352756443941\\
35	0.00264381134905952\\
40	0.00152055286734253\\
};

\addplot [name path = B, color=blue, line width=\lw pt, forget plot, opacity=0.1]
  table[row sep=crcr]{%
-10	0.463851825819878\\
-5	0.260843050653979\\
0	0.146682827181283\\
5	0.0824858161485828\\
10	0.0463851844584005\\
15	0.026084305368742\\
20	0.0146682832575589\\
25	0.00824858253082451\\
30	0.00463851938479833\\
35	0.00260843188484151\\
40	0.00146683035895964\\
};

\addplot[fill = blue, area legend, fill opacity=0.3] fill between[of=A and B];

\addplot [color=blue, dashed, line width= \lw pt]
  table[row sep=crcr]{%
-10	0.46385175946177\\
-5	0.260843013107607\\
0	0.146682805657575\\
5	0.0824858033161191\\
10	0.0463851759461769\\
15	0.0260843013107607\\
20	0.0146682805657575\\
25	0.00824858033161192\\
30	0.0046385175946177\\
35	0.00260843013107607\\
40	0.00146682805657575\\
};

\addplot [color=mycolor2, line width=0.5 pt, mark size=\ms pt, mark=square, mark options={solid, mycolor2}]
  table[row sep=crcr]{%
-10	1.41094049650167\\
-5	0.793430319226072\\
0	0.446178960638193\\
5	0.250905407665257\\
10	0.141308969230765\\
15	0.0793432565475281\\
20	0.0451970436953675\\
25	0.0266125419783915\\
30	0.0163089695720525\\
35	0.0113222932930645\\
40	0.00881093641369633\\
};

\addplot [name path = C, color=mycolor2, line width=\lw pt, forget plot, opacity=0.1]
  table[row sep=crcr]{%
-10	1.41382250120092\\
-5	0.795051388227346\\
0	0.447091263655317\\
5	0.251419694015662\\
10	0.143426656936634\\
15	0.0795058881712097\\
20	0.0490327844527703\\
25	0.0320502390017011\\
30	0.0264944056506079\\
35	0.0253549474443013\\
40	0.0243120164399581\\
};
\addplot [name path = D, color=mycolor2, line width=\lw pt, forget plot, opacity=0.1]
  table[row sep=crcr]{%
-10	1.40982193169965\\
-5	0.792801133378014\\
0	0.445824839998201\\
5	0.250705731402047\\
10	0.140982193362251\\
15	0.0792801133688868\\
20	0.0445824840550986\\
25	0.0250705732805157\\
30	0.0140982195453822\\
35	0.0079280117504204\\
40	0.00445824920318653\\
};

\addplot[fill = mycolor2, area legend, fill opacity=0.3] fill between[of=C and D];

\addplot [color=mycolor2, dashed, line width= \lw pt]
  table[row sep=crcr]{%
-10	1.40982170131915\\
-5	0.792801003801922\\
0	0.445824767090212\\
5	0.250705690328188\\
10	0.140982170131915\\
15	0.0792801003801922\\
20	0.0445824767090212\\
25	0.0250705690328188\\
30	0.0140982170131915\\
35	0.00792801003801921\\
40	0.00445824767090212\\
};

\addplot [color=green, line width=0.5 pt, mark size=\ms pt, mark=diamond, mark options={solid, green}]
  table[row sep=crcr]{%
-10	0.0297236413315579\\
-5	0.0167148474705228\\
0	0.00939947724403736\\
5	0.00528576386659769\\
10	0.00297250038640752\\
15	0.00167150529922671\\
20	0.000940004067078501\\
25	0.000528717746447751\\
30	0.000297496643690854\\
35	0.000167636854332825\\
40	9.48383301037748e-05\\
};

\addplot [name path = E, color=green, line width=\lw pt, forget plot, opacity=0.1]
  table[row sep=crcr]{%
-10	0.0297838668143296\\
-5	0.0167487522049146\\
0	0.00941860988508817\\
5	0.00529664138565854\\
10	0.00297883404432702\\
15	0.00167494507277914\\
20	0.000942089699603583\\
25	0.000530255565922185\\
30	0.000298684918703948\\
35	0.000169125520961238\\
40	9.70976026137785e-05\\
};
\addplot [name path = F, color=mycolor2, line width=\lw pt, forget plot, opacity=0.1]
  table[row sep=crcr]{%
-10	0.0297000742049604\\
-5	0.0167015790888257\\
0	0.00939198812141681\\
5	0.00528150305318937\\
10	0.002970007438055\\
15	0.00167015791172119\\
20	0.000939198817189528\\
25	0.000528150326371522\\
30	0.000297000776976797\\
35	0.000167015846434773\\
40	9.39200517279419e-05\\
};
\addplot[fill = green, area legend, fill opacity=0.3] fill between[of=E and F];

\addplot [color=green, dashed, line width= \lw pt]
  table[row sep=crcr]{%
-10	0.0297000726130359\\
-5	0.0167015781914642\\
0	0.00939198761295821\\
5	0.00528150276044225\\
10	0.00297000726130359\\
15	0.00167015781914642\\
20	0.000939198761295819\\
25	0.000528150276044224\\
30	0.000297000726130359\\
35	0.000167015781914642\\
40	9.39198761295817e-05\\
};

\end{axis}

\end{tikzpicture}%
    \vspace{-1.cm}
    \centerline{\ \ \ \footnotesize{(b) CFO}} \medskip
\end{minipage}
\hfill
\begin{minipage}[b]{0.48\linewidth}
  \centering
%
%
\newcommand\ms{0.8}
\newcommand\lw{0.3}
\definecolor{mycolor1}{rgb}{0.00000,1.00000,1.00000}%
\definecolor{mycolor2}{rgb}{1.00000,0.00000,1.00000}%

\pgfplotsset{every tick label/.append style={font=\tiny}}

\begin{tikzpicture}[font=\footnotesize]
\begin{axis}[%
width=3cm,
height=2cm,
at={(0cm,0cm)},
scale only axis,
xmin=0,
xmax=40,
xlabel style={font=\scriptsize, yshift=-0.5 ex},
xlabel={$P$ [dBm]},
ymode=log,
ymin=1e-04,
ymax=1,
ylabel style={font=\tiny, yshift=0 ex},
ylabel={}, axis background/.style={fill=white},
xmajorgrids,
ymajorgrids,
legend columns=3, 
legend style={font=\scriptsize, at={(-0.3,1.53)}, anchor=south west, legend cell align=left, align=left, draw=white!15!black}
]

\addplot [color=blue, line width=0.5 pt, mark size=\ms pt, mark=o, mark options={solid, blue}]
  table[row sep=crcr]{%
-10	0.463812345124953\\
-5	0.260828273453048\\
0	0.146688050302113\\
5	0.0825125643923226\\
10	0.04644264958052\\
15	0.0261918805275564\\
20	0.0148613154975617\\
25	0.00858617997630762\\
30	0.00520589578890382\\
35	0.00349394647311672\\
40	0.00270863980703059\\
};

\addplot [name path = A, color=blue, line width=\lw pt, forget plot, opacity=0.1]
  table[row sep=crcr]{%
-10	0.465467188254353\\
-5	0.261754201701196\\
0	0.147213143968834\\
5	0.0828317446717151\\
10	0.0466655201092993\\
15	0.0264209522418235\\
20	0.0153210639760886\\
25	0.00939631977176734\\
30	0.00647887925183709\\
35	0.00522629656660068\\
40	0.00476529109226827\\
};

\addplot [name path = B, color=blue, line width=\lw pt, forget plot, opacity=0.1]
  table[row sep=crcr]{%
-10	0.462275045606302\\
-5	0.259960942928501\\
0	0.146197161325127\\
5	0.0822293732028325\\
10	0.0462645551612271\\
15	0.0260443035321709\\
20	0.0146779876596155\\
25	0.00826799309611463\\
30	0.00466936891668778\\
35	0.00266072111697213\\
40	0.00155694570434468\\
};

\addplot[fill = blue, area legend, fill opacity=0.3] fill between[of=A and B];

\addplot [color=blue, dashed, line width= \lw pt]
  table[row sep=crcr]{%
-10	0.46385175946177\\
-5	0.260843013107607\\
0	0.146682805657575\\
5	0.0824858033161191\\
10	0.0463851759461769\\
15	0.0260843013107607\\
20	0.0146682805657575\\
25	0.00824858033161192\\
30	0.0046385175946177\\
35	0.00260843013107607\\
40	0.00146682805657575\\
};

\addplot [color=mycolor2, line width=0.5 pt, mark size=\ms pt, mark=square, mark options={solid, mycolor2}]
  table[row sep=crcr]{%
-10	1.40974831724247\\
-5	0.792884470346124\\
0	0.446094824169377\\
5	0.251252117061399\\
10	0.141983742387754\\
15	0.0810474781685219\\
20	0.047584312472771\\
25	0.029822456720853\\
30	0.0209083615739522\\
35	0.0167566614480155\\
40	0.014973354277154\\
};

\addplot [name path = C, color=mycolor2, line width=\lw pt, forget plot, opacity=0.1]
  table[row sep=crcr]{%
-10	1.41484601337017\\
-5	0.796040294953826\\
0	0.448713133130494\\
5	0.25424949629116\\
10	0.146281412038151\\
15	0.0878429872513433\\
20	0.0582971946703148\\
25	0.0450251146603721\\
30	0.0398826645972871\\
35	0.0381116965990637\\
40	0.0375623532589804\\
};
\addplot [name path = D, color=mycolor2, line width=\lw pt, forget plot, opacity=0.1]
  table[row sep=crcr]{%
-10	1.4049348632011\\
-5	0.79018844395466\\
0	0.444462172213344\\
5	0.249985840850154\\
10	0.140643239683073\\
15	0.0791060554210797\\
20	0.0445044628106124\\
25	0.0250663705034813\\
30	0.014163196067007\\
35	0.00808653612797591\\
40	0.00475418556961096\\
};

\addplot[fill = mycolor2, area legend, fill opacity=0.3] fill between[of=C and D];

\addplot [color=mycolor2, dashed, line width= \lw pt]
  table[row sep=crcr]{%
-10	1.40982170131915\\
-5	0.792801003801922\\
0	0.445824767090212\\
5	0.250705690328188\\
10	0.140982170131915\\
15	0.0792801003801922\\
20	0.0445824767090212\\
25	0.0250705690328188\\
30	0.0140982170131915\\
35	0.00792801003801921\\
40	0.00445824767090212\\
};

\addplot [color=green, line width=0.5 pt, mark size=\ms pt, mark=diamond, mark options={solid, green}]
  table[row sep=crcr]{%
-10	0.0296971768353918\\
-5	0.0166999497775507\\
0	0.00939107189071492\\
5	0.0052809878170947\\
10	0.0029697176948028\\
15	0.00166999498228797\\
20	0.000939107193469394\\
25	0.000528098788575833\\
30	0.000296971781809602\\
35	0.00016699952252941\\
40	9.39107610434106e-05\\
};

\addplot [name path = E, color=green, line width=\lw pt, forget plot, opacity=0.1]
  table[row sep=crcr]{%
-10	0.0298027471703083\\
-5	0.0167593163555834\\
0	0.00942445616947472\\
5	0.00529976117719344\\
10	0.00298027473508712\\
15	0.00167593164084922\\
20	0.000942445622806877\\
25	0.000529976126186562\\
30	0.000298027495020733\\
35	0.00016759319808132\\
40	9.42446182639295e-05\\
};
\addplot [name path = F, color=mycolor2, line width=\lw pt, forget plot, opacity=0.1]
  table[row sep=crcr]{%
-10	0.029598790440386\\
-5	0.0166446230463033\\
0	0.00935995939409684\\
5	0.0052634919778744\\
10	0.00295987907471383\\
15	0.00166446230941771\\
20	0.000935995948351047\\
25	0.000526349213377102\\
30	0.000295987935762685\\
35	0.000166446289695713\\
40	9.35997009593804e-05\\
};
\addplot[fill = green, area legend, fill opacity=0.3] fill between[of=E and F];

\addplot [color=green, dashed, line width= \lw pt]
  table[row sep=crcr]{%
-10	0.0297000726130359\\
-5	0.0167015781914642\\
0	0.00939198761295821\\
5	0.00528150276044225\\
10	0.00297000726130359\\
15	0.00167015781914642\\
20	0.000939198761295819\\
25	0.000528150276044224\\
30	0.000297000726130359\\
35	0.000167015781914642\\
40	9.39198761295817e-05\\
};

\end{axis}

\end{tikzpicture}%
    \vspace{-1.cm}
  \centerline{\ \ \ \footnotesize{(c) MC}} \medskip
\end{minipage}
\hfill
\hspace{0.2cm}
\begin{minipage}[b]{0.48\linewidth}
  \centering
%
%
\newcommand\ms{0.8}
\newcommand\lw{0.3}
\definecolor{mycolor1}{rgb}{0.00000,1.00000,1.00000}%
\definecolor{mycolor2}{rgb}{1.00000,0.00000,1.00000}%

\pgfplotsset{every tick label/.append style={font=\tiny}}

\begin{tikzpicture}[font=\footnotesize]
\begin{axis}[%
width=3cm,
height=2cm,
at={(0cm,0cm)},
scale only axis,
xmin=0,
xmax=40,
xlabel style={font=\scriptsize, yshift=-0.5 ex},
xlabel={$P$ [dBm]},
ymode=log,
ymin=1e-04,
ymax=1,
ylabel style={font=\tiny, yshift=-0.5 ex},
xmajorgrids,
ymajorgrids,
legend columns=3, 
legend style={font=\scriptsize, at={(-0.3,1.53)}, anchor=south west, legend cell align=left, align=left, draw=white!15!black}
]

\addplot [color=blue, line width=0.5 pt, mark size=\ms pt, mark=o, mark options={solid, blue}]
  table[row sep=crcr]{%
-10	0.463832265470201\\
-5	0.260845550497184\\
0	0.146708325406404\\
5	0.0825430569802229\\
10	0.0464938526071376\\
15	0.0262801083531113\\
20	0.0150144913231332\\
25	0.0088431112159239\\
30	0.00560410614231191\\
35	0.00403613860987153\\
40	0.00335612716729785\\
};

\addplot [name path = A, color=blue, line width=\lw pt, forget plot, opacity=0.1]
  table[row sep=crcr]{%
-10	0.464262307991107\\
-5	0.261080002763039\\
0	0.146826222296182\\
5	0.0826539451825694\\
10	0.046682468825103\\
15	0.0266163289696434\\
20	0.0155865205262618\\
25	0.00979039857482851\\
30	0.00701913982933577\\
35	0.00587781920659544\\
40	0.00546776033820098\\
};

\addplot [name path = B, color=blue, line width=\lw pt, forget plot, opacity=0.1]
  table[row sep=crcr]{%
-10	0.463161639888025\\
-5	0.260474206211993\\
0	0.146507890258593\\
5	0.0824267219356592\\
10	0.0463569447791377\\
15	0.0260771288532137\\
20	0.0146804830814143\\
25	0.00828283610345296\\
30	0.00470767213603795\\
35	0.00273391038020447\\
40	0.001681739950414\\
};

\addplot[fill = blue, area legend, fill opacity=0.3] fill between[of=A and B];

\addplot [color=blue, dashed, line width= \lw pt]
  table[row sep=crcr]{%
-10	0.46385175946177\\
-5	0.260843013107607\\
0	0.146682805657575\\
5	0.0824858033161191\\
10	0.0463851759461769\\
15	0.0260843013107607\\
20	0.0146682805657575\\
25	0.00824858033161192\\
30	0.0046385175946177\\
35	0.00260843013107607\\
40	0.00146682805657575\\
};

\addplot [color=mycolor2, line width=0.5 pt, mark size=\ms pt, mark=square, mark options={solid, mycolor2}]
  table[row sep=crcr]{%
-10	1.40977541927634\\
-5	0.79282781858252\\
0	0.445933369986769\\
5	0.250933587896543\\
10	0.141402916557613\\
15	0.080038544679918\\
20	0.0459062197052736\\
25	0.0273039588419279\\
30	0.0176147419791103\\
35	0.0129662026846309\\
40	0.0109563210561599\\
};

\addplot [name path = C, color=mycolor2, line width=\lw pt, forget plot, opacity=0.1]
  table[row sep=crcr]{%
-10	1.41096858317574\\
-5	0.793486962831787\\
0	0.446373990283442\\
5	0.251770768016418\\
10	0.142941356387196\\
15	0.0827627656353174\\
20	0.0505477761969713\\
25	0.0346053726635768\\
30	0.0276978109291012\\
35	0.0251121449746598\\
40	0.0242500379329036\\
};
\addplot [name path = D, color=mycolor2, line width=\lw pt, forget plot, opacity=0.1]
  table[row sep=crcr]{%
-10	1.40769684546181\\
-5	0.79162147274574\\
0	0.445186717939689\\
5	0.250387613166015\\
10	0.140876068690631\\
15	0.0792821250033355\\
20	0.0446229756228974\\
25	0.0251302144662619\\
30	0.0141982721560652\\
35	0.0081002288364417\\
40	0.00475372813054275\\
};

\addplot[fill = mycolor2, area legend, fill opacity=0.3] fill between[of=C and D];

\addplot [color=mycolor2, dashed, line width= \lw pt]
  table[row sep=crcr]{%
-10	1.40982170131915\\
-5	0.792801003801922\\
0	0.445824767090212\\
5	0.250705690328188\\
10	0.140982170131915\\
15	0.0792801003801922\\
20	0.0445824767090212\\
25	0.0250705690328188\\
30	0.0140982170131915\\
35	0.00792801003801921\\
40	0.00445824767090212\\
};

\addplot [color=green, line width=0.5 pt, mark size=\ms pt, mark=diamond, mark options={solid, green}]
  table[row sep=crcr]{%
-10	0.0296979643846639\\
-5	0.0167003926513955\\
0	0.00939132094944389\\
5	0.00528112789041452\\
10	0.00296979649695026\\
15	0.00167003929327643\\
20	0.000939132123174632\\
25	0.000528112828094808\\
30	0.000296979709143132\\
35	0.000167004037103482\\
40	9.39134045138109e-05\\
};

\addplot [name path = E, color=green, line width=\lw pt, forget plot, opacity=0.1]
  table[row sep=crcr]{%
-10	0.0297261375252519\\
-5	0.0167162355364218\\
0	0.00940023004990072\\
5	0.00528613783670916\\
10	0.00297261375005078\\
15	0.00167162360574589\\
20	0.00094002307501726\\
25	0.000528613809346071\\
30	0.000297261377832821\\
35	0.00016716235583805\\
40	9.40023171671304e-05\\
};
\addplot [name path = F, color=mycolor2, line width=\lw pt, forget plot, opacity=0.1]
  table[row sep=crcr]{%
-10	0.0296537083488077\\
-5	0.0166755056462833\\
0	0.00937732595074575\\
5	0.00527325792843422\\
10	0.0029653708606504\\
15	0.00166755059801762\\
20	0.000937732609754755\\
25	0.000527325829817099\\
30	0.000296537107917677\\
35	0.00016675510717318\\
40	9.3773344528436e-05\\
};
\addplot[fill = green, area legend, fill opacity=0.3] fill between[of=E and F];

\addplot [color=green, dashed, line width= \lw pt]
  table[row sep=crcr]{%
-10	0.0297000726130359\\
-5	0.0167015781914642\\
0	0.00939198761295821\\
5	0.00528150276044225\\
10	0.00297000726130359\\
15	0.00167015781914642\\
20	0.000939198761295819\\
25	0.000528150276044224\\
30	0.000297000726130359\\
35	0.000167015781914642\\
40	9.39198761295817e-05\\
};

\end{axis}

\end{tikzpicture}%
    \vspace{-1.cm}
    \centerline{\ \ \ \footnotesize{(d) AGE}} \medskip
\end{minipage}
\begin{minipage}[b]{0.48\linewidth}
  \centering
%
%
\newcommand\ms{0.8}
\newcommand\lw{0.3}
\definecolor{mycolor1}{rgb}{0.00000,1.00000,1.00000}%
\definecolor{mycolor2}{rgb}{1.00000,0.00000,1.00000}%

\pgfplotsset{every tick label/.append style={font=\tiny}}

\begin{tikzpicture}[font=\footnotesize]
\begin{axis}[%
width=3cm,
height=2cm,
at={(0cm,0cm)},
scale only axis,
xmin=0,
xmax=40,
xlabel style={font=\scriptsize, yshift=-0.5 ex},
xlabel={$P$ [dBm]},
ymode=log,
ymin=1e-04,
ymax=1,
ylabel style={font=\tiny, yshift=0 ex},
ylabel={}, axis background/.style={fill=white},
xmajorgrids,
ymajorgrids,
legend columns=3, 
legend style={font=\scriptsize, at={(-0.3,1.53)}, anchor=south west, legend cell align=left, align=left, draw=white!15!black}
]

\addplot [color=blue, line width=0.5 pt, mark size=\ms pt, mark=o, mark options={solid, blue}]
  table[row sep=crcr]{%
-10	0.463876912921091\\
-5	0.260885284669145\\
0	0.146755952485747\\
5	0.0826151644012105\\
10	0.046613528783556\\
15	0.0264856980914595\\
20	0.0153587663838181\\
25	0.00937931799136437\\
30	0.00634072256340499\\
35	0.00492036827922106\\
40	0.00431237440529507\\
};

\addplot [name path = A, color=blue, line width=\lw pt, forget plot, opacity=0.1]
  table[row sep=crcr]{%
-10	0.464028918668218\\
-5	0.261037276846511\\
0	0.14703002014377\\
5	0.0831069290652835\\
10	0.047479668118327\\
15	0.027990268038193\\
20	0.0178382695758646\\
25	0.0130732462429773\\
30	0.0111445371474247\\
35	0.0104682727395574\\
40	0.0102429891721407\\
};

\addplot [name path = B, color=blue, line width=\lw pt, forget plot, opacity=0.1]
  table[row sep=crcr]{%
-10	0.463760900309422\\
-5	0.260800595689721\\
0	0.146665235675551\\
5	0.0824771233115037\\
10	0.0463819808643252\\
15	0.0260859012330532\\
20	0.0146749829745737\\
25	0.00826043722986726\\
30	0.00465930494365925\\
35	0.00264528153501313\\
40	0.00153147682125791\\
};

\addplot[fill = blue, area legend, fill opacity=0.3] fill between[of=A and B];

\addplot [color=blue, dashed, line width= \lw pt]
  table[row sep=crcr]{%
-10	0.46385175946177\\
-5	0.260843013107607\\
0	0.146682805657575\\
5	0.0824858033161191\\
10	0.0463851759461769\\
15	0.0260843013107607\\
20	0.0146682805657575\\
25	0.00824858033161192\\
30	0.0046385175946177\\
35	0.00260843013107607\\
40	0.00146682805657575\\
};

\addplot [color=mycolor2, line width=0.5 pt, mark size=\ms pt, mark=square, mark options={solid, mycolor2}]
  table[row sep=crcr]{%
-10	1.41001599327854\\
-5	0.793136198978137\\
0	0.446420842427902\\
5	0.251757698249162\\
10	0.142839583652174\\
15	0.0825169200993948\\
20	0.0500217151992031\\
25	0.0335422199416631\\
30	0.0259465001824789\\
35	0.0228025131192336\\
40	0.021620465950558\\
};

\addplot [name path = C, color=mycolor2, line width=\lw pt, forget plot, opacity=0.1]
  table[row sep=crcr]{%
-10	1.41048464955386\\
-5	0.793850071895281\\
0	0.447618116351511\\
5	0.253844564273354\\
10	0.146494241220998\\
15	0.0886997131958725\\
20	0.0597190041124025\\
25	0.0469935002741604\\
30	0.0421650640105427\\
35	0.0405367648501026\\
40	0.040002514706117\\
};
\addplot [name path = D, color=mycolor2, line width=\lw pt, forget plot, opacity=0.1]
  table[row sep=crcr]{%
-10	1.40977981006953\\
-5	0.792792809557719\\
0	0.44585136529793\\
5	0.250766339768985\\
10	0.141091896071134\\
15	0.0794726744985361\\
20	0.0449211821583309\\
25	0.0256645538773357\\
30	0.0151279608780686\\
35	0.00964126925144983\\
40	0.00707550761723912\\
};

\addplot[fill = mycolor2, area legend, fill opacity=0.3] fill between[of=C and D];

\addplot [color=mycolor2, dashed, line width= \lw pt]
  table[row sep=crcr]{%
-10	1.40982170131915\\
-5	0.792801003801922\\
0	0.445824767090212\\
5	0.250705690328188\\
10	0.140982170131915\\
15	0.0792801003801922\\
20	0.0445824767090212\\
25	0.0250705690328188\\
30	0.0140982170131915\\
35	0.00792801003801921\\
40	0.00445824767090212\\
};

\addplot [color=green, line width=0.5 pt, mark size=\ms pt, mark=diamond, mark options={solid, green}]
  table[row sep=crcr]{%
-10	0.0296999051497777\\
-5	0.0167014840246178\\
0	0.00939193466777209\\
5	0.00528147300436227\\
10	0.00296999055703105\\
15	0.00167014842141718\\
20	0.000939193487047771\\
25	0.000528147326170801\\
30	0.000296999097037127\\
35	0.000167014921810661\\
40	9.39194905913063e-05\\
};

\addplot [name path = E, color=green, line width=\lw pt, forget plot, opacity=0.1]
  table[row sep=crcr]{%
-10	0.0297070359127525\\
-5	0.016705493960178\\
0	0.0093941896258428\\
5	0.00528274104288281\\
10	0.0029707036007818\\
15	0.00167054939626188\\
20	0.000939418986690528\\
25	0.000528274103707686\\
30	0.000297070359729097\\
35	0.000167054952600605\\
40	9.39419050484191e-05\\
};
\addplot [name path = F, color=mycolor2, line width=\lw pt, forget plot, opacity=0.1]
  table[row sep=crcr]{%
-10	0.0296944265790537\\
-5	0.0166984031669933\\
0	0.00939020216092967\\
5	0.00528049872877454\\
10	0.00296944265153139\\
15	0.00166984031715297\\
20	0.000939020220651943\\
25	0.000528049891749743\\
30	0.000296944267280325\\
35	0.000166984033404102\\
40	9.39020286149581e-05\\
};
\addplot[fill = green, area legend, fill opacity=0.3] fill between[of=E and F];

\addplot [color=green, dashed, line width= \lw pt]
  table[row sep=crcr]{%
-10	0.0297000726130359\\
-5	0.0167015781914642\\
0	0.00939198761295821\\
5	0.00528150276044225\\
10	0.00297000726130359\\
15	0.00167015781914642\\
20	0.000939198761295819\\
25	0.000528150276044224\\
30	0.000297000726130359\\
35	0.000167015781914642\\
40	9.39198761295817e-05\\
};

\end{axis}

\end{tikzpicture}%
    \vspace{-1.cm}
    \centerline{\ \ \ \footnotesize{(e) ADE}} \medskip
\end{minipage}
\hfill
\hspace{0.1cm}
\begin{minipage}[b]{0.48\linewidth}
  \centering
%
%
\newcommand\ms{0.8}
\newcommand\lw{0.3}
\definecolor{mycolor1}{rgb}{0.00000,1.00000,1.00000}%
\definecolor{mycolor2}{rgb}{1.00000,0.00000,1.00000}%

\pgfplotsset{every tick label/.append style={font=\tiny}}

\begin{tikzpicture}[font=\footnotesize]
\begin{axis}[%
width=3cm,
height=2cm,
at={(0cm,0cm)},
scale only axis,
xmin=0,
xmax=40,
xlabel style={font=\scriptsize, yshift=-0.5 ex},
xlabel={$P$ [dBm]},
ymode=log,
ymin=1e-04,
ymax=1,
ylabel style={font=\tiny, yshift=0 ex},
ylabel={}, axis background/.style={fill=white},
xmajorgrids,
ymajorgrids,
legend columns=3, 
legend style={font=\scriptsize, at={(-0.3,1.53)}, anchor=south west, legend cell align=left, align=left, draw=white!15!black}
]

\addplot [color=blue, line width=0.5 pt, mark size=\ms pt, mark=o, mark options={solid, blue}]
  table[row sep=crcr]{%
-10	0.463614945225226\\
-5	0.260709882806062\\
0	0.146608004043523\\
5	0.0824438575875739\\
10	0.0463617965118974\\
15	0.0260710396726045\\
20	0.0146608929368006\\
25	0.00824454584749313\\
30	0.00463646891690889\\
35	0.00260768852104159\\
40	0.00146709357947779\\
};

\addplot [name path = A, color=blue, line width=\lw pt, forget plot, opacity=0.1]
  table[row sep=crcr]{%
-10	0.482234583431361\\
-5	0.271180513337549\\
0	0.152496100652232\\
5	0.0857550305742664\\
10	0.0482239070239892\\
15	0.0271181217545063\\
20	0.0152497467985502\\
25	0.00857574163930522\\
30	0.00482279803023768\\
35	0.00271264084529666\\
40	0.00152644922178313\\
};

\addplot [name path = B, color=blue, line width=\lw pt, forget plot, opacity=0.1]
  table[row sep=crcr]{%
-10	0.452101263520995\\
-5	0.254235314325501\\
0	0.142967163239591\\
5	0.0803966018803717\\
10	0.0452107681962927\\
15	0.0254236412714517\\
20	0.0142969002303387\\
25	0.00804000015915508\\
30	0.00452167809791029\\
35	0.0025435103942391\\
40	0.00143171148290953\\
};

\addplot[fill = blue, area legend, fill opacity=0.3] fill between[of=A and B];

\addplot [color=blue, dashed, line width= \lw pt]
  table[row sep=crcr]{%
-10	0.46385175946177\\
-5	0.260843013107607\\
0	0.146682805657575\\
5	0.0824858033161191\\
10	0.0463851759461769\\
15	0.0260843013107607\\
20	0.0146682805657575\\
25	0.00824858033161192\\
30	0.0046385175946177\\
35	0.00260843013107607\\
40	0.00146682805657575\\
};

\addplot [color=mycolor2, line width=0.5 pt, mark size=\ms pt, mark=square, mark options={solid, mycolor2}]
  table[row sep=crcr]{%
-10	1.40910170101877\\
-5	0.792396158341398\\
0	0.445597500729553\\
5	0.250578050117192\\
10	0.140910935232026\\
15	0.0792397903884767\\
20	0.0445599765111129\\
25	0.0250582778212948\\
30	0.0140919644413367\\
35	0.0079254972920239\\
40	0.00445882773687361\\
};

\addplot [name path = C, color=mycolor2, line width=\lw pt, forget plot, opacity=0.1]
  table[row sep=crcr]{%
-10	1.46569312023478\\
-5	0.824220515766743\\
0	0.463492866263762\\
5	0.260642014211292\\
10	0.146570804795735\\
15	0.0824222418705651\\
20	0.0463497591114112\\
25	0.0260651676220309\\
30	0.0146584414557114\\
35	0.00824489270581292\\
40	0.00464121369345676\\
};
\addplot [name path = D, color=mycolor2, line width=\lw pt, forget plot, opacity=0.1]
  table[row sep=crcr]{%
-10	1.37410757993341\\
-5	0.772717639625441\\
0	0.43453142959059\\
5	0.244355636695287\\
10	0.137412435422907\\
15	0.0772722211773171\\
20	0.043453627475507\\
25	0.0244364263484764\\
30	0.013743170614875\\
35	0.00773020557909832\\
40	0.00434945493801647\\
};

\addplot[fill = mycolor2, area legend, fill opacity=0.3] fill between[of=C and D];

\addplot [color=mycolor2, dashed, line width= \lw pt]
  table[row sep=crcr]{%
-10	1.40982170131915\\
-5	0.792801003801922\\
0	0.445824767090212\\
5	0.250705690328188\\
10	0.140982170131915\\
15	0.0792801003801922\\
20	0.0445824767090212\\
25	0.0250705690328188\\
30	0.0140982170131915\\
35	0.00792801003801921\\
40	0.00445824767090212\\
};

\addplot [color=green, line width=0.5 pt, mark size=\ms pt, mark=diamond, mark options={solid, green}]
  table[row sep=crcr]{%
-10	0.0296852215163493\\
-5	0.0166936371785746\\
0	0.00938824959739877\\
5	0.00528069553692503\\
10	0.00297185267710054\\
15	0.00167517413168248\\
20	0.000949069558681677\\
25	0.000545880117135631\\
30	0.000326927130941715\\
35	0.000213643094153205\\
40	0.00015926894438511\\
};

\addplot [name path = E, color=green, line width=\lw pt, forget plot, opacity=0.1]
  table[row sep=crcr]{%
-10	0.0308769644057421\\
-5	0.0173635151205349\\
0	0.00976437922049169\\
5	0.00549118928380433\\
10	0.00308841247981292\\
15	0.00173754885108915\\
20	0.000994526350864248\\
25	0.000607185680352586\\
30	0.000415759645546923\\
35	0.000332203367000843\\
40	0.000301521731354133\\
};
\addplot [name path = F, color=mycolor2, line width=\lw pt, forget plot, opacity=0.1]
  table[row sep=crcr]{%
-10	0.0289479767828978\\
-5	0.0162790577921117\\
0	0.00915512614985831\\
5	0.00514962778503989\\
10	0.002898182206937\\
15	0.0016337635685033\\
20	0.000920587941186973\\
25	0.000519114163145057\\
30	0.000294405854833582\\
35	0.000166815256128712\\
40	9.38776706273424e-05\\
};
\addplot[fill = green, area legend, fill opacity=0.3] fill between[of=E and F];

\addplot [color=green, dashed, line width= \lw pt]
  table[row sep=crcr]{%
-10	0.0297000726130359\\
-5	0.0167015781914642\\
0	0.00939198761295821\\
5	0.00528150276044225\\
10	0.00297000726130359\\
15	0.00167015781914642\\
20	0.000939198761295819\\
25	0.000528150276044224\\
30	0.000297000726130359\\
35	0.000167015781914642\\
40	9.39198761295817e-05\\
};

\end{axis}

\end{tikzpicture}%
    \vspace{-1.cm}
    \centerline{\ \ \ \footnotesize{(f) IQI}} \medskip
\end{minipage}
\vspace{-0.8cm}
\caption{LBs of channel parameter estimation under different types of impairment with multiple realizations: (a) Phase noise, (b) Carrier frequency offset, (c) Mutual coupling, (d) Array gain error, (e) Antenna displacement error, (f) IQ-imbalance.}
\label{fig:single_impairment_on_localization}
\end{figure}

To understand the effect of different types of~\acp{hwi}, we study the LB for AOA, AOD, and delay estimation by considering one type of \acp{hwi} at a time. The results are shown in Fig.~\ref{fig:single_impairment_on_localization} for (a) PN, (b) CFO, (c) MC, (d) AGE, (e) ADE and (f) IQI. The effect of PA will be separately discussed in Sec.~\ref{sec:sim_pa}. Considering we define the HWIs as random variables with a fixed impairment level as shown in Table~\ref{table:Simulation_parameters}, we perform multiple hardware realizations with a fixed pilot signal and plot all the resultant LBs in the shaded regions. We can see that different types of the \acp{hwi} affect angle and delay estimation differently. The PN and IQI introduce noise on the symbols across different subcarriers and hence affect delay estimation\footnote{{If multiple RFCs or several local oscillators are adopted in the array, PN may have a larger effect on angle estimation.}}. The phase changes introduced by the CFO increase with time (see~\eqref{eq:cfo}), and hence, angle estimation (relying on multiple transmissions) will be affected more than delay estimation. The rest of the impairments, namely, the MC, AGE, and ADE distort the steering vectors and therefore have a more significant effect on the angle estimation.
For all the HWIs, the negative effect on the performance occurs when the transmit power is high.

\begin{figure}[htb]
\centering
\begin{minipage}[b]{0.48\linewidth}
\vspace{-0.3cm}
  \centering
%
%
\newcommand\ms{0.8}

\pgfplotsset{every tick label/.append style={font=\tiny}}
\begin{tikzpicture}[font=\footnotesize]
\begin{axis}[%
width=3 cm,
height=2 cm,
at={(0cm,0cm)},
scale only axis,
xmin=0,
xmax=50,
xlabel style={},
xlabel={$P$ [dBm]},
ymode=log,
ymin=0.4e-03,
ymax=1,
ylabel style={},
ylabel={Angle Error [$^\circ$]},
axis background/.style={fill=white},
xmajorgrids,
ymajorgrids,
legend columns=2, 
legend style={font=\tiny, at={(0., 1.05)}, anchor=south west, legend cell align=left, align=left, draw=white!15!black}
]

\addplot [color=blue, line width=1.0pt, mark=o,mark size=\ms pt, mark options={solid, blue}]
  table[row sep=crcr]{%
0	0.146851078023138\\
5	0.0825830013717114\\
10	0.0464451760436049\\
15	0.0261150380261128\\
20	0.0146903796442206\\
25	0.00826513143186858\\
30	0.00465518465177554\\
35	0.00262444738256177\\
40	0.00149007867372746\\
45	0.000861905346344179\\
50	0.000522468479975213\\
};
\addlegendentry{BS 8x8, UE 4x4, BS first\ \ }

\addplot [color=red, line width=1.0pt, mark=square,mark size=\ms pt, mark options={solid, red}]
  table[row sep=crcr]{%
0	0.146815312128769\\
5	0.0825672495156457\\
10	0.0464363029012226\\
15	0.0261625887028336\\
20	0.0147444849900374\\
25	0.00831341416203485\\
30	0.00474164814317204\\
35	0.00279048125728068\\
40	0.00172610380477795\\
45	0.00116422151603912\\
50	0.00088940899264942\\
};
\addlegendentry{BS 8x8, UE 4x4, UE first\ \ }

\addplot [color=blue, dashed, line width=1.0pt, mark=diamond, mark size=\ms pt, mark options={solid, blue}]
  table[row sep=crcr]{%
0	0.719749925627036\\
5	0.404742583021253\\
10	0.227603893346621\\
15	0.127992548487235\\
20	0.0719762502020503\\
25	0.0404784363690385\\
30	0.0227628012771963\\
35	0.0128004843049271\\
40	0.00720134695434857\\
45	0.00405084118774251\\
50	0.00227981479656024\\
};
\addlegendentry{BS 4x4, UE 8x8, BS first\ \ }

\addplot [color=red, dashed, line width=1.0pt, mark=triangle, mark size=\ms pt, mark options={solid, red}]
  table[row sep=crcr]{%
0	0.720261327276394\\
5	0.405033460937044\\
10	0.227774372131663\\
15	0.12808282651419\\
20	0.0720354451318349\\
25	0.0405455866639008\\
30	0.0228270454474863\\
35	0.0129473563813705\\
40	0.00742497648650047\\
45	0.00440629027125088\\
50	0.00276156644046626\\
};
\addlegendentry{BS 4x4, UE 8x8, UE first}

\end{axis}

\end{tikzpicture}%
    \vspace{-1.cm}
  \centerline{\footnotesize{(a) AALB (average)}} \medskip
\end{minipage}
\hfill
\begin{minipage}[b]{0.48\linewidth}
\vspace{-0.3cm}
  \centering
%
%

\newcommand\ms{0.8}

\pgfplotsset{every tick label/.append style={font=\tiny}}
\begin{tikzpicture}[font=\footnotesize]
\begin{axis}[%
width=3 cm,
height=2 cm,
at={(0in,0in)},
scale only axis,
xmin=0,
xmax=50,
xlabel style={},
xlabel={$P$ [dBm]},
ymode=log,
ymin=0.4e-03,
ymax=1,
ylabel style={},
ylabel={Angle Error [$^\circ$]},
axis background/.style={fill=white},
xmajorgrids,
ymajorgrids,
legend columns=2, 
legend style={font=\footnotesize, at={(-0.1,1.13)}, anchor=south west, legend cell align=left, align=left, draw=white!15!black}
]

\addplot [color=blue, line width=1.0pt, mark=o, mark size=\ms pt, mark options={solid, blue}]
  table[row sep=crcr]{%
0	0.446330412960188\\
5	0.250990847465891\\
10	0.142061795045027\\
15	0.0793702750229344\\
20	0.0468869039404292\\
25	0.0282034831750928\\
30	0.019017639457115\\
35	0.0141834351718573\\
40	0.0116488010950092\\
45	0.0105218524905289\\
50	0.0100077972618669\\
};

\addplot [color=red, line width=1.0pt, mark=square, mark size=\ms pt, mark options={solid, red}]
  table[row sep=crcr]{%
0	0.446284704790947\\
5	0.251002252666615\\
10	0.141308302312206\\
15	0.0798554593440703\\
20	0.0452674307837112\\
25	0.0260540340388361\\
30	0.0157918949923337\\
35	0.0103481044939352\\
40	0.00752971975834958\\
45	0.0061730758401784\\
50	0.00551509315997026\\
};

\addplot [color=blue, dashed, line width=1.0pt, mark size=\ms pt, mark=diamond, mark options={solid, blue}]
  table[row sep=crcr]{%
0	0.308455647187184\\
5	0.173635435518509\\
10	0.097944214945771\\
15	0.0557246640944808\\
20	0.0322936004511254\\
25	0.019630241542835\\
30	0.0129736526111325\\
35	0.00970645496118632\\
40	0.0080623214352033\\
45	0.00729538729823298\\
50	0.00694321725279122\\
};

\addplot [color=red, dashed, line width=1.0pt, mark size=\ms pt, mark=triangle, mark options={solid, red}]
  table[row sep=crcr]{%
0	0.308475255047439\\
5	0.173468702400333\\
10	0.0975631739000068\\
15	0.0548556202338969\\
20	0.0308843982268944\\
25	0.017404495145706\\
30	0.00984699202221588\\
35	0.00564408590155383\\
40	0.00333319536323026\\
45	0.00208541084589008\\
50	0.00142956600939572\\
};

\end{axis}

\end{tikzpicture}%
    \vspace{-1.cm}
  \centerline{\footnotesize{(b) ADLB (average)}} \medskip
\end{minipage}
\vspace{-0.2cm}
\caption{The effect of CFO on channel geometrical parameters with different sweeping strategies. The `BS first' strategy (blue curves) works better for AOA estimation, while the `UE first' strategy (red curves) works better for AOD estimation.}
\label{fig:sim_cfo}
\end{figure}
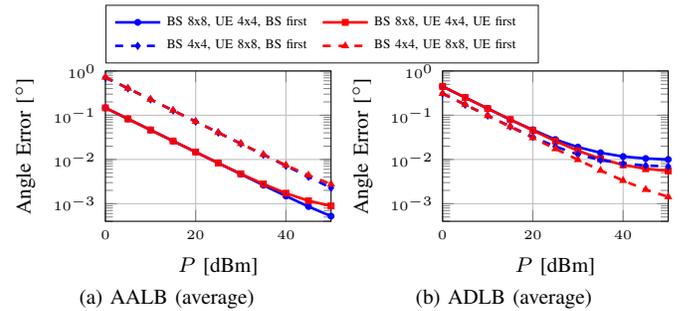

One special observation is that the effect of CFO on the AOA is less pronounced than on AOD in Fig.~\ref{fig:single_impairment_on_localization} (b). This is because the sweeping strategy is `BS-first'. For a system with analog arrays, the estimation of AOA/AOD relies on phase shifts across consecutive beams over time, meaning the angle cannot be estimated from a single receive beam, like in a digital array. If the BS sweeps across different beams while the UE is using a fixed beam, the AOA can be estimated in one BS sweep, and the effect of CFO will be minor. However, the AOD estimation requires multiple BS sweeps, which increases the effect of CFO. To verify the explanation, we further changed the sweeping strategy from `BS-first' to `UE-first,' and the results with different array sizes can be found in Fig.~\ref{fig:sim_cfo}. We can see that the AOA is less affected if the sweeping is `BS-first' (blue curves in (a)) as shown in~\eqref{eq:cfo}. Similarly, the AODs are less affected if the sweeping is `UE-first' (dashed red curves in (b)) with a large UE array size. However, when the array size is small, sweeping order will have less impact (i.e., the gaps are small between the dashed curves in (a) and the solid curves in (b)).


\subsubsection{The Effect of PA with Different Pilot Signals}
\label{sec:sim_pa}
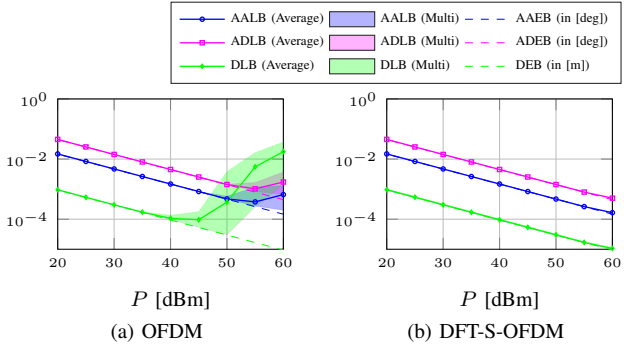
\begin{figure}
\begin{minipage}[b]{0.48\linewidth}
%
%
\newcommand\ms{0.8}
\newcommand\lw{0.3}
\definecolor{mycolor1}{rgb}{0.00000,1.00000,1.00000}%
\definecolor{mycolor2}{rgb}{1.00000,0.00000,1.00000}%

\pgfplotsset{every tick label/.append style={font=\tiny}}

\begin{tikzpicture}[font=\footnotesize]
\begin{axis}[%
width=3cm,
height=2.cm,
at={(0in,0in)},
scale only axis,
xmin=20,
xmax=60,
xlabel style={font=\footnotesize, yshift=-0.5 ex},
xlabel={$P$ [dBm]},
ymode=log,
ymin=1e-05,
ymax=1,
ylabel style={font=\footnotesize, yshift=-0.5 ex},
axis background/.style={fill=white},
xmajorgrids,
ymajorgrids,
legend columns=3, 
legend style={font=\tiny, at={(0.5,1.1)}, anchor=south west, legend cell align=left, align=left, draw=white!15!black}
]

\addplot [color=blue, line width=0.5 pt, mark size=\ms pt, mark=o, mark options={solid, blue}]
  table[row sep=crcr]{%
10	0.0468484455651716\\
15	0.0263441574466023\\
20	0.0148137495175772\\
25	0.00832972553506102\\
30	0.00468349264704901\\
35	0.00263306842734199\\
40	0.00148003638786792\\
45	0.000831652312495442\\
50	0.000473061799261907\\
55	0.000375977074732039\\
60	0.000659255423270108\\
};
\addlegendentry{AALB (Average)}

\addplot [name path = A, color=blue, line width=\lw pt, forget plot, opacity=0.1]
  table[row sep=crcr]{%
10	0.0468485487193113\\
15	0.0263442604676915\\
20	0.0148138522959537\\
25	0.00832982815533538\\
30	0.00468359490057115\\
35	0.00263316981552953\\
40	0.00148013593794023\\
45	0.000831745446453119\\
50	0.000584508928420386\\
55	0.00116860323278007\\
60	0.00131202730690288\\
};

\addplot [name path = B, color=blue, line width=\lw pt, forget plot, opacity=0.1]
  table[row sep=crcr]{%
10	0.0468483211605318\\
15	0.0263440332665837\\
20	0.0148136257247343\\
25	0.00832960206310472\\
30	0.00468336917674164\\
35	0.00263294614523458\\
40	0.00147991586450169\\
45	0.000831530527613485\\
50	0.000467021594113033\\
55	0.000270374560127239\\
60	0.000198500371364835\\
};

\addplot[fill = blue, area legend, fill opacity=0.3] fill between[of=A and B];
\addlegendentry{AALB (Multi)}

\addplot [color=blue, dashed, line width= \lw pt]
  table[row sep=crcr]{%
10	0.046385175946177\\
15	0.0260843013107607\\
20	0.0146682805657575\\
25	0.00824858033161192\\
30	0.0046385175946177\\
35	0.00260843013107607\\
40	0.00146682805657575\\
45	0.000824858033161192\\
50	0.00046385175946177\\
55	0.000260843013107607\\
60	0.000146682805657575\\
};
\addlegendentry{AAEB (in [deg])}

\addplot [color=mycolor2, line width=0.5 pt, mark size=\ms pt, mark=square, mark options={solid, mycolor2}]
  table[row sep=crcr]{%
10	0.142390222641662\\
15	0.0800699010971597\\
20	0.045024613155606\\
25	0.0253172016896879\\
30	0.014234915976681\\
35	0.00800289976063371\\
40	0.00449839928686034\\
45	0.00252771579087495\\
50	0.00142159573715833\\
55	0.0010245317671807\\
60	0.00172448745669892\\
};
\addlegendentry{ADLB (Average)}

\addplot [name path = C, color=mycolor2, line width=\lw pt, forget plot, opacity=0.1]
  table[row sep=crcr]{%
10	0.142390535729571\\
15	0.0800702149015349\\
20	0.0450249234965014\\
25	0.0253175184888886\\
30	0.0142352299360607\\
35	0.00800320325854968\\
40	0.00449869961124463\\
45	0.0025280029824789\\
50	0.00145062548702389\\
55	0.00166731967478525\\
60	0.00374355924949837\\
};
\addplot [name path = D, color=mycolor2, line width=\lw pt, forget plot, opacity=0.1]
  table[row sep=crcr]{%
10	0.142389844179289\\
15	0.0800695157891916\\
20	0.0450242348978837\\
25	0.0253168232465415\\
30	0.0142345387990144\\
35	0.00800252550722497\\
40	0.00449803765355232\\
45	0.00252734217035611\\
50	0.00141938478168043\\
55	0.000835795332488516\\
60	0.000658480890619529\\
};

\addplot[fill = mycolor2, area legend, fill opacity=0.3] fill between[of=C and D];
\addlegendentry{ADLB (Multi)}

\addplot [color=mycolor2, dashed, line width= \lw pt]
  table[row sep=crcr]{%
10	0.140982170131915\\
15	0.0792801003801922\\
20	0.0445824767090212\\
25	0.0250705690328188\\
30	0.0140982170131915\\
35	0.00792801003801922\\
40	0.00445824767090212\\
45	0.00250705690328189\\
50	0.00140982170131915\\
55	0.000792801003801922\\
60	0.000445824767090212\\
};
\addlegendentry{ADEB (in [deg])}

\addplot [color=green, line width=0.5 pt, mark size=\ms pt, mark=diamond, mark options={solid, green}]
  table[row sep=crcr]{%
10	0.00299967049917662\\
15	0.0016867986486842\\
20	0.000948526665068077\\
25	0.000533422698674313\\
30	0.000300308927318848\\
35	0.000170950353640703\\
40	0.000106528108005709\\
45	9.57329284668593e-05\\
50	0.000369925716701168\\
55	0.00550306260416332\\
60	0.017696276944472\\
};
\addlegendentry{DLB (Average)}

\addplot [name path = E, color=green, line width=\lw pt, forget plot, opacity=0.1]
  table[row sep=crcr]{%
10	0.002999676911829\\
15	0.00168680696249483\\
20	0.00094856272513493\\
25	0.000533636090198699\\
30	0.000301521774640682\\
35	0.000177493338700985\\
40	0.000136257021765272\\
45	0.0001791945250496\\
50	0.00385682969276594\\
55	0.0164402020626146\\
60	0.037049786067913\\
};
\addplot [name path = F, color=mycolor2, line width=\lw pt, forget plot, opacity=0.1]
  table[row sep=crcr]{%
10	0.00299966226840776\\
15	0.0016867892293386\\
20	0.000948510044279283\\
25	0.000533348129517769\\
30	0.00029988338849955\\
35	0.00016859674649829\\
40	9.47708280931065e-05\\
45	5.32657506345616e-05\\
50	2.9950840101703e-05\\
55	0.000258029436252925\\
60	0.00144872674306122\\
};
\addplot[fill = green, area legend, fill opacity=0.3] fill between[of=E and F];
\addlegendentry{DLB (Multi)}

\addplot [color=green, dashed, line width= \lw pt]
  table[row sep=crcr]{%
10	0.00297000726130359\\
15	0.00167015781914642\\
20	0.000939198761295819\\
25	0.000528150276044224\\
30	0.000297000726130358\\
35	0.000167015781914642\\
40	9.39198761295817e-05\\
45	5.28150276044225e-05\\
50	2.97000726130359e-05\\
55	1.67015781914642e-05\\
60	9.39198761295819e-06\\
};
\addlegendentry{DEB (in [m])}

\end{axis}

\end{tikzpicture}%
    \vspace{-1. cm}
  \centerline{\footnotesize{(a) OFDM}} \medskip
\end{minipage}
\begin{minipage}[b]{0.48\linewidth}
%
%
\newcommand\ms{0.8}
\newcommand\lw{0.3}
\definecolor{mycolor1}{rgb}{0.00000,1.00000,1.00000}%
\definecolor{mycolor2}{rgb}{1.00000,0.00000,1.00000}%

\pgfplotsset{every tick label/.append style={font=\tiny}}

\begin{tikzpicture}[font=\footnotesize]
\begin{axis}[%
width=3cm,
height=2.cm,
at={(0in,0in)},
scale only axis,
xmin=20,
xmax=60,
xlabel style={font=\footnotesize, yshift=-0.5 ex},
xlabel={$P$ [dBm]},
ymode=log,
ymin=1e-05,
ymax=1,
ylabel style={font=\footnotesize, yshift=-0.5 ex},
axis background/.style={fill=white},
xmajorgrids,
ymajorgrids,
legend columns=3, 
legend style={font=\tiny, at={(1,1)}, anchor=south east, legend cell align=left, align=left, draw=white!15!black}
]

\addplot [color=blue, line width=0.5 pt, mark size=\ms pt, mark=o, mark options={solid, blue}]
  table[row sep=crcr]{%
10	0.0468488114751426\\
15	0.0263445231266665\\
20	0.0148141147193604\\
25	0.00833009006396722\\
30	0.00468385591465781\\
35	0.00263342931321622\\
40	0.00148039243853543\\
45	0.000831997008218403\\
50	0.000467386049183028\\
55	0.000262365370090212\\
60	0.000164682530912057\\
};

\addplot [name path = A, color=blue, line width=\lw pt, forget plot, opacity=0.1]
  table[row sep=crcr]{%
10	0.0468488114754188\\
15	0.0263445231270948\\
20	0.0148141147194459\\
25	0.00833009006400434\\
30	0.00468385591466595\\
35	0.00263342931321884\\
40	0.00148039243853639\\
45	0.000831997008218695\\
50	0.000467386049183129\\
55	0.000262365370090236\\
60	0.000164682530912071\\
};

\addplot [name path = B, color=blue, line width=\lw pt, forget plot, opacity=0.1]
  table[row sep=crcr]{%
10	0.0468488114748578\\
15	0.026344523126454\\
20	0.0148141147192995\\
25	0.00833009006393937\\
30	0.00468385591465077\\
35	0.00263342931321419\\
40	0.00148039243853469\\
45	0.000831997008217964\\
50	0.000467386049182927\\
55	0.000262365370090194\\
60	0.000164682530912046\\
};

\addplot[fill = blue, area legend, fill opacity=0.3] fill between[of=A and B];

\addplot [color=blue, dashed, line width= \lw pt]
  table[row sep=crcr]{%
10	0.046385175946177\\
15	0.0260843013107607\\
20	0.0146682805657575\\
25	0.00824858033161192\\
30	0.0046385175946177\\
35	0.00260843013107607\\
40	0.00146682805657575\\
45	0.000824858033161192\\
50	0.00046385175946177\\
55	0.000260843013107607\\
60	0.000146682805657575\\
};

\addplot [color=mycolor2, line width=0.5 pt, mark size=\ms pt, mark=square, mark options={solid, mycolor2}]
  table[row sep=crcr]{%
10	0.14239133462685\\
15	0.080071013329691\\
20	0.04502572210014\\
25	0.0253183080727189\\
30	0.0142360173905567\\
35	0.00800399204903294\\
40	0.00449947498040703\\
45	0.00252875496037782\\
50	0.00142056375036001\\
55	0.00079742802497291\\
60	0.000500532769730955\\
};

\addplot [name path = C, color=mycolor2, line width=\lw pt, forget plot, opacity=0.1]
  table[row sep=crcr]{%
10	0.142391334627362\\
15	0.080071013330131\\
20	0.0450257221003304\\
25	0.0253183080728014\\
30	0.0142360173905732\\
35	0.00800399204903913\\
40	0.00449947498040841\\
45	0.00252875496037832\\
50	0.00142056375036018\\
55	0.00079742802497297\\
60	0.000500532769730971\\
};
\addplot [name path = D, color=mycolor2, line width=\lw pt, forget plot, opacity=0.1]
  table[row sep=crcr]{%
10	0.14239133462605\\
15	0.0800710133293943\\
20	0.0450257220999505\\
25	0.0253183080726632\\
30	0.0142360173905313\\
35	0.00800399204902926\\
40	0.00449947498040552\\
45	0.00252875496037745\\
50	0.00142056375035983\\
55	0.000797428024972848\\
60	0.000500532769730935\\
};

\addplot[fill = mycolor2, area legend, fill opacity=0.3] fill between[of=C and D];

\addplot [color=mycolor2, dashed, line width= \lw pt]
  table[row sep=crcr]{%
10	0.140982170131915\\
15	0.0792801003801922\\
20	0.0445824767090212\\
25	0.0250705690328188\\
30	0.0140982170131915\\
35	0.00792801003801922\\
40	0.00445824767090212\\
45	0.00250705690328189\\
50	0.00140982170131915\\
55	0.000792801003801922\\
60	0.000445824767090212\\
};

\addplot [color=green, line width=0.5 pt, mark size=\ms pt, mark=diamond, mark options={solid, green}]
  table[row sep=crcr]{%
10	0.00302389141389325\\
15	0.00170042685774403\\
20	0.000956188063127644\\
25	0.000537671864626209\\
30	0.000302322966971052\\
35	0.000169976655513348\\
40	9.5553032043293e-05\\
45	5.37018662867974e-05\\
50	3.01677804963385e-05\\
55	1.69345681338941e-05\\
60	1.06295565578392e-05\\
};

\addplot [name path = E, color=green, line width=\lw pt, forget plot, opacity=0.1]
  table[row sep=crcr]{%
10	0.00324069778500694\\
15	0.00182234372773415\\
20	0.00102474464658052\\
25	0.000576221756095207\\
30	0.00032399885952202\\
35	0.000182163608287717\\
40	0.000102403974517962\\
45	5.75521721205934e-05\\
50	3.23307440815779e-05\\
55	1.81487394651197e-05\\
60	1.13916724106989e-05\\
};
\addplot [name path = F, color=mycolor2, line width=\lw pt, forget plot, opacity=0.1]
  table[row sep=crcr]{%
10	0.00285881496080393\\
15	0.00160759930675308\\
20	0.00090398905451831\\
25	0.000508319962659554\\
30	0.00028581893417221\\
35	0.000160697505054698\\
40	9.03367218480382e-05\\
45	5.07702419665018e-05\\
50	2.85208992032366e-05\\
55	1.60100976223876e-05\\
60	1.00492812587928e-05\\
};
\addplot[fill = green, area legend, fill opacity=0.3] fill between[of=E and F];

\addplot [color=green, dashed, line width= \lw pt]
  table[row sep=crcr]{%
10	0.00299396571351279\\
15	0.00168363064691125\\
20	0.000946775089115158\\
25	0.000532410758270229\\
30	0.000299396571351279\\
35	0.000168363064691125\\
40	9.46775089115158e-05\\
45	5.32410758270229e-05\\
50	2.99396571351279e-05\\
55	1.68363064691125e-05\\
60	9.46775089115159e-06\\
};

\end{axis}

\end{tikzpicture}%
    \vspace{-1. cm}
  \centerline{\footnotesize{(b) DFT-S-OFDM}} \medskip
\end{minipage}
\vspace{-0.3cm}
\caption{The effect of PA on channel parameters estimation using (a) OFDM, and (b) DFT-S-OFDM.}
\label{fig:pa_ofdm}
\vspace{-0.cm}
\end{figure}

High \ac{papr} is one of the critical issues in implementing the OFDM signals, and a promising alternative is to use DFT-S-OFDM~\cite{berardinelli2017generalized}. When increasing the transmit power, the PAN is more likely to happen, as can be seen in Fig.~\ref{fig:pa_ofdm} (a). While delay estimation exploits phase changes across subcarriers within an OFDM symbol, angle estimation relies on phase/amplitude changes across multiple symbols in an OFDM frame for analog arrays, where beam sweeping is performed over time (i.e., different beams for different symbols). In addition, signals at different antenna elements experience similar distortions with identical PAs adopted in this work (see Fig. 1). Therefore, the effect of signal distortion due to PAN is less pronounced (at the same level of transmit power) for angle estimation than for delay estimation. We compare using the random OFDM symbols and the FFT version of the benchmark symbols (a special case of DFT-S-OFDM by choosing an identity mapping matrix~\cite{berardinelli2017generalized}), and the results are shown in Fig.~\ref{fig:pa_ofdm}. Due to the reduced \ac{papr} by DFT-S-OFDM, the localization performance can be improved, as shown in the right figure. 

\begin{figure}
\begin{minipage}[b]{0.48\linewidth}
  \centering
%
%
\newcommand\ms{2}
\newcommand\lw{1}
\definecolor{mycolor1}{rgb}{0.00000,1.00000,1.00000}%
\definecolor{mycolor2}{rgb}{1.00000,0.00000,1.00000}%

\pgfplotsset{every tick label/.append style={font=\tiny}}

\begin{tikzpicture}[font=\footnotesize]
\begin{axis}[%
width=3 cm,
height=2.2 cm,
at={(0in,0in)},
scale only axis,
xmin=-1.1,
xmax=1.1,
xlabel style={font=\footnotesize, yshift=-0.5 ex},
xlabel={$10\text{log}(c_\text{HWI})$},
ymode=log,
ymin=1e-6,
ymax=1,
ylabel style={font=\footnotesize, yshift=-0.5 ex},
ylabel={PALB [m]}, axis background/.style={fill=white},
xmajorgrids,
ymajorgrids,
legend columns=7, 
legend style={font=\tiny, at={(-0.1, 1.05)}, anchor=south west, legend cell align=left, align=left, draw=white!15!black}
]
\addplot [color=black, line width=\lw pt]
  table[row sep=crcr]{%
-1	0.000106641546929495\\
-0.6	0.000366626548888829\\
-0.2	0.000720203798761509\\
0.2	0.00186785568769502\\
0.6	0.00530952451996656\\
1	0.0465630510549273\\
};
\addlegendentry{ALL}

\addplot [color=red, line width=\lw pt, mark=square, mark options={solid, red}, mark size=\ms pt]
  table[row sep=crcr]{%
-1	3.74775827681621e-05\\
-0.6	0.000108501621064923\\
-0.2	0.000267342493306586\\
0.2	0.000468274561652524\\
0.6	0.00254746934001569\\
1	0.00637887514238846\\
};
\addlegendentry{PN}

\addplot [color=blue, line width=\lw pt, mark=o, mark options={solid, blue}, mark size=\ms pt]
  table[row sep=crcr]{%
-1	3.23816602379467e-06\\
-0.6	2.1176377969982e-05\\
-0.2	4.99056172413719e-05\\
0.2	0.000612442094288223\\
0.6	0.00652883149425217\\
1	0.0528203286893003\\
};
\addlegendentry{CFO}

\addplot [color=green, line width=\lw pt, mark=diamond, mark options={solid, green}, mark size=\ms pt]
  table[row sep=crcr]{%
-1	5.22107691160377e-05\\
-0.6	0.000109163751397785\\
-0.2	0.000355871671183371\\
0.2	0.000770810031315166\\
0.6	0.00223639879379148\\
1	0.00548645541355731\\
};
\addlegendentry{MC}

\addplot [color=black, dashed, line width=\lw pt, mark=triangle, mark options={solid, rotate=270, black}, mark size=\ms pt]
  table[row sep=crcr]{%
-1	9.85307288440783e-06\\
-0.6	2.46478008719546e-05\\
-0.2	7.33681629335129e-05\\
0.2	0.000136562128255398\\
0.6	0.000429865487962871\\
1	0.000867857931969222\\
};
\addlegendentry{AGE}

\addplot [color=mycolor1, dashed, line width=\lw pt, mark=asterisk, mark options={solid, mycolor1}, mark size=\ms pt]
  table[row sep=crcr]{%
-1	8.29850792219036e-05\\
-0.6	0.000183979775629682\\
-0.2	0.00063054963262438\\
0.2	0.00137844156330949\\
0.6	0.00266900938576278\\
1	0.0103975050081308\\
};
\addlegendentry{ADE}

\addplot [color=mycolor2, dashed, line width=\lw pt, mark=+, mark options={solid, mycolor2}, mark size=\ms pt]
  table[row sep=crcr]{%
-1	1.4294303352433e-05\\
-0.6	3.63294843451538e-05\\
-0.2	0.000124163918300523\\
0.2	0.00026198752058562\\
0.6	0.000952644501566435\\
1	0.0023442615123014\\
};
\addlegendentry{IQI}

\end{axis}

\end{tikzpicture}%
  \vspace{-1. cm}
  \centerline{\footnotesize{(a) PALB}} \medskip
\end{minipage}
\hfill
\begin{minipage}[b]{0.48\linewidth}
  \centering
%
%
\newcommand\ms{2}
\newcommand\lw{1}
\definecolor{mycolor1}{rgb}{0.00000,1.00000,1.00000}%
\definecolor{mycolor2}{rgb}{1.00000,0.00000,1.00000}%

\pgfplotsset{every tick label/.append style={font=\tiny}}

\begin{tikzpicture}[font=\footnotesize]
\begin{axis}[%
width=3cm,
height=2.2 cm,
at={(0in,0in)},
scale only axis,
xmin=-1.1,
xmax=1.1,
xlabel style={font=\footnotesize, yshift=-0.5 ex},
xlabel={$10\text{log}(c_\text{HWI})$},
ymode=log,
ymin=1e-6,
ymax=1,
ylabel style={font=\footnotesize, yshift=-0.5 ex},
ylabel={OALB}, axis background/.style={fill=white},
xmajorgrids,
ymajorgrids,
legend columns=4, 
legend style={font=\tiny, at={(0, 1)}, anchor=north west, legend cell align=left, align=left, draw=white!15!black}
]

\addplot [color=black, line width=\lw pt]
  table[row sep=crcr]{%
-1	8.82530747752241e-05\\
-0.6	0.000261062176864613\\
-0.2	0.00059061643231928\\
0.2	0.00167694798130391\\
0.6	0.0048041730402506\\
1	0.0388317213477869\\
};

\addplot [color=red, line width=\lw pt, mark=square, mark options={solid, red}, mark size=\ms pt]
  table[row sep=crcr]{%
-1	3.75373219316799e-05\\
-0.6	0.000107866861979389\\
-0.2	0.000286035532106155\\
0.2	0.000566569280053386\\
0.6	0.00167726370306021\\
1	0.00606764168814244\\
};

\addplot [color=blue, line width=\lw pt, mark=o, mark options={solid, blue}, mark size=\ms pt]
  table[row sep=crcr]{%
-1	2.72087394215248e-06\\
-0.6	1.828841715934e-05\\
-0.2	6.03108679726461e-05\\
0.2	0.000496519728464762\\
0.6	0.00552254361754834\\
1	0.0457578915226193\\
};

\addplot [color=green, line width=\lw pt, mark=diamond, mark options={solid, green}, mark size=\ms pt]
  table[row sep=crcr]{%
-1	4.21975788142959e-05\\
-0.6	8.66844877460693e-05\\
-0.2	0.000232119078196646\\
0.2	0.000467596379891688\\
0.6	0.00177936964947278\\
1	0.00455646664424414\\
};

\addplot [color=black, dashed, line width=\lw pt, mark=triangle, mark options={solid, rotate=270, black}, mark size=\ms pt]
  table[row sep=crcr]{%
-1	8.2473471305633e-06\\
-0.6	2.09838930320672e-05\\
-0.2	6.09185476303509e-05\\
0.2	0.000115427074382168\\
0.6	0.00033569719361104\\
1	0.000735231534205738\\
};

\addplot [color=mycolor1, dashed, line width=\lw pt, mark=asterisk, mark options={solid, mycolor1}, mark size=\ms pt]
  table[row sep=crcr]{%
-1	5.74050528996395e-05\\
-0.6	0.00013530329452216\\
-0.2	0.000477148510929837\\
0.2	0.00109486527692759\\
0.6	0.00210417424892739\\
1	0.00887278367163802\\
};

\addplot [color=mycolor2, dashed, line width=\lw pt, mark=+, mark options={solid, mycolor2}, mark size=\ms pt]
  table[row sep=crcr]{%
-1	1.23956600699415e-05\\
-0.6	4.95718801580836e-05\\
-0.2	9.06068258981207e-05\\
0.2	0.000253339158811983\\
0.6	0.000821443382575282\\
1	0.00183131751253995\\
};

\end{axis}

\end{tikzpicture}%
  \vspace{-1. cm}
  \centerline{\footnotesize{(b) OALB}} \medskip
\end{minipage}
\vspace{-0.3 cm}
\caption{An example of ALB with different levels of impairments: (a) PALB, (b) OALB. The ALBs of the position and orientation affected by the HWIs increase with $c_\text{HWI}$ (reflecting the impairment level).}
\label{fig-different_HWI_level_all}
\end{figure}
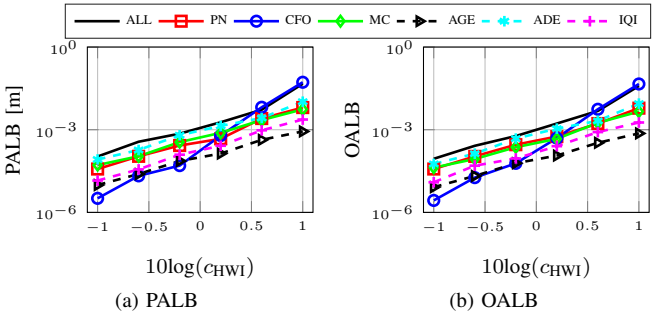

\subsubsection{Evaluation of HWIs with Different Impairment Levels}
We further evaluate the position and orientation \acp{alb} with different levels of HWIs by defining an impairment coefficient $c_\text{HWI}$. With different values of $c_\text{HWI}$, the position ALB and orientation ALB, by considering all the HWIs, and individual HWIs, are shown in Fig.~\ref{fig-different_HWI_level_all} (a) and (b). {All the results indicate the 75th percentile of the total 100 realizations.} We notice that the effect of PN, MC, AGE, ADE, and IQI on the localization increases approximately in a linear trend with impairment level. The CFO has a larger effect in high impairment levels as the error residue accumulates over time.
Based on Fig.~\ref{fig-different_HWI_level_all}, we can quantize the contribution of individual HWIs (e.g., if the ALBs are much smaller than the current CRB, the negative contribution of HWI on localization is negligible). In addition, it can also identify dominant impairment factors for further compensation (e.g., ADE is one of the dominant factors under current system parameters).

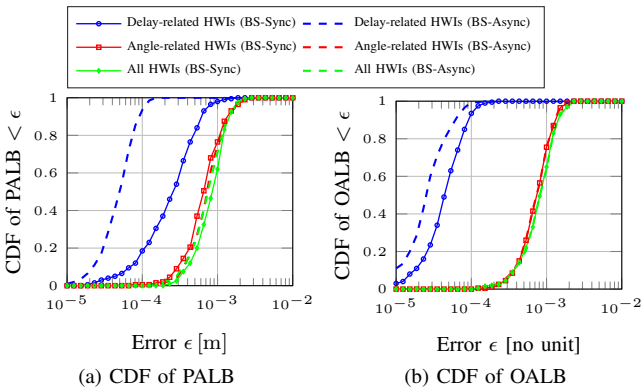
\begin{figure}
\vspace{0.3 cm}
\begin{minipage}[b]{0.48\linewidth}
\vspace{-0.5 cm}
%
%
\newcommand\ms{0.8}
\newcommand\lw{0.3}
\definecolor{mycolor1}{rgb}{0.00000,1.00000,1.00000}%
\definecolor{mycolor2}{rgb}{1.00000,0.00000,1.00000}%

\pgfplotsset{every tick label/.append style={font=\tiny}}

\begin{tikzpicture}[font=\footnotesize]
\begin{axis}[%
width=3cm,
height=2.5cm,
at={(0in,0in)},
scale only axis,
xmin=1e-5,
xmax=0.01,
xmode=log,
xlabel style={font=\footnotesize, yshift=-0.5 ex},
xlabel={Error $\unit[\epsilon]{[m]}$},
ymin=0,
ymax=1,
ylabel style={font=\footnotesize, yshift=-0.5 ex},
ylabel={CDF of $\text{PALB}<\epsilon$},
axis background/.style={fill=white},
xmajorgrids,
ymajorgrids,
legend columns=2, 
legend style={font=\tiny, at={(0, 1.05)}, anchor=south west, legend cell align=left, align=left, draw=white!15!black}
]

\addplot [color=blue, line width=0.5 pt, mark size=\ms pt, mark=o, mark options={solid, blue}]
  table[row sep=crcr]{%
1e-05	0\\
1.23284673944207e-05	0\\
1.51991108295293e-05	0\\
1.87381742286038e-05	0.005\\
2.31012970008316e-05	0.015\\
2.8480358684358e-05	0.03\\
3.51119173421513e-05	0.04\\
4.32876128108306e-05	0.05\\
5.33669923120631e-05	0.0649999999999999\\
6.57933224657568e-05	0.095\\
8.11130830789687e-05	0.12\\
0.0001	0.185\\
0.000123284673944207	0.23\\
0.000151991108295293	0.305\\
0.000187381742286038	0.37\\
0.000231012970008316	0.46\\
0.00028480358684358	0.54\\
0.000351119173421513	0.665\\
0.000432876128108306	0.765\\
0.000533669923120631	0.84\\
0.000657933224657568	0.93\\
0.000811130830789687	0.965\\
0.001	0.98\\
0.00123284673944207	0.99\\
0.00151991108295293	0.995\\
0.00187381742286039	1\\
0.00231012970008316	1\\
0.0028480358684358	1\\
0.00351119173421513	1\\
0.00432876128108306	1\\
0.00533669923120631	1\\
0.00657933224657568	1\\
0.00811130830789687	1\\
0.01	1\\
};
\addlegendentry{Delay-related HWIs (BS-Sync)}

\addplot [color=blue, dashed, line width=0.8 pt]
  table[row sep=crcr]{%
1e-05	0.01\\
1.23284673944207e-05	0.02\\
1.51991108295293e-05	0.035\\
1.87381742286038e-05	0.0649999999999999\\
2.31012970008316e-05	0.11\\
2.8480358684358e-05	0.17\\
3.51119173421513e-05	0.285\\
4.32876128108306e-05	0.41\\
5.33669923120631e-05	0.545\\
6.57933224657568e-05	0.72\\
8.11130830789687e-05	0.855\\
0.0001	0.935\\
0.000123284673944207	0.985\\
0.000151991108295293	1\\
0.000187381742286038	1\\
0.000231012970008316	1\\
0.00028480358684358	1\\
0.000351119173421513	1\\
0.000432876128108306	1\\
0.000533669923120631	1\\
0.000657933224657568	1\\
0.000811130830789687	1\\
0.001	1\\
0.00123284673944207	1\\
0.00151991108295293	1\\
0.00187381742286039	1\\
0.00231012970008316	1\\
0.0028480358684358	1\\
0.00351119173421513	1\\
0.00432876128108306	1\\
0.00533669923120631	1\\
0.00657933224657568	1\\
0.00811130830789687	1\\
0.01	1\\
};
\addlegendentry{Delay-related HWIs (BS-Async)}

\addplot [color=red, line width=0.5 pt, mark size=\ms pt, mark=square, mark options={solid, red}]
  table[row sep=crcr]{%
1e-05	0\\
1.23284673944207e-05	0\\
1.51991108295293e-05	0\\
1.87381742286038e-05	0\\
2.31012970008316e-05	0\\
2.8480358684358e-05	0\\
3.51119173421513e-05	0\\
4.32876128108306e-05	0\\
5.33669923120631e-05	0\\
6.57933224657568e-05	0\\
8.11130830789687e-05	0.005\\
0.0001	0.005\\
0.000123284673944207	0.005\\
0.000151991108295293	0.015\\
0.000187381742286038	0.02\\
0.000231012970008316	0.045\\
0.00028480358684358	0.09\\
0.000351119173421513	0.145\\
0.000432876128108306	0.205\\
0.000533669923120631	0.37\\
0.000657933224657568	0.505\\
0.000811130830789687	0.68\\
0.001	0.765\\
0.00123284673944207	0.875\\
0.00151991108295293	0.93\\
0.00187381742286039	0.965\\
0.00231012970008316	0.99\\
0.0028480358684358	1\\
0.00351119173421513	1\\
0.00432876128108306	1\\
0.00533669923120631	1\\
0.00657933224657568	1\\
0.00811130830789687	1\\
0.01	1\\
};
\addlegendentry{Angle-related HWIs (BS-Sync)}

\addplot [color=red, dashed, line width=0.8 pt]
  table[row sep=crcr]{%
1e-05	0\\
1.23284673944207e-05	0\\
1.51991108295293e-05	0\\
1.87381742286038e-05	0\\
2.31012970008316e-05	0\\
2.8480358684358e-05	0\\
3.51119173421513e-05	0\\
4.32876128108306e-05	0\\
5.33669923120631e-05	0\\
6.57933224657568e-05	0\\
8.11130830789687e-05	0.005\\
0.0001	0.005\\
0.000123284673944207	0.005\\
0.000151991108295293	0.005\\
0.000187381742286038	0.015\\
0.000231012970008316	0.025\\
0.00028480358684358	0.04\\
0.000351119173421513	0.095\\
0.000432876128108306	0.17\\
0.000533669923120631	0.27\\
0.000657933224657568	0.405\\
0.000811130830789687	0.61\\
0.001	0.75\\
0.00123284673944207	0.855\\
0.00151991108295293	0.925\\
0.00187381742286039	0.965\\
0.00231012970008316	0.99\\
0.0028480358684358	1\\
0.00351119173421513	1\\
0.00432876128108306	1\\
0.00533669923120631	1\\
0.00657933224657568	1\\
0.00811130830789687	1\\
0.01	1\\
};
\addlegendentry{Angle-related HWIs (BS-Async)}

\addplot [color=green, line width=0.5 pt, mark size=\ms pt, mark=diamond, mark options={solid, green}]
  table[row sep=crcr]{%
1e-05	0\\
1.23284673944207e-05	0\\
1.51991108295293e-05	0\\
1.87381742286038e-05	0\\
2.31012970008316e-05	0\\
2.8480358684358e-05	0\\
3.51119173421513e-05	0\\
4.32876128108306e-05	0\\
5.33669923120631e-05	0\\
6.57933224657568e-05	0\\
8.11130830789687e-05	0\\
0.0001	0\\
0.000123284673944207	0\\
0.000151991108295293	0\\
0.000187381742286038	0\\
0.000231012970008316	0.01\\
0.00028480358684358	0.025\\
0.000351119173421513	0.075\\
0.000432876128108306	0.12\\
0.000533669923120631	0.2\\
0.000657933224657568	0.325\\
0.000811130830789687	0.465\\
0.001	0.62\\
0.00123284673944207	0.83\\
0.00151991108295293	0.925\\
0.00187381742286039	0.975\\
0.00231012970008316	0.995\\
0.0028480358684358	1\\
0.00351119173421513	1\\
0.00432876128108306	1\\
0.00533669923120631	1\\
0.00657933224657568	1\\
0.00811130830789687	1\\
0.01	1\\
};
\addlegendentry{All HWIs (BS-Sync)}

\addplot [color=green, dashed, line width=0.8 pt]
  table[row sep=crcr]{%
1e-05	0\\
1.23284673944207e-05	0\\
1.51991108295293e-05	0\\
1.87381742286038e-05	0\\
2.31012970008316e-05	0\\
2.8480358684358e-05	0\\
3.51119173421513e-05	0\\
4.32876128108306e-05	0\\
5.33669923120631e-05	0\\
6.57933224657568e-05	0\\
8.11130830789687e-05	0\\
0.0001	0\\
0.000123284673944207	0.005\\
0.000151991108295293	0.005\\
0.000187381742286038	0.02\\
0.000231012970008316	0.03\\
0.00028480358684358	0.05\\
0.000351119173421513	0.105\\
0.000432876128108306	0.16\\
0.000533669923120631	0.26\\
0.000657933224657568	0.405\\
0.000811130830789687	0.565\\
0.001	0.71\\
0.00123284673944207	0.845\\
0.00151991108295293	0.925\\
0.00187381742286039	0.98\\
0.00231012970008316	0.995\\
0.0028480358684358	1\\
0.00351119173421513	1\\
0.00432876128108306	1\\
0.00533669923120631	1\\
0.00657933224657568	1\\
0.00811130830789687	1\\
0.01	1\\
};
\addlegendentry{All HWIs (BS-Async)}





\end{axis}

\end{tikzpicture}%
    \vspace{-1. cm}
  \centerline{{\footnotesize{(a) CDF of PALB}}} \medskip
\end{minipage}
\begin{minipage}[b]{0.48\linewidth}
\vspace{-0.5 cm}
%
%
\newcommand\ms{0.8}
\newcommand\lw{0.3}
\definecolor{mycolor1}{rgb}{0.00000,1.00000,1.00000}%
\definecolor{mycolor2}{rgb}{1.00000,0.00000,1.00000}%

\pgfplotsset{every tick label/.append style={font=\tiny}}

\begin{tikzpicture}[font=\footnotesize]
\begin{axis}[%
width=3cm,
height=2.5cm,
at={(0in,0in)},
scale only axis,
xmin=1e-5,
xmax=0.01,
xlabel style={font=\footnotesize, yshift=-0.5 ex},
xlabel={Error $\epsilon$ [no unit]},
xmode=log,
ymin=0,
ymax=1,
ylabel style={font=\footnotesize, yshift=-0.5 ex},
ylabel={CDF of $\text{OALB} < \epsilon$},
axis background/.style={fill=white},
xmajorgrids,
ymajorgrids,
legend columns=2, 
legend style={font=\tiny, at={(1.0,1)}, anchor=south east, legend cell align=left, align=left, draw=white!15!black}
]

\addplot [color=blue, line width=0.5 pt, mark size=\ms pt, mark=o, mark options={solid, blue}]
  table[row sep=crcr]{%
1e-05	0.03\\
1.23284673944207e-05	0.04\\
1.51991108295293e-05	0.08\\
1.87381742286038e-05	0.11\\
2.31012970008316e-05	0.16\\
2.8480358684358e-05	0.235\\
3.51119173421513e-05	0.34\\
4.32876128108306e-05	0.49\\
5.33669923120631e-05	0.63\\
6.57933224657568e-05	0.74\\
8.11130830789687e-05	0.85\\
0.0001	0.935\\
0.000123284673944207	0.975\\
0.000151991108295293	0.99\\
0.000187381742286038	0.995\\
0.000231012970008316	1\\
0.00028480358684358	1\\
0.000351119173421513	1\\
0.000432876128108306	1\\
0.000533669923120631	1\\
0.000657933224657568	1\\
0.000811130830789687	1\\
0.001	1\\
0.00123284673944207	1\\
0.00151991108295293	1\\
0.00187381742286039	1\\
0.00231012970008316	1\\
0.0028480358684358	1\\
0.00351119173421513	1\\
0.00432876128108306	1\\
0.00533669923120631	1\\
0.00657933224657568	1\\
0.00811130830789687	1\\
0.01	1\\
};

\addplot [color=blue, dashed, line width=0.8 pt]
  table[row sep=crcr]{%
1e-05	0.11\\
1.23284673944207e-05	0.135\\
1.51991108295293e-05	0.2\\
1.87381742286038e-05	0.285\\
2.31012970008316e-05	0.43\\
2.8480358684358e-05	0.6\\
3.51119173421513e-05	0.71\\
4.32876128108306e-05	0.795\\
5.33669923120631e-05	0.86\\
6.57933224657568e-05	0.92\\
8.11130830789687e-05	0.965\\
0.0001	0.995\\
0.000123284673944207	0.995\\
0.000151991108295293	1\\
0.000187381742286038	1\\
0.000231012970008316	1\\
0.00028480358684358	1\\
0.000351119173421513	1\\
0.000432876128108306	1\\
0.000533669923120631	1\\
0.000657933224657568	1\\
0.000811130830789687	1\\
0.001	1\\
0.00123284673944207	1\\
0.00151991108295293	1\\
0.00187381742286039	1\\
0.00231012970008316	1\\
0.0028480358684358	1\\
0.00351119173421513	1\\
0.00432876128108306	1\\
0.00533669923120631	1\\
0.00657933224657568	1\\
0.00811130830789687	1\\
0.01	1\\
};

\addplot [color=red, line width=0.5 pt, mark size=\ms pt, mark=square, mark options={solid, red}]
  table[row sep=crcr]{%
1e-05	0\\
1.23284673944207e-05	0\\
1.51991108295293e-05	0\\
1.87381742286038e-05	0\\
2.31012970008316e-05	0\\
2.8480358684358e-05	0\\
3.51119173421513e-05	0\\
4.32876128108306e-05	0\\
5.33669923120631e-05	0\\
6.57933224657568e-05	0\\
8.11130830789687e-05	0\\
0.0001	0\\
0.000123284673944207	0.005\\
0.000151991108295293	0.005\\
0.000187381742286038	0.01\\
0.000231012970008316	0.025\\
0.00028480358684358	0.045\\
0.000351119173421513	0.085\\
0.000432876128108306	0.15\\
0.000533669923120631	0.26\\
0.000657933224657568	0.415\\
0.000811130830789687	0.565\\
0.001	0.755\\
0.00123284673944207	0.87\\
0.00151991108295293	0.965\\
0.00187381742286039	0.99\\
0.00231012970008316	1\\
0.0028480358684358	1\\
0.00351119173421513	1\\
0.00432876128108306	1\\
0.00533669923120631	1\\
0.00657933224657568	1\\
0.00811130830789687	1\\
0.01	1\\
};

\addplot [color=red, dashed, line width=0.8 pt]
  table[row sep=crcr]{%
1e-05	0\\
1.23284673944207e-05	0\\
1.51991108295293e-05	0\\
1.87381742286038e-05	0\\
2.31012970008316e-05	0\\
2.8480358684358e-05	0\\
3.51119173421513e-05	0\\
4.32876128108306e-05	0\\
5.33669923120631e-05	0\\
6.57933224657568e-05	0\\
8.11130830789687e-05	0\\
0.0001	0\\
0.000123284673944207	0.005\\
0.000151991108295293	0.005\\
0.000187381742286038	0.015\\
0.000231012970008316	0.02\\
0.00028480358684358	0.055\\
0.000351119173421513	0.09\\
0.000432876128108306	0.15\\
0.000533669923120631	0.265\\
0.000657933224657568	0.425\\
0.000811130830789687	0.565\\
0.001	0.755\\
0.00123284673944207	0.875\\
0.00151991108295293	0.975\\
0.00187381742286039	0.99\\
0.00231012970008316	1\\
0.0028480358684358	1\\
0.00351119173421513	1\\
0.00432876128108306	1\\
0.00533669923120631	1\\
0.00657933224657568	1\\
0.00811130830789687	1\\
0.01	1\\
};

\addplot [color=green, line width=0.5 pt, mark size=\ms pt, mark=diamond, mark options={solid, green}]
  table[row sep=crcr]{%
1e-05	0\\
1.23284673944207e-05	0\\
1.51991108295293e-05	0\\
1.87381742286038e-05	0\\
2.31012970008316e-05	0\\
2.8480358684358e-05	0\\
3.51119173421513e-05	0\\
4.32876128108306e-05	0\\
5.33669923120631e-05	0\\
6.57933224657568e-05	0\\
8.11130830789687e-05	0\\
0.0001	0\\
0.000123284673944207	0.005\\
0.000151991108295293	0.015\\
0.000187381742286038	0.02\\
0.000231012970008316	0.025\\
0.00028480358684358	0.045\\
0.000351119173421513	0.09\\
0.000432876128108306	0.14\\
0.000533669923120631	0.22\\
0.000657933224657568	0.33\\
0.000811130830789687	0.495\\
0.001	0.65\\
0.00123284673944207	0.83\\
0.00151991108295293	0.93\\
0.00187381742286039	0.97\\
0.00231012970008316	1\\
0.0028480358684358	1\\
0.00351119173421513	1\\
0.00432876128108306	1\\
0.00533669923120631	1\\
0.00657933224657568	1\\
0.00811130830789687	1\\
0.01	1\\
};

\addplot [color=green, dashed, line width=0.8 pt]
  table[row sep=crcr]{%
1e-05	0\\
1.23284673944207e-05	0\\
1.51991108295293e-05	0\\
1.87381742286038e-05	0\\
2.31012970008316e-05	0\\
2.8480358684358e-05	0\\
3.51119173421513e-05	0\\
4.32876128108306e-05	0\\
5.33669923120631e-05	0\\
6.57933224657568e-05	0\\
8.11130830789687e-05	0\\
0.0001	0.005\\
0.000123284673944207	0.01\\
0.000151991108295293	0.01\\
0.000187381742286038	0.02\\
0.000231012970008316	0.03\\
0.00028480358684358	0.045\\
0.000351119173421513	0.09\\
0.000432876128108306	0.14\\
0.000533669923120631	0.205\\
0.000657933224657568	0.34\\
0.000811130830789687	0.495\\
0.001	0.64\\
0.00123284673944207	0.83\\
0.00151991108295293	0.93\\
0.00187381742286039	0.97\\
0.00231012970008316	1\\
0.0028480358684358	1\\
0.00351119173421513	1\\
0.00432876128108306	1\\
0.00533669923120631	1\\
0.00657933224657568	1\\
0.00811130830789687	1\\
0.01	1\\
};

\end{axis}

\end{tikzpicture}%
    \vspace{-1. cm}
  \centerline{{\footnotesize{(b) CDF of OALB}}} \medskip
\end{minipage}
\vspace{-0.3cm}
\caption{{The effect of HWIs on different localization systems.}}
\label{fig:CDF_PALB_OALB}
\vspace{-0.3cm}
\end{figure}

{
\subsubsection{The Effect of HWIs on Different Localization Systems}

We compare the default BS-synchronized system (i.e., BS-Sync) with the system without BS synchronization (i.e., BS-Async). Both systems require AOA estimations for positioning, but the BS-Sync system provides extra TDOA information. We first plot the CDF of PALB of two systems considering different types of HWIs (similar to Fig.~\ref{fig-different_HWI_level_all}) $c_\text{HWI} =1$, as shown in Fig.~\ref{fig:CDF_PALB_OALB} (a). We can see the angle-based system performs slightly better with all types of HWIs  (see the dashed green curve and solid green curve with diamond markers), and much better when introducing delay-related HWIs (as shown in the blue curves in Fig.~\ref{fig:CDF_PALB_OALB} (a)). This is because the erroneous delay estimation will affect the positioning performance of the BS-Sync system, indicating that the extra TDOA measurement under HWI may not always contribute to localization performance. If we only consider angle-related HWIs (i.e., CFO, MC, AGE, ADE), the BS-Sync system performs slightly better (as shown in the two red curves in Fig.~\ref{fig:CDF_PALB_OALB} (a)), indicating the synchronization does not contribute too much when the existing of large angle errors. The gap between blue curves and red curves shows that in this MIMO system, the angle-related HWIs affect localization more than communications (also see the results related to CFO, MC, AGE, ADE in Fig.~\ref{fig:sim_commun_individual} and Fig.~\ref{fig:single_impairment_on_localization}). 

Regarding the orientation, it depends on the quality of position and AOD estimates at the UE and hence shows a similar pattern with position CDF in Fig.~\ref{fig:CDF_PALB_OALB}~(a) (e.g., a better position performance indicates a better orientation performance). However, due to the contribution of AOD estimations being independent of BS synchronization, the gaps between the dashed curves and the corresponding solid curves in Fig.~\ref{fig:CDF_PALB_OALB}~(b) are reduced compared to the gaps in (a).}

\subsection{Summary of the Effects of HWIs}
From the simulation, we found that the HWIs affect both localization and {communications}, especially at high transmit power. However, different types of HWIs affect localization and {communications} differently. 
{For {communications}, the HWIs distort the transmitted and received signal and hence affect SER. Based on Fig.~\ref{fig:sim_commun_individual}, CFO, MC, AGE, and ADE have a limited effect on {communications}. The distortion introduced by CFO is a fixed phase shift that accumulates with time and can be mitigated by more frequent online compensation. Since {communications} does not exploit the phase relationship between antennas (e.g., no sweeping is needed once the communication link is established), MC, AGE, and ADE also have less impact on the SER.}

{
As for localization, the distortion of signals affects the channel parameter extraction from the estimated channel. More specifically, a bias will be introduced based on the MCRB analysis (as shown in \eqref{eq:LB_channel_parameters}), which is caused by the mismatch between the \ac{tm} and the assumed \ac{mm} (i.e., the one used to develop the algorithm). 
Such a bias will not affect the localization performance too much when the SNR is low or when the accuracy requirement is not stringent; however, it cannot be ignored in high-accuracy localization systems (see the saturation of the performance in Fig.~\ref{fig:sim_vs_crb} and Fig.~\ref{fig:sim_vs_crb_localization}).
For the angle estimation for localization, the performance is strongly affected by CFO, MC, AGE, and ADE. When talking about the TOA, it is mainly affected by the factors (e.g., PN and IQI) that also affect the SER in {communications}, as shown in Fig.~\ref{fig:single_impairment_on_localization}. 

It should be noted that the effect of CFO on AOA and AOD estimation depends on the number of transmissions and sweeping order (e.g., `BS first' preferred if AOA is more important), while the effect of PA depends on the transmit power and the nonlinear region of the amplifier (e.g., DFT-S-OFDM is preferred for a lower \ac{papr}). Since different localization systems (e.g., BS-Sync or BS-Async) and scenarios (3D or 6D localization) may treat angle and delay estimation differently, the selection of hardware (e.g., a receiver with a lower PN level) and compensation algorithms should be considered when performing localization.}
The effect of the individual impairment on angle/delay estimation and {communications} (i.e., SER) is summarized in Table~\ref{table_2:effect_of_hwi_com_loc} (H/L denotes High/Low).

\begin{table}[h]
\vspace{-0.1cm}
\scriptsize
    \centering
    \caption{The effects of HWIs on localization and {communications}}
    \vspace{-0.2cm}
    \renewcommand{\arraystretch}{1.}
    \begin{tabular}{ c|c|c|c|c } 
    \hline
    Type of HWI & AOD & AOA & TOA & SER\\
    \hline
    Phase Noise & L & L & H & H
    \\
    \hline
    Carrier Frequency Offset & H$^*$ & H$^*$ & L & L
    \\
    \hline
    Mutual Coupling & H & H & L & L
    \\
    \hline
    Power Amplifier Nonlinearity & H$^*$ & H$^*$ & H$^*$ & H$^*$
    \\
    \hline
    Array Gain Error & H & H & L & L
    \\
    \hline
    Antenna Displacement Error & H & H & L & L
    \\
    \hline
    IQ Imbalance & L & L & H & H
    \\
    \hline
    \end{tabular}
\vspace{0.2cm}
\\
\scriptsize{\raggedright 
{$^*${The effect of CFO on angle estimations depends on the sweeping order and number of transmissions. The effect of PAN depends on the transmit power and the nonlinear region of the amplifier.}} \par}
\par
\label{table_2:effect_of_hwi_com_loc}
\end{table}



\vspace{-0.1in}
\section{Conclusion}
\label{sec:conc}
As the requirements on localization and communication performance are more stringent to support new applications, \acp{hwi} become a prominent factor affecting the performance in 5G/6G systems. We have modeled different types of \acp{hwi} and utilized the \ac{mcrb} to evaluate the localization error caused by the model mismatch. The effects of HWIs on angle/delay and position/orientation estimation are evaluated. 
We found that PN and IQI have a stronger effect on delay estimation, while CFO, MC, AGE, and ADE have a more significant effect on angle estimation. The PAN affects both angle and delay, which is determined by the transmit power (or amplitude) of the signals. Furthermore, we evaluated the effect of individual HWIs on communication performance in terms of \ac{ser}. The dominant impairments that degrade SER (i.e., PN, PAN, and IQI) are in good agreement with the factors that affect delay estimation. 

In summary, the localization and communication performance that improves with transmit power in an ideal model will be saturated due to the effect of HWIs. To fully realize the potential of {localization in 5G/6G communication systems}, a dedicated pilot signal design and algorithms for estimating and mitigating HWI are needed. {Depending on the type of localization system (e.g., delay- or angle-based), the beam-sweeping order and weighting factor of delay/angle estimation should be considered.} Further works can consider the effect of HWIs in multipath and reconfigurable intelligent surface-aided scenarios, as well as the development of calibration and mitigation algorithms (including learning-based methods) to address the performance loss caused by the identified dominant impairment factors.



\appendices
\bibliographystyle{IEEEtran}
\bibliography{IEEEabrv, ref}

\end{document}